\documentclass[11pt,a4paper,oneside]{article}
\pdfoutput=1

\usepackage[affil-it]{authblk}
\usepackage[a4paper,left=2.5cm,right=2.5cm,top=2.8cm,bottom=3.7cm]{geometry}
	
\usepackage[]{amsmath}
	\numberwithin{equation}{section}
\usepackage[]{amssymb}
\usepackage{amsfonts}
\usepackage{mathrsfs}

\DeclareMathOperator{\R}{\mathbb{R}}
\DeclareMathOperator{\C}{\mathbb{C}}

\newcommand{\mN}{\mathcal{N}}
\newcommand{\dd}{\mathrm{d}}
\newcommand{\YM}{Yang--Mills}
\newcommand{\CS}{Chern--Simons}

\usepackage[utf8]{inputenc}
\usepackage[english]{babel}

\usepackage[pdftex]{graphicx}
\usepackage[dvipsnames]{xcolor}

\usepackage[linktocpage=true,colorlinks=true,citecolor=Violet,linkcolor=Violet,urlcolor=Violet]{hyperref}

\usepackage[toc,page]{appendix}
\usepackage[nottoc]{tocbibind}
\usepackage{caption}
\usepackage{enumerate}

\usepackage{tikz}
		\usetikzlibrary{er,positioning,arrows.meta}

\usepackage[numbers,compress,square,comma]{natbib}	

\begin{document}
\bibliographystyle{myJHEP}
\captionsetup[figure]{labelfont={bf,small},labelformat={default},labelsep=period,font=small}

{\pagenumbering{roman} 

		\renewcommand*{\thefootnote}{\fnsymbol{footnote}}
	\title{\huge\textbf{Phases of five-dimensional supersymmetric gauge theories}}

	\author{Leonardo Santilli\footnote{lsantilli@fc.ul.pt}}
	\affil{\small Grupo de F\'{\i}sica Matem\'{a}tica, Departamento de Matem\'{a}tica, Faculdade de Ci\^{e}ncias, Universidade de Lisboa, Campo Grande, Edif\'{\i}cio C6, 1749-016 Lisboa, Portugal.}

	\date{ \hspace{1pt} }

	\maketitle
	\thispagestyle{empty}

	\begin{abstract}
		Five-dimensional $\mathcal{N}=1$ theories with gauge group $U(N)$, $SU(N)$, $USp(2N)$ and $SO(N)$ are studied at large rank through localization on a large sphere. The phase diagram of theories with fundamental hypermultiplets is universal and characterized by third order phase transitions, with the exception of $U(N)$, that shows both second and third order transitions. The phase diagram of theories with adjoint or (anti-)symmetric hypermultiplets is also determined and found to be universal. Moreover, Wilson loops in fundamental and antisymmetric representations of any rank are analyzed in this limit. Quiver theories are discussed as well. All the results substantiate the $\mathcal{F}$-theorem.
	\end{abstract}

	\clearpage
	\tableofcontents
	\thispagestyle{empty}
}

	\clearpage
	\pagenumbering{arabic}
	\setcounter{page}{1}
		\renewcommand*{\thefootnote}{\arabic{footnote}}
		\setcounter{footnote}{0}

	\section{Introduction}
\begin{figure}[bht]
	\centering
		\includegraphics[width=0.28\textwidth]{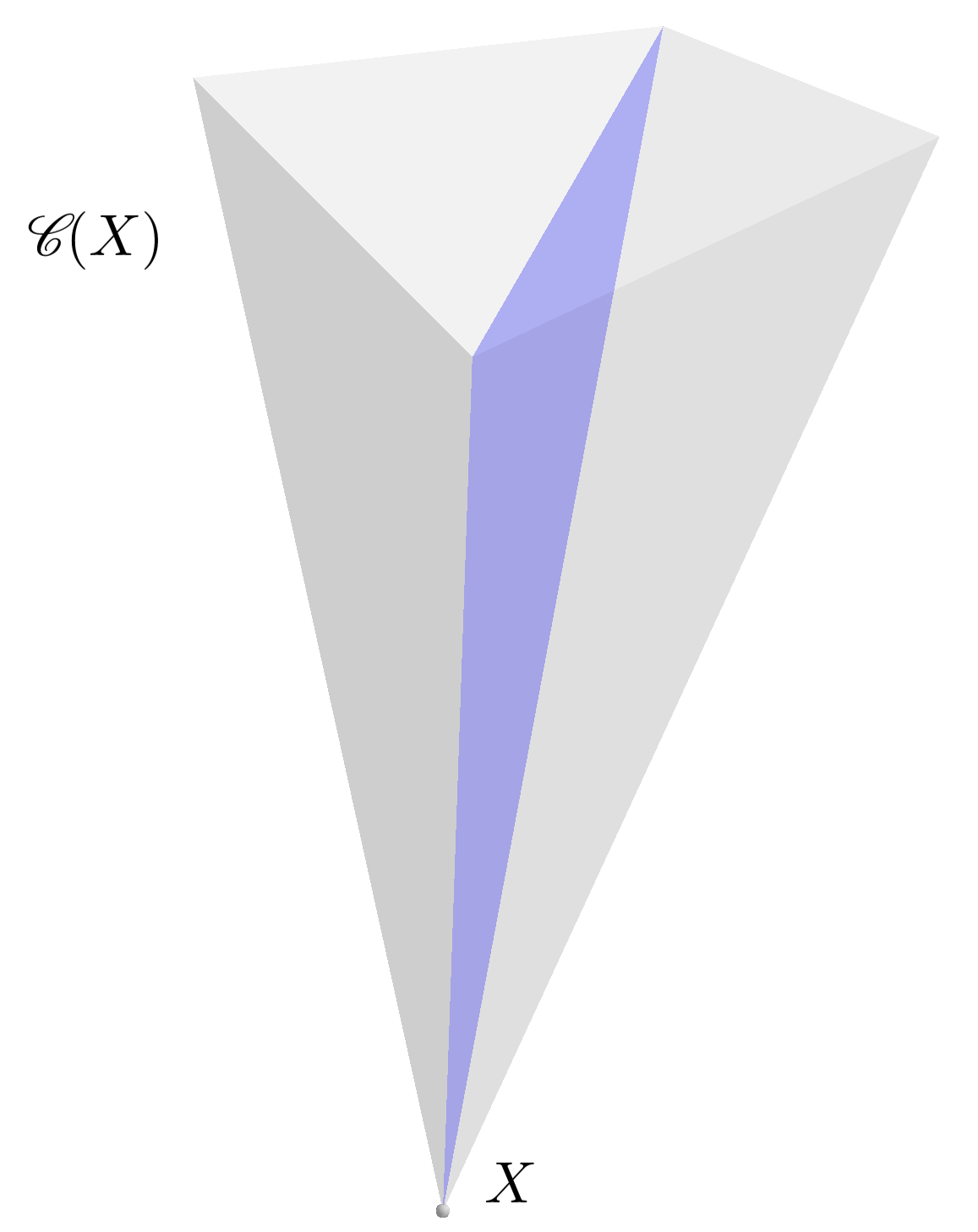}
	\caption{Schematic illustration of the extended K\"{a}hler cone $\mathscr{C} (X)$ of a singular Calabi--Yau threefold $X$. Across the blue wall separating distinct chambers, a state becomes massless.}
	\label{fig:Kcone}
	\end{figure}

	Supersymmetric quantum field theories in five and six dimensions are valuable windows onto the dynamics of interacting systems: they are constrained enough to be treated analytically, yet they are rich enough to uncover new phenomena. Five-dimensional $\mathcal{N}=1$ field theories admit UV completion at superconformal fixed points \cite{Seiberg96}, which are necessarily isolated \cite{Cordova:2016xhm} and strongly coupled \cite{Chang:2018xmx}. $5d$ $\mathcal{N}=1$ gauge theories, which are the main characters of the present work, sit in the IR of such superconformal field theories (SCFTs) and are connected to them by a renormalization group (RG) flow. The Coulomb branches of these theories have a geometric meaning inherited from M-theory compactified on a singular Calabi--Yau threefold $X$ \cite{Witten}. The extended K\"{a}hler cone of $X$, that we denote $\mathscr{C} (X)$, is the union of chambers that parametrize different crepant resolutions of $X$, as sketched in figure \ref{fig:Kcone}. In the gauge theory description, the extended  K\"{a}hler cone $\mathscr{C} (X)$ is identified with the extended Coulomb branch and the walls separating two chambers correspond to codimension-one loci on the Coulomb branch at which a state becomes massless \cite{Witten,Morrison:1996xf,Intriligator:1997pq}.\par
	\par
	Five-dimensional $\mathcal{N}=1$ \YM~theories with classical gauge group descend from $6d$ $\mathcal{N}=(1,0)$ SCFTs compactified on a circle, with subsequent RG flows triggered by massive deformations. The combination of geometric and field theoretic perspectives, integrated with new combinatorial tools \cite{Hayashi:2014kca,FF0}, yields a firm grasp of the gauge theories consistently realized within this framework \cite{Morrison:1996xf,Intriligator:1997pq,Hayashi:2014kca,FF0,a,FF1,FF2,Bhardwaj:2020gyu,Eckhard:2020jyr,c,Bhardwaj:2020avz}.\par
	The importance of weakly coupled, Lagrangian gauge theories resides in the fact that supersymmetry protected quantities carry information on the strongly interacting UV fixed point.\par
	The supersymmetric localization program \cite{Pestun:2016jze} aims at reducing the path integral description of supersymmetric observables, as the sphere partition function or the vacuum expectation value of Wilson loops, to finite-dimensional integrals. In recent years, a wealth of exact results has been obtained from localization on a broad variety of compact manifolds. Localization on the five-sphere has been carried out in \cite{KZ,KQZ,Hosomichi:2012ek,Kim2,Lockhart:2012vp,Mezei:2018url}, see also the review article \cite{Mreview}.\par
	\medskip
	The goal of the present work is to analyze the phase structure of the sphere partition function and of half-BPS Wilson loops at large rank of the gauge group. There exists a vast literature discussing the large $N$ behaviour of $5d$ $\mN=1$ theories on the sphere \cite{JafferisPufu,Kallen:2012zn,Minahan:2013jwa,Assel:2012nf,Giasemidis:2013oea,Chang:2017mxc,Fluder:2018chf,Crichigno:2018adf,Uhlemann,UhlemannWL}, mostly pivoting around the match with the holographic dual. A thorough analysis of the large $N$ phases of certain theories on the five-sphere appears in the work of Minahan and Nedelin \cite{Minahan:2014hwa,Minahan:2020ifb}.\par
	A fruitful approach to study the phase diagram consists in putting the theory on a very large sphere. This procedure, named decompactification limit, has been successfully applied to supersymmetric theories in $4d$ \cite{Russo:2012ay,Russo:2013qaa,Russo:2013kea,Chen:2014vka,Zarembo:2014ooa,Chen-Lin:2015dfa,Hollowood:2015oma,Russo:2019ipg} and $3d$ \cite{Barranco,Russo:2014bda,Anderson:2014hxa,Anderson:2015ioa,Anderson:2017xrv,STWL}. The first realization of a decompactification limit in $5d$ is in \cite{Nedelin}. Curvature effects are negligible from this vantage point, thus providing a reliable approximation of flat space dynamics without spoiling the computability guaranteed by localization.\par
	In this work, we undertake a systematic study of the phases of five-dimensional $\mN=1$ gauge theories in the decompactification limit. We discuss both the sphere partition function and the vacuum expectation value of Wilson loops, for various choices of gauge group and matter content. It is worthwhile to emphasize that phase transitions are a signature of systems with infinitely many degrees of freedom, whilst localization on $\mathbb{S}^5$ reduces the observables to matrix integrals over zero-modes. For this reason, the large $N$ limit is instrumental for the ensuing analysis and crucial for the appearance of critical loci in parameter space: it will be this limit, rather than the large sphere limit, to give rise to a non-analytic behaviour.\par
	\bigskip
	The paper is organized as follows. In the rest of this introductory section, we list concisely our main results and mention potential avenues for future research.\par
	The next section is the core of the subsequent analysis. In subsection \ref{sec:largeNlimit}, the most general solution to the sphere free energy is obtained, for theories with fundamental hypermultiplets. After that, we discuss half-BPS Wilson loops in the fundamental and in the antisymmetric representation and find the most general solution for these observables in subsection \ref{sec:WLlimit}. Subsection \ref{sec:adjhyper} extends the results to theories with hypermultiplets in representations of higher dimension.\par
	The sections that follow are devoted to a detailed analysis of various gauge theories with gauge group $U(N)$ in section \ref{sec:UN}, $SU(N)$ in section \ref{sec:SuN} and the other classical groups in section \ref{sec:SpN}. We study unitary quiver gauge theories in section \ref{sec:quiverGT}. In subsection \ref{sec:quivershort} we present two examples of quivers with gauge group $U(N) \times U(N)$. Long quivers, in which the number of nodes is taken large, are dealt with in subsection \ref{sec:quivers}, although without completely determining their phase structure. The text is complemented with three appendices.

	\subsection{Summary of results and outlook}
	
		Before entering the body of the paper, we summarize our main results and present related open problems.\par
		In the study of gauge theories with simple gauge group we find a rich phase diagram, with the models undergoing a phase transition each time a mass parameter is decreased below or above a characteristic scale.
		\begin{itemize}
			\item For gauge group $U(N)$ and hypermultiplets in the fundamental representation, the phase transitions are generically second order. There are, however, exceptions of two types: 
				\begin{enumerate}[(i)]
					\item\label{item:symF2} In the theory with symmetric assignment of masses the phase transitions are third order;
					\item\label{item:noYM} In absence of a \YM~term all the phase transitions are third order.
				\end{enumerate}
			\item For gauge groups $SU(N)$, $USp (2N)$, $SO(2N)$ or $SO(2N+1)$ with fundamental hypermultiplets the phase transitions are always third order.
			\item Expectation values of Wilson loops in the fundamental representation follow a perimeter law. Moreover, 
				\begin{enumerate}[--]
					\item In $U(N)$ theories, their derivative is discontinuous;
					\item They have second order discontinuities when the gauge group is $SU(N)$, $USp(2N)$, $SO(2N)$ or $SO(2N+1)$, and for unitary group in case (\ref{item:symF2}) above.
				\end{enumerate}
			\item Wilson loops in the antisymmetric representation follow a perimeter law and have discontinuities in the first derivative in $U(N)$ theories, and in the second derivative in all other theories.
			\item For any gauge group and an adjoint hypermultiplet, the theory has a second order phase transition.
		\end{itemize}
		We conclude that all gauge theories with a known UV SCFT completion belong to the same universality class. On the contrary, the critical behaviour of $U(N)$ gauge theories depends on the deformation pattern. This raises the question of what kind of UV completion they admit, if at all. It might be that only balanced theories possess a UV fixed point. Another possible explanation is that every $U(N)$ gauge theory descends from a bona fide SCFT, but the Abelian factor introduces some subtlety in the order of limits, namely strong coupling and large $N$ limit do not commute. Point (\ref{item:noYM}) above would fit in this scenario, but other puzzles would remain. In either case, a deeper understanding of these models by diverse and more refined methods is highly desirable.\par
		\medskip
		There are various directions worth pursuing to extend the results of the present work. An intriguing one is to understand the phase structure of theories with defects. A formulation of these theories on the sphere may entail a generalization of \cite{Mezei:2018url,Pan:2017zie,Wang:2020seq} along the lines of \cite{Qiu:2013pta,SST}.\par
		Gauge theories with eight supercharges can be localized on a $d$-dimensional sphere \cite{Minahan:2015any,Gorantis:2017vzz}, extending the setup of the present work away from $d=5$. It is possible to show that the continuation is analytic in $4<d<6$, but the method of subsection \ref{sec:largeNlimit} as it stands does not yield a consistent solution for non-integer $d$. In $4<d<5$, fundamental hypermultiplets become massless in real codimension higher than one, thus no phase transition is expected in that case. To investigate further the phase diagram of more general theories in non-integer dimension is an interesting open problem.\par
		Finally, a systematic understanding of the relation between phase transitions and one-form symmetries, elaborating on the observations in subsection \ref{sec:1formSSB}, is left for future work.

	\section{Gauge theories with large rank on the five-sphere}
	\label{sec:CBloc}
	
	\subsection{Coulomb branch localization and large $N$ limit}
	
	The moduli spaces of supersymmetric vacua of five-dimensional $\mN=1$ gauge theories consist of various branches. Among them, the Coulomb branch is parametrized by the zero-mode of the real scalar $\phi$ in the $\mN=1$ vector multiplet, conjugated in a Cartan subalgebra. For the sake of clarity, the ensuing exposition is based on gauge group $U(N)$, but the aspects we review hold for any compact semi-simple Lie group $G$.\par
	The Coulomb branch is a wedge inside $\R^{\text{rank} (G)}$ fixed by the choice of Weyl chamber:
	\begin{equation}
	\label{eq:defCgauge}
		 \mathscr{C}_{\text{gauge}} = \R^{^{\text{rank} (G)}} / \text{Weyl} (G) .
	\end{equation}
	It is convenient to consider the extended Coulomb branch of the theory, 
	\begin{equation}
	\label{eq:defCext}
		 \mathscr{C} (X) = \mathscr{C}_{\text{gauge}} \times \mathscr{C}_{\text{flavour}}  .
	\end{equation}
	In the left-hand side we have adopted the notation $\mathscr{C} (X) $ from M-theory on the singular Calabi--Yau threefold $X$, and on the right-hand side we have split the extended Coulomb branch into the gauge part, defined in \eqref{eq:defCgauge} and parametrized by the dynamical scalar $\phi$, and a flavour part, parametrized by the real scalar fields $\left\{ m_{\alpha} \right\}$ in a background vector multiplet for the flavour symmetry group. More generally, one may think of $\mathscr{C} (X)$ as a $\mathscr{C}_{\text{gauge}}$-fibration over the parameter space $\mathscr{C}_{\text{flavour}}$ \cite{a}.\par
	Hypermultiplet modes are massive at generic points of the Coulomb branch and become massless at codimension-one loci inside the extended Coulomb branch.\par			
	\medskip
	In this work, we analyze the phases of the $5d$ $\mN=1$ gauge theories looking at the matrix model obtained from localization on $\mathbb{S}^5$ \cite{KZ,KQZ,Kim2} (for a review, see \cite{Mreview}). 
	The partition function of the theory in its Coulomb branch localized on $\mathbb{S}^5$ is 
	\begin{equation}
	\label{eq:Zloc}
		\mathcal{Z}_{\mathbb{S}^5 } = \frac{1}{N!} \int_{- \infty} ^{+ \infty} \dd  \phi_1 \cdots \int_{- \infty} ^{+ \infty} \dd  \phi_N ~Z_{\text{class}} (\phi) Z_{\text{1-loop}}^{\text{vec}} (\phi)  Z_{\text{1-loop}}^{\text{hyp}} (\phi)   Z_{\text{inst}} (\phi) 
	\end{equation}
	where $Z_{\text{class}}$ is the classical contribution by the BPS field configuration, $Z_{\text{1-loop}}$ are the one-loop determinants and $ Z_{\text{inst}}$ contains the non-perturbative contributions from instantons on $\mathbb{P}^2 \subset \mathbb{S}^5$ \cite{KZ,KQZ,Kim2}. The integration domain has been extended from $\mathscr{C}_{\text{gauge}} \cong  \R^N / S_N$ to the whole $\R^N$ using the Weyl invariance of the integral, at the cost of a factor $\frac{1}{N!}$.\par
	The classical piece is 
	\begin{subequations}
	\begin{align}
		Z_{\text{class}} (\phi)  & = \prod_{a=1} ^{N} e^{-  V (\phi_a )} , \\
		V (\phi) & =  \frac{\pi r^3 k }{3} \phi ^3  +  \frac{ 8 \pi^3 r^3 }{g_{\text{\tiny YM}}^2 } \phi^2    .  \label{eq:Vphi}
	\end{align}
	\end{subequations}
	 $k$ is the \CS~level and $g_{\text{\tiny YM}}$ is the \YM~coupling, and we will henceforth use the notation 
	\begin{equation}
	\label{eq:defh}
		h = \frac{ 8 \pi^2 }{g_{\text{\tiny YM}} ^2} 
	\end{equation}
	for the inverse gauge coupling, with mass dimension one. The $5d$ $\mN=1$ gauge theories we study, with the exception of those with gauge group $U(N)$, are massive deformations of a UV SCFT, with $h$ determining the scale of such deformation. If the UV completion is a $6d$ $\mN=(1,0)$ theory compactified on a circle of radius $\beta$, then $h \propto \beta^{-1}$.\par
\begin{figure}[bht]
				\centering
					\includegraphics[width=0.4\textwidth]{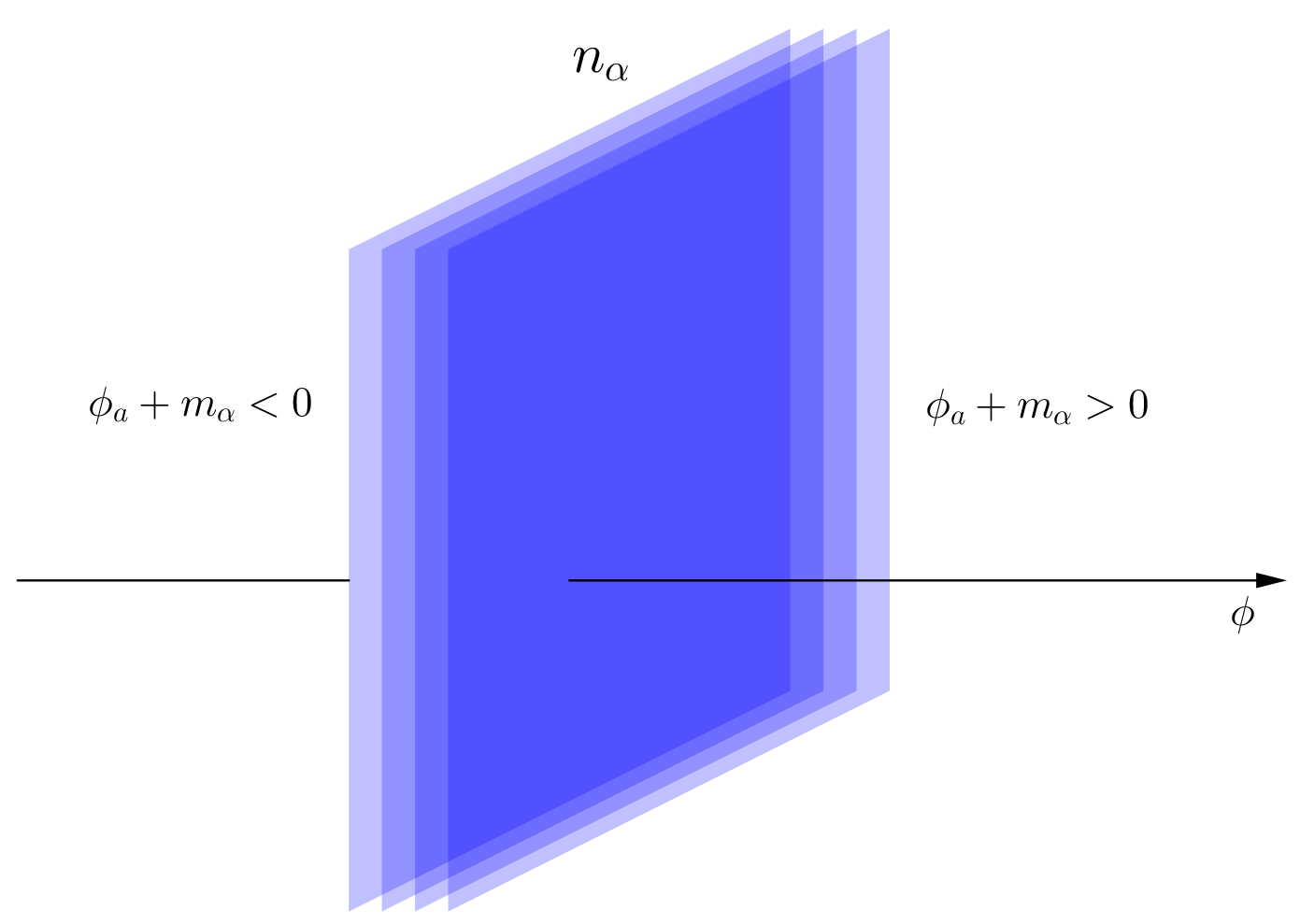}
				\caption{When the masses are degenerate, $n_{\alpha}$ walls inside the Coulomb branch collide.}
				\label{fig:degnalpha}
				\end{figure}

	The one-loop determinants for gauge group $U(N)$ or $SU(N)$ are \cite{KQZ}: 
	\begin{subequations}
	\label{eq:Z1loopfull}
	\begin{align}
		Z_{\text{1-loop}}^{\text{vec}} (\phi) & = \prod_{1 \le a < b \le N } \left[ \sinh \pi r \left( \phi_a - \phi_b \right)~ e^{\frac{1}{2} f \left( i r \left( \phi_a - \phi_b \right) \right)} \right]^{2} \label{eq:Z1loopvec} , \\
		Z_{\text{1-loop}}^{\text{hyp}} (\phi) & = \prod_{\alpha=1}^{F} \prod_{a=1}^{N}  \left[ \cosh \pi r \left(\phi_a + m_{\alpha} \right) ~ e^{- f  \left( \frac{1}{2} - i r \left( \phi_a+m_{\alpha} \right) \right) - f  \left( \frac{1}{2} + i r \left( \phi_a+m_{\alpha} \right) \right) } \right]^{\frac{n_{\alpha}}{4}} . \label{eq:Z1loophyp} 
	\end{align}
	\end{subequations}
	Here we have assumed that the matter content consists of $N_f$ fundamental hypermultiplets with degenerate masses, so that $n_{\alpha}$ of them have equal mass $m_{\alpha}$, and 
	\begin{equation}
		N_f = \sum_{\alpha=1}^{F} n_{\alpha}  .
	\end{equation}
	This choice of masses is non-generic, and the singular loci of the Coulomb branch degenerate into walls having $n_{\alpha}$ layers, as in figure \ref{fig:degnalpha}.\par
			\par
				
	The function $f (x)$ appearing in the one-loop determinants \eqref{eq:Z1loopfull} has been defined in \cite{KZ,KQZ} and comes from the zeta function regularization of the infinite product 
	\begin{equation}
		\prod_{n=1} ^{\infty} \left( 1- \frac{x^2}{n^2} \right)^{n^2} .
	\end{equation}
	For our purposes, it suffices to say that it is manifestly even, $f(-x)=f(x)$, and its derivative satisfies 
	\begin{equation}
		\label{eq:df}
			\frac{\dd f (x)}{\dd x} = \pi x^2 \cot \left( \pi x \right) .
	\end{equation}\par
	We do not discuss the non-perturbative contributions, since $Z_{\text{inst}} \to 1$ exponentially fast in the setup of this work.\footnote{Equivalently, in the 't Hooft limit taken below, instantons become infinitely massive and decouple.}\par
	When the gauge group is $SU(N)$, the scalar $\phi$ must satisfy the constraint 
	\begin{equation}
		\sum_{a=1} ^{N} \phi_a= 0 ,
	\end{equation}
	that can be enforced adding a linear term in the potential \eqref{eq:Vphi} and imposing the independence of the partition function from the Lagrange multiplier. Writing this linear shift as 
	\begin{equation}
		V (\phi) \mapsto  V (\phi ) + 4 \pi r^2 \xi \phi 
	\end{equation}
	with $V(\phi)$ as in \eqref{eq:Vphi}, we recognize in the Lagrange multiplier $\xi$ a Fayet--Iliopoulos parameter. A geometric reduction from a $U(N)$ to a $SU(N)$ factor in the gauge group has been described in \cite{c}, and we revisit their argument in the matrix model language in appendix \ref{app:specialunitary}.\par
		Before proceeding we notice that, for the integral representation of the partition function to be convergent, one has to impose 
		\begin{equation}
		\label{eq:condNnf}
			N - \vert k \vert - \frac{1}{2} \sum_{\alpha=1} ^{F} n_{\alpha} \ge 0 ,
		\end{equation}
		which is a necessary (and believed sufficient) condition for the theory to descend from a non-trivial SCFT in the UV \cite{Jefferson:2017ahm}.

	\subsection{Large $N$ and decompactification limit}
	\label{sec:decompactification}
	
		We will take the large $N$ limit of the matrix model \eqref{eq:Zloc} and then compute its decompactification limit $r \to \infty$. Writing the partition function in the form 
		\begin{equation}
			\mathcal{Z}_{\mathbb{S}^5 } = \int \dd \phi ~ e^{- S_{\text{eff}} (\phi )} ,
		\end{equation}
		we see that the leading contributions in the large $N$ and large $r$ limit come from the stationary points of $S_{\text{eff}}$, while away from these points the integrand is damped as $ e^{- r^3 N^2 \left( \cdots \right) }$. Therefore, the problem is reduced to finding the solutions $\phi^{\ast}$ to the saddle point equations (SPEs)
		\begin{equation}
			\left. \frac{ \partial S_{\text{eff}} (\phi )}{\partial \phi_a} \right\rvert_{\phi = \phi^{\ast}} =0 , \qquad a=1, \dots, N .
		\end{equation}\par
		Let us now look into the simplifications brought in by the large radius limit. Using the parity of $f$, property \eqref{eq:df} and retaining only the leading contribution at large $r$, we see that the hypermultiplet and vector multiplet one-loop determinants contribute to the SPE respectively 
		\begin{subequations}
		\begin{align}
			 - & \frac{\pi r^3}{2} \sum_{\alpha=1} ^{N_f}  \left( \phi_a ^{\ast} +  m_{\alpha} \right)^2  \text{sign} \left( \phi_a ^{\ast} + m_{\alpha} \right)  , \\ 
			& \pi r^3 \sum_{b \ne a} \left( \phi_a ^{\ast} - \phi_b ^{\ast} \right)^2  \text{sign} \left( \phi_a ^{\ast} - \phi_b ^{\ast} \right) . 
		\end{align}
		\end{subequations}
		Putting these terms together with the derivative of the classical piece we arrive at the system of $N$ SPEs 
		\begin{equation}
		\label{eq:SPE}
			k \left( \phi_a ^{\ast} \right)^2 + 2 h \phi_a ^{\ast} + \check{\xi} - \frac{1}{2} \sum_{\alpha =1} ^{F} n_{\alpha} \left( \phi_a ^{\ast} + m_{\alpha} \right)^2 \text{sign} \left( \phi_a ^{\ast} + m_{\alpha} \right) = -  \sum_{b \ne a} \left( \phi_a ^{\ast} - \phi_b ^{\ast} \right)^2  \text{sign} \left( \phi_a ^{\ast} - \phi_b ^{\ast} \right) 
		\end{equation}
		for $a=1, \dots, N$. In the latter expression we have introduced the scaled quantity $\check{\xi} = \frac{ 4 \xi}{r}$, to keep track of the Lagrange multiplier at large radius.\footnote{The linear coupling between $\xi$ and $\phi$ actually comes from a mixed \CS~term, so it should scale with $r^3$, as the pure \CS~term. The introduction of a new variable $\check{\xi}$ is an artefact of the normalization, not an additional scaling that we impose.}\par
		Integrating the SPE \eqref{eq:SPE} and summing over $a$, we arrive at twice the prepotential of \cite{Intriligator:1997pq}. The factor of two is predicted from the equivariant localization on the round sphere: the partition function only receives contributions from small neighbourhoods of the two fixed points of an isometry rotating a $\mathbb{P}^2$ inside $\mathbb{S}^5$. A slightly different prepotential has been derived in \cite{a}, and in appendix \ref{app:otherdeco} we relate it to a different decompactification limit.\par

		\subsection{Solution}
		\label{sec:largeNlimit}
			
			Our goal is to solve the SPE \eqref{eq:SPE} in a large $N$ 't Hooft limit, with 
			\begin{equation}
			\label{eq:tHooft}
				\frac{N}{h} = \lambda \text{ fixed}, \qquad   \qquad \frac{N}{k} = t \text{ fixed}.
			\end{equation}
			We moreover consider a Veneziano limit, in which the number $N_f$ of fundamental hypermultiplets grows linearly with $N$, hence we introduce the Veneziano parameters 
			\begin{equation}
			\label{eq:Veneziano}
				\frac{n_{\alpha}}{N} = \zeta_{\alpha} \text{ fixed}, \qquad \qquad  \forall ~ \alpha = 1, \dots, F .
			\end{equation}
			We also keep $\tilde{\xi} = \frac{\check{\xi}}{N}$ fixed. The convergence condition \eqref{eq:condNnf} in this limit becomes 
			\begin{equation}
			\label{eq:tzetale1}
				\frac{1}{\vert t \vert } + \frac{1}{2} \sum_{\alpha=1}^{F} \zeta_{\alpha}  \le 1 . 
			\end{equation}\par
			The large $N$ limit of a $5d$ $\mathcal{N}=1$ $U(N)$ Yang--Mills theory in the decompactification regime has been addressed in \cite{Nedelin}, but only for a very special choice of masses and no Chern--Simons term. We now derive the phase structure of the most general consistent gauge theory with simple gauge group in the decompactification limit.\par
			Let us introduce the eigenvalue density $\rho (\phi )$, which is normalized: 
			\begin{equation}
			\label{eq:rhonorm}
				\int \dd \phi \rho (\phi) = 1 
			\end{equation}
			and has compact support. The effective action $S_{\text{eff}} (\phi)$ is not an even function, therefore we do not expect the support of the eigenvalue density to be symmetric. Moreover, $\rho (\phi)$ is not required to be a function and, in general, it is sufficient that $\rho (\phi) \dd \phi$ is a measure on the union of intervals along a selected integration cycle. Throughout all this work, the integration cycle is the real axis and the measure is supported on a compact interval,
			\begin{equation}
				\text{supp} \rho = [A, B ]  \subset \R .
			\end{equation}\par
			In the scaling limit \eqref{eq:tHooft}-\eqref{eq:Veneziano} the system of $N$ saddle point equations is recast into a single integral equation
			\begin{equation}
			\label{eq:LargeNspe}
				- \int_A ^{B}  \dd  \psi \rho (\psi)~ \left( \phi^{\star} - \psi \right)^2 \text{sign} \left( \phi^{\star} - \psi \right)  =  \frac{1}{t} \left( \phi^{\star} \right)^2 + \frac{2}{\lambda} \phi^{\star}  + \tilde{\xi} - \sum_{\alpha=1} ^{F} \frac{\zeta_{\alpha}}{2} \left( \phi^{\star} + m_{\alpha} \right)^2 \text{sign}  \left( \phi^{\star} + m_{\alpha} \right)  
			\end{equation}
			to be satisfied by every $\phi^{\star} \in [A,B]$. Here we have denoted $\phi^{\star}$ the variable running over the continuous spectrum of eigenvalues of $\phi^{\ast}$, being $\phi^{\ast}$ the solution to the SPE \eqref{eq:SPE}. We have done so in the hope of avoiding confusion between the original $N$-dimensional integration variable $\phi= \left( \phi_1, \dots, \phi_N \right)$, the fixed $N$-dimensional saddle point $\phi^{\ast} = \left( \phi_1 ^{\ast}, \dots, \phi_N ^{\ast} \right)$, and the one-dimensional real variable $\phi^{\star} \in [A,B]$. Henceforth, we will simply use $\phi$ instead of $\phi^{\star}$ to reduce clutter.\par
			\medskip
				The mechanism triggering the phase transitions is read off from \eqref{eq:LargeNspe}: the right-hand side changes when a mass parameter crosses $A$ or $B$, leading to a new eigenvalue density.\par
				Taking three derivatives, we find that the generic solution to the SPE \eqref{eq:LargeNspe} is 
				\begin{equation}
				\label{eq:genericrho}
					\rho (\phi) = c_A \delta \left( \phi - A \right) + c_B \delta \left( \phi - B \right) + \sum_{\alpha=1} ^{F} c_{\alpha} \delta \left( \phi + m_{\alpha} \right) ,
				\end{equation}
				with coefficients 
				\begin{equation}
					c_{\alpha} = \begin{cases} \frac{\zeta_{\alpha}}{2}  & - m_{\alpha} \in [A,B] \\ 0 & \text{otherwise} \end{cases} \qquad \alpha =1, \dots, F .
				\end{equation}
				Throughout this work, we define the $\delta$-functions centered at the endpoints of $\text{supp} \rho$ taking the limit from inside the support \cite{Nedelin}, 
					\begin{equation}
						 \delta \left( \phi - A  \right) = \lim_{\varepsilon \to 0^{+}}  \delta \left( \phi - \left( A+ \varepsilon \right)  \right) ; \qquad  \delta \left( \phi - B  \right) = \lim_{\varepsilon \to 0^{+}}  \delta \left( \phi - \left( B- \varepsilon \right)  \right) .
					\end{equation}\par
				The solution \eqref{eq:genericrho} is given in terms of two coefficients $c_A$, $c_B$ and two endpoints $A$, $B$ to be determined. Plugging \eqref{eq:genericrho} back into the cubic equation \eqref{eq:LargeNspe} yields a system of three equations:
				\begin{subequations}
				\label{eq:system3eqsUN}
				\begin{align}
					-c_A + c_B - \sum_{\alpha=1} ^{F} c_{\alpha} \tilde{s}_{\alpha}  & = \frac{1}{t} - \sum_{\alpha=1} ^{F} \frac{\zeta_{\alpha}}{2} \tilde{s}_{\alpha} \label{eq:system3eqsUNa} \\
					c_A A - c_B B - \sum_{\alpha=1} ^{F} c_{\alpha} \tilde{s}_{\alpha}  m_{\alpha} & = \frac{1}{\lambda} - \sum_{\alpha=1} ^{F} \frac{\zeta_{\alpha}}{2} \tilde{s}_{\alpha} m_{\alpha} \label{eq:system3eqsUNb} \\
					- c_A A^2 + c_B B^2 - \sum_{\alpha=1} ^{F} c_{\alpha} \tilde{s}_{\alpha}  m_{\alpha} ^2 & = \tilde{\xi} - \sum_{\alpha=1} ^{F} \frac{\zeta_{\alpha}}{2} \tilde{s}_{\alpha} m_{\alpha} ^2  \label{eq:system3eqsUNc}
				\end{align}
				\end{subequations}
				in which we have introduced the shorthand notation 
				\begin{equation}
					\tilde{s}_{\alpha}  = \frac{1}{2} \left[ \text{sign} \left(  A+ m_{\alpha} \right) + \text{sign} \left(  B+ m_{\alpha} \right) \right] = \begin{cases} -1 & -m_{\alpha} > B \\ 0 & A \le -m_{\alpha} \le B \\ +1 & -m_{\alpha} < A . \end{cases}
				\end{equation}
				The normalization condition \eqref{eq:rhonorm} applied to \eqref{eq:genericrho} imposes 
				\begin{equation}
				\label{eq:normcondgenrhoUN}
					c_A + c_B + \sum_{\alpha=1}^{F} c_{\alpha} =1 .
				\end{equation}
				Therefore the two coefficients $c_A$ and $c_B$ and the two endpoints $A$ and $B$ of $\text{supp} \rho$ are determined as functions of the gauge theoretical parameters $t$, $\lambda$, $\left\{ \zeta_{\alpha}, m_{\alpha} \right\}$, from the system \eqref{eq:system3eqsUN} completed by the normalization \eqref{eq:normcondgenrhoUN}.\par
				Solving \eqref{eq:system3eqsUNa} together with \eqref{eq:normcondgenrhoUN} yields
				\begin{subequations}
				\begin{align}
					c_A & = \frac{1}{2} \left[  1 - \frac{1}{t} - \sum_{\alpha=1} ^{F} \left( c_{\alpha} - \tilde{s}_{\alpha} \left( \frac{\zeta_{\alpha}}{2} - c_{\alpha} \right)  \right)  \right] \\
					c_B & = \frac{1}{2} \left[  1 + \frac{1}{t} - \sum_{\alpha=1} ^{F} \left( c_{\alpha} + \tilde{s}_{\alpha} \left( \frac{\zeta_{\alpha}}{2} - c_{\alpha} \right)  \right)  \right] .
				\end{align}
				\end{subequations}
				To find the endpoints $A$ and $B$ we plug these values in \eqref{eq:system3eqsUNb}-\eqref{eq:system3eqsUNc}. The system is quadratic in the variables $A$ and $B$, thus we find a pair of solutions: at each point in the parameter space, we should retain the one consistent with $A<B$, which must hold by construction. We stress that \eqref{eq:genericrho} has been derived taking derivatives with respect to $\phi$, thus working under the assumption that the interior of $\text{supp} \rho $ is not empty. Whenever a consistent pair of endpoints $A$, $B$ cannot be found, we should drop this assumption and take into account solutions supported at a single point, $\rho (\phi) = \delta ( \phi )$.\par
				\medskip
				Phase transitions in the theory are signalled by non-analyticities in the free energy 
				\begin{equation}
					\label{eq:freeenergy}
						\mathcal{F}_{\mathbb{S}^5} = - \frac{1}{\pi r^3 N^2} \log \left\vert \mathcal{Z}_{\mathbb{S}^5}  \right\vert  .
				\end{equation}
				In the decompactification and large $N$ 't Hooft limit it becomes 
				\begin{equation}
					\mathcal{F}_{\mathbb{S}^5} = \frac{1}{6} \int \dd  \phi \rho (\phi)\int \dd  \psi \rho (\psi)~\vert \phi - \psi \vert^3   + \int \dd  \phi \rho (\phi) \left[ \frac{1}{3t} \phi^3 + \frac{1}{\lambda} \phi^2 - \sum_{\alpha=1} ^{F} \frac{\zeta_{\alpha}}{6}   \vert \phi + m_{\alpha} \vert^3  \right]  .
				\end{equation}
				The linear term proportional to $\tilde{\xi}$ does not contribute by construction. Using the solution \eqref{eq:genericrho} for $\rho (\phi)$, $\mathcal{F}_{\mathbb{S}^5}$ is found to be 
				\begin{align}
					\mathcal{F}_{\mathbb{S}^5} &= \frac{1}{3} \left[ c_A c_B \vert B-A \vert^3 +\sum_{\alpha=1}^{F}  c_{\alpha} \left( c_A  \vert A + m_{\alpha} \vert^3 + c_B  \vert B + m_{\alpha} \vert^3  + \sum_{\alpha^{\prime}=1} ^{F}  \frac{c_{\alpha^{\prime}}}{2} \vert m_{\alpha} - m_{\alpha^{\prime}} \vert^3  \right)  \right] \label{eq:freeenergygeneral} \\
						& + \frac{1}{3t} \left[  c_A A^3 + c_B B^3 - \sum_{\alpha=1} ^{F} c_{\alpha} m_{\alpha} ^3 \right] + \frac{1}{\lambda} \left[  c_A A^2 + c_B B^2 + \sum_{\alpha=1} ^{F} c_{\alpha} m_{\alpha} ^2 \right] \notag \\
						& - \sum_{\alpha=1} ^{F} \frac{\zeta_{\alpha}}{6} \left[ c_A \vert A+ m_{\alpha} \vert^3 + c_B \vert B+ m_{\alpha} \vert^3 + \sum_{\alpha^{\prime}=1}^{F}   c_{\alpha^{\prime}}  \vert  m_{\alpha} -  m_{\alpha^{\prime}} \vert^3  \right] .  \notag
				\end{align}\par
				Recall that the coefficients $c_{\alpha}$ vanish unless $A<-m_{\alpha}<B$, in which case $c_{\alpha}= \frac{\zeta_{\alpha}}{2}$. This implies that whenever $A<-m_{\alpha}<B$ the one-loop contribution of the hypermultiplets of mass $m_{\alpha}$ is almost entirely cancelled between the first and the last line in \eqref{eq:freeenergygeneral}. This is consistent with the mass $m_{\alpha}$ being below the characteristic energy scale of the problem: the hypermultiplet cannot be integrated out, whence no one-loop effect is generated. The cancellation of the one-loop effects between the first and third line of \eqref{eq:freeenergygeneral} when $A<-m_{\alpha} <B$ leaves behind a contribution 
				\begin{equation}
				\label{eq:malphambetaint}
					- \sum_{\alpha^{\prime}=1} ^{F} \frac{\zeta_{\alpha} \zeta_{\alpha^{\prime}} }{12} \vert m_{\alpha} - m_{\alpha^{\prime}} \vert^3 .
				\end{equation}
				It reproduces the one-loop contribution of the massive W-bosons in the background vector multiplet for the flavour symmetry broken by the solution in the phase considered.\par
				The continuity of $A$ and $B$ at each critical surface and the jump by $\frac{\zeta_{\alpha}}{2}$ of $c_A$ when $-m_{\alpha}$ crosses $A$, or of $c_B$ when $-m_{\alpha}$ crosses $B$, guarantee the continuity of the free energy at each transition point. Furthermore, the continuity of $\rho (\phi)$ can be used to prove that the transition must be at least second order. This is confirmed by the explicit computations in each case.\par
				In sections \ref{sec:UN} and \ref{sec:SuN} we consider gauge theories with gauge group $U(N)$ and $SU(N)$ respectively, and with the other classical groups in section \ref{sec:SpN}, and present their large $N$ phase structure explicitly.\par

				\subsubsection{$\mathcal{F}$-theorem}
				\label{sec:Ftheorem}
					The sphere partition function measures the degrees of freedom of a field theory in odd dimensions \cite{Klebanov:2011gs}. In $5d$ and with normalization \eqref{eq:freeenergy}, the $\mathcal{F}$-theorem states \cite{Klebanov:2011gs} 
					\begin{equation}
					\label{eq:Ftheo}
						\mathcal{F}_{\mathbb{S}^5} ^{(\mathrm{IR})} > \mathcal{F}_{\mathbb{S}^5} ^{(\mathrm{UV})} .
					\end{equation}
					Compelling evidence for this claim has been presented, for instance, in \cite{Chang:2017cdx,Fluder:2020pym}. Inequality \eqref{eq:Ftheo} holds when both sides are evaluated at fixed points but, under favourable circumstances, the free energy can be shown to be monotonic all along the RG flow connecting the UV and the IR fixed points. Expression \eqref{eq:freeenergygeneral} can be used to provide new support for the $\mathcal{F}$-theorem.\par
					For fixed values of the masses, $\lambda \to 0 $ drives the theory to the IR. From \eqref{eq:freeenergygeneral} and using the dependence of $A$ and $B$ on $\lambda$ through \eqref{eq:system3eqsUN}, it follows that \eqref{eq:Ftheo} is satisfied between any two points on the RG flow. A direct proof of \eqref{eq:Ftheo} is less obvious from \eqref{eq:freeenergygeneral} at fixed $\lambda$ and increasing masses, but it can nevertheless be confirmed using the explicit results in section \ref{sec:SuN}.

				\subsubsection{Remarks on the decompactification limit}
				\label{sec:remarksdeco}
					
					We follow the standard nomenclature denoting the large sphere limit as decompactification limit, but it ought to be remarked that the localization procedure requires a (equivariantly) compact topology, and the limit $r \to \infty$ should be really  meant as the zero curvature limit $\frac{1}{r} \to 0$. As already emphasized in the introduction, this allows to neglect curvature effects in a controlled way, but only after putting the localization machinery at work on a compact manifold.\par
					A further remark concerns the sign of the \YM~'t Hooft coupling $\lambda$. We will consider $\lambda^{-1} \in \mathbb{R}$. The interpretation of this may be puzzling from a field theoretic viewpoint, because then instanton corrections would contribute exponentially (instead of being exponentially suppressed) for $\lambda^{-1} <0$. Moreover, as reviewed in appendix \ref{sec:geometric}, the parameter $h$ in \eqref{eq:defh} has the meaning of a volume, thus it should not become negative. Nevertheless, the perturbative partition function can be analytically continued letting $h \in \R $ in \eqref{eq:Zloc} but keeping the non-perturbative quantities, such as instanton masses, as functions of $\vert h \vert$. For a thorough discussion on negative \YM~coupling, see \cite{Nedelin,Minahan:2020ifb}. Besides, the SPE \eqref{eq:LargeNspe} may be likewise analytically continued to negative values of the Veneziano parameters $\zeta_{\alpha}$.\par
					One last comment is about flavour symmetry. For $U(N)$, the mass parameters belong to a background $SU(N_f)$ vector multiplet. In order not to violate the flavour symmetry, we will assume $N_f = 1 + \sum_{\alpha=1} ^{F} n_{\alpha}$ and give the extra hypermultiplet a mass 
					\begin{equation}
						m_{N_f} = - \sum_{\alpha=1} ^{F} n_{\alpha} m_{\alpha} .
					\end{equation}
					Its contribution is suppressed in the Veneziano limit \eqref{eq:Veneziano} and drops out of the SPE.

				\subsubsection{Remarks on phase transitions and matrix models}
					
					As already mentioned, integrals over matrix degrees of freedom do not admit a notion of phase diagram, unless the number $N$ of eigenvalues is sent to infinity. From a field theoretical perspective, a phase structure may originate from the infinite volume limit as well. Phase transitions among distinct chambers of $\mathscr{C} (X)$ in flat space belong to this latter class, while the phase transitions we are concerned with are instead of the first type.\par
					The presence of a phase transition in the decompactification limit does not automatically imply that the transition exists at large $N$ but finite radius. In fact, this implication fails in $3d$ \cite{Barranco}. Nevertheless, we will now argue that the situation is different in $5d$ and the transitions discovered with the aid of the decompactification limit persist at finite radius.\par
					A generic effect of finite $\frac{1}{r^2}$ corrections is to smoothen the $\delta$-singularities into peaked curves of finite height and width. In $3d$ Chern--Simons theories with fundamental hypermultiplets, the solution at large $N$ but finite $r$ is given by a deformation of the pure Chern--Simons eigenvalue density, on top of which a peak forms each time a mass parameter is decreased \cite{Barranco,STWL}. The shape of the eigenvalue density is changed without breaking its support \cite{Barranco}, therefore there is no phase transition at finite radius.\par
					On the contrary, in $5d$ we do not have a distribution on top of which the peaks are formed, and we expect that each new peak will produce a new cut in the support of $\rho ( \phi )$. Let us elaborate further on this statement. Starting with a pure gauge theory and assuming a very small size of the support, the $\sinh (\phi_a - \phi_b) \approx (\phi_a - \phi_b)$ in \eqref{eq:Z1loopvec} will dominate against the $e^{f}$ term, leading to the equilibrium equation of a cubic matrix model. The generic solution is supported on two intervals and the two pieces degenerate into $\delta (\phi-A)$ and $\delta (\phi - B)$ as $\frac{1}{r} \to 0 $. A phase transition when the two intervals merge was observed in \cite{Minahan:2014hwa}. In turn, we can work with finite but large enough $r$ to guarantee that the model remains in the two-cut phase. Most importantly, in the one-cut phase $\text{supp} \rho$ is moved away from the real axis \cite{Minahan:2014hwa}, thus such solution is discarded by the procedure adopted in the present work.\par
					Decreasing the masses of the hypermultiplets from infinity, new peaks will form on top of the finite radius solution. However, as these new peaks are moved away from one endpoint, they will break the support and produce additional intervals, until they reach the other endpoint and the two intervals merge. In conclusion, the phase transitions uncovered throughout this work are expected to be genuine large $N$ phase transitions, associated with splitting of $\text{supp} \rho$, and not a consequence of the large $r$ approximation.\par
					\medskip
					Phase transitions as the ones observed are ubiquitous in gauge theories with an underlying cohomological structure, that allows to reduce the observables to a matrix model. Prototypical in this respect is the Gross--Witten--Wadia third order phase transition \cite{Gross:1980he,Wadia:1980cp,Wadia:2012fr} in $2d$. We ought to emphasize that the integrals from localization of $5d$ $\mathcal{N}=1$ gauge theories are not of standard random matrix type, meaning that there seem to be no change of variables to recast the vector multiplet one-loop determinant in the form of a Vandermode determinant. As a consequence, the mechanism underlying the phase transitions is inherently technically different from the Gross--Witten--Wadia transition. Nonetheless, a recurrent theme is that phase transitions are triggered by states becoming massless. In the present setup the light states come from the matter sector. Conversely, in pure $2d$ Yang--Mills theory there are no propagating perturbative particles, thus the transition is induced by instantons \cite{Neuberger:1989kd}.

				\subsection{Wilson loops}
				\label{sec:WLlimit}
				
						The eigenvalue density \eqref{eq:genericrho} can be exploited to compute the vacuum expectation value (vev) of Wilson loops on $\mathbb{S}^5$ that preserve half of the supercharges (that is, are half-BPS) in the large $r$ and large $N$ limit. The contribution to the effective action from a Wilson loop in a representation of fixed size is sub-leading and does not alter the eigenvalue density in the large $N$ limit. We conclude that the vev of a Wilson loop in the fundamental representation $\mathsf{F}$ is 
						\begin{equation}
							\left.\langle \mathcal{W}_{\mathsf{F}} \right.\rangle  = \int_A ^{B} \dd  \phi \rho (\phi) e^{2 \pi r \phi }  = c_A e^{2 \pi r A} + c_B e^{2 \pi r B} + \sum_{\alpha=1} ^{F} c_{\alpha} e^{2 \pi r m_{\alpha}} .
						\end{equation}
						The continuity of this expression follows from the continuity of $A$ and $B$ at the critical values, together with the jump by $\frac{\zeta_{\alpha}}{2}$ of $c_A$ or $c_B$ when $-m_{\alpha}$ crosses $A$ or $B$ respectively. The Wilson loop vev follows a perimeter law, $\log \left.\langle \mathcal{W}_{\mathsf{F}} \right.\rangle  \approx (2 \pi B) r $, as expected and in agreement with \cite{Assel:2012nf}.\par
						For classical gauge group, it is proven in subsection \ref{sec:SuNWL} that the Wilson loop is differentiable, as a consequence of the scalar $\phi$ being traceless.

					\subsubsection{Wilson loops in large antisymmetric representations}
					
						Expectation values of Wilson loops in a given representation whose size grows with $N$ deserve further consideration. Let $\mathsf{A}_{K}$ be the rank-$K$ antisymmetric representation of the gauge group. This implies $0 \le K \le N$ for $U(N)$ and $0 \le K \le N-1$ for $SU(N)$. We consider a Wilson loop in the representation $\mathsf{A}_K$ along a great circle inside $\mathbb{S}^5$.\par
\begin{figure}[htb]
							\centering
							\includegraphics[width=0.2\textwidth]{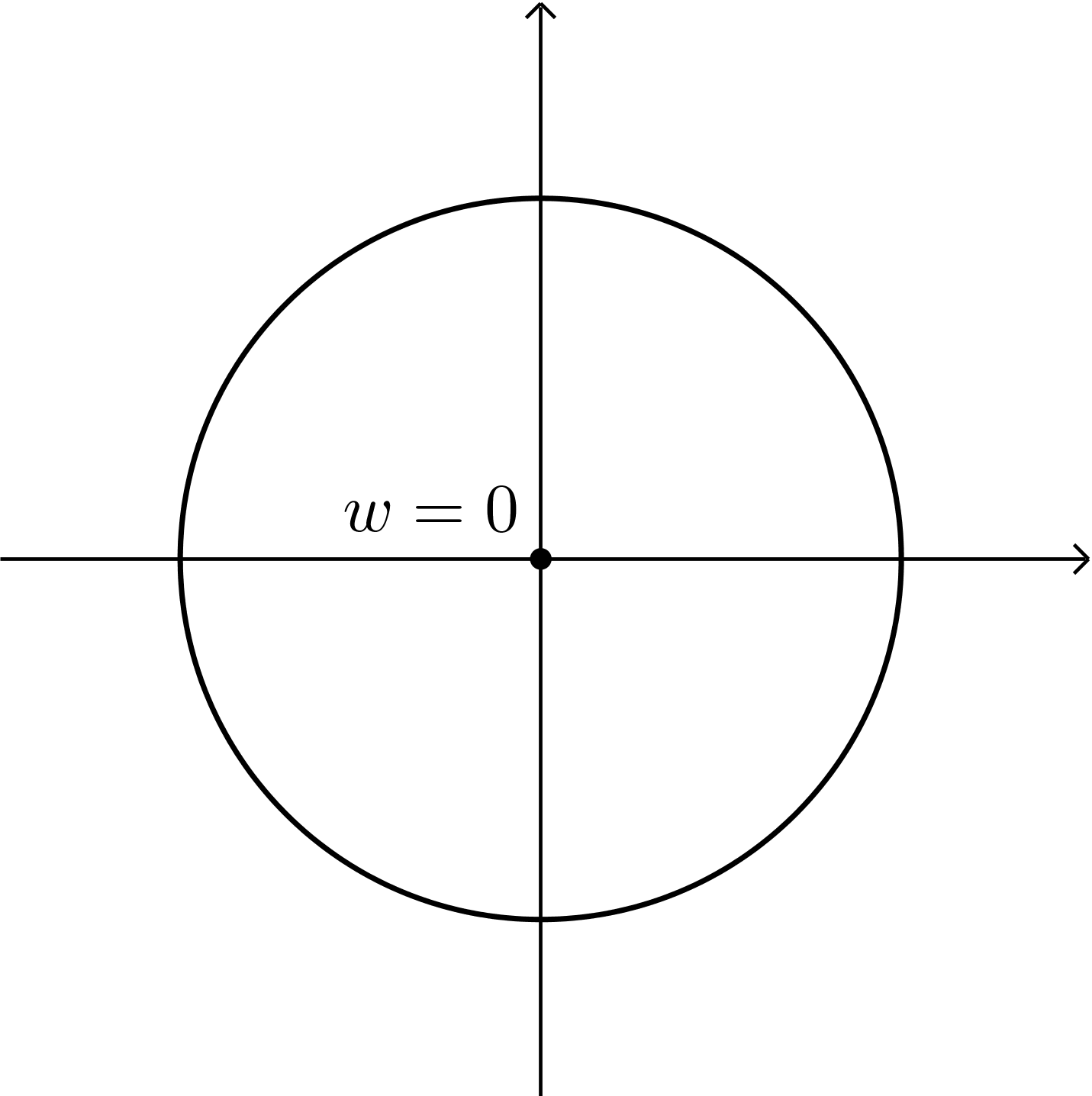}%
							\hspace{0.1\textwidth}
							\includegraphics[width=0.45\textwidth]{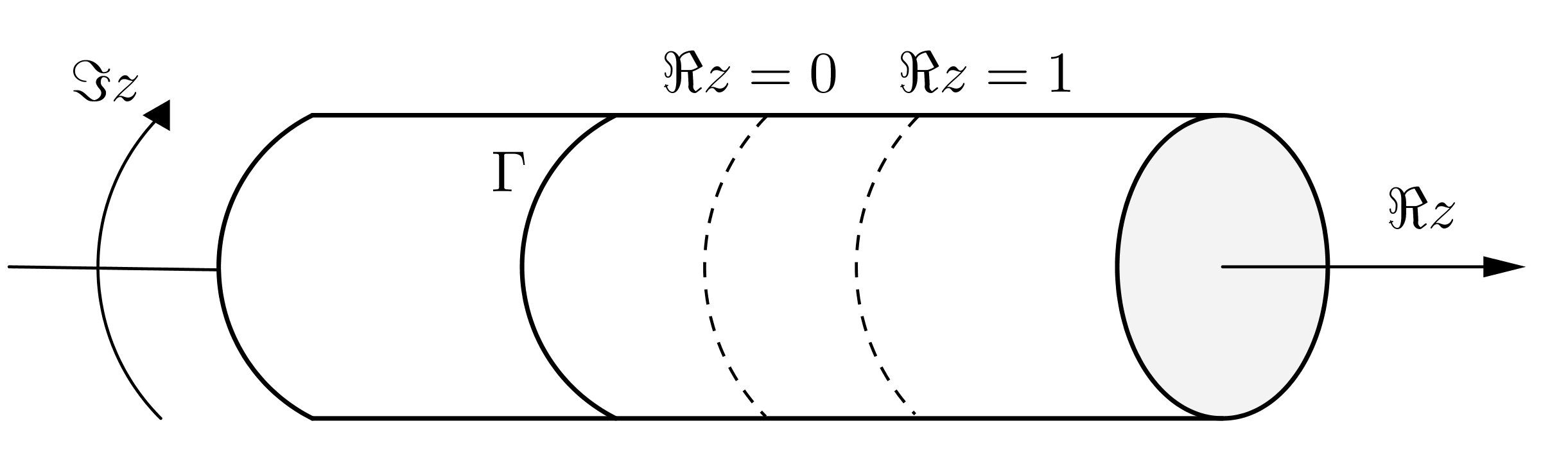}
							\caption{Left: integration cycle in the $w$-plane. Right: the exponential map sends $\mathbb{C}$ to a cover of the cylinder, and the circle around $w=0$ to a circle $\Gamma$ wrapping the cylinder once.}
							\label{fig:Ctocylmap}
						\end{figure}

						The formalism to study the vev of such loop operators in the large $N$ limit, with $K$ growing with $N$, was developed in \cite{Hartnoll:2006} for $4d$ $\mathcal{N}=4$ \YM, and applied to $4d$ $\mathcal{N}=2$ in \cite{Russo:2017ngf} and to $3d$ \CS~theories in \cite{STWL}. The derivation of \cite{Hartnoll:2006} does not depend on the specific theory, as long as the Wilson loop vev is localized to a finite-dimensional integral, and directly extends to the present five-dimensional setting, with a few improvements to accommodate a non-even eigenvalue density. The central idea is to introduce the generating function 
						\begin{equation}
						\label{eq:defPhiA}
							\Phi_{\mathsf{A}} (w) = \left\langle \prod_{a=1} ^{N} \left( w+ e^{r \phi_a} \right) \right\rangle
						\end{equation}
						and notice that the expectation value $\left\langle \mathcal{W}_{\mathsf{A}_K} \right\rangle$ is extracted as 
						\begin{equation}
						\label{eq:Wantysymoint}
							\left\langle \mathcal{W}_{\mathsf{A}_K} \right\rangle = \oint \frac{dw}{2 \pi i } \frac{  \Phi_{\mathsf{A}} (w) }{ w^{N-K+1} } ,
						\end{equation}
						with integration cycle a closed loop in $ \mathbb{C}$ around $w =0$. We are interested in the large $N$ with the ratio 
						\begin{equation}
						\label{eq:defkappaWL}
							\kappa = \frac{K}{N} \text{ fixed}, \qquad 0 \le \kappa \le 1 .
						\end{equation}
						From the definition of $\Phi_{\mathsf{A}} (w)$ in \eqref{eq:defPhiA} we have, at large $N$, 
						\begin{equation}
							\Phi_{\mathsf{A}} (w)  = \exp \left[ N  \int_{A} ^{B} \dd  \phi \rho (\phi) \log \left( w+e^{r \phi} \right)   \right] ,
						\end{equation}
						with the eigenvalue density $\rho (\phi)$ as in \eqref{eq:genericrho}. It is convenient to map the complex $w$-plane to a cylinder \cite{Hartnoll:2006}. We use a change of variables 
						\begin{equation}
							\log w = r \left[ A + z \left( B-A  \right) \right],
						\end{equation}
						with $z$ the holomorphic variable on a multiply-sheeted cover of the cylinder. The original integration cycle in \eqref{eq:Wantysymoint} is mapped onto a circle $\Gamma$ wrapping the cylinder at fixed $\Re z$, and shrinking the original cycle around $w=0$ pushes $\Gamma$ towards $\Re z \to - \infty $, see figure \ref{fig:Ctocylmap}.\par
						\par
						
						Following \cite{Hartnoll:2006,STWL} we arrive at 
						\begin{equation}
						\label{eq:WantisymlargeN}
							\left\langle \mathcal{W}_{\mathsf{A}_{K}} \right\rangle = e^{r A K } r (B-A) \oint_{\Gamma} \frac{ dz}{2 \pi i } \exp \left\{ N  r \left[  \int_A ^B  \frac{ \dd  \phi }{r}  \rho (\phi) \log \left( 1+ e^{r \left( \phi -A - z (B-A) \right) }  \right)  + \kappa z (B-A)  \right] \right\} 
						\end{equation}
						in the large $N$ approximation. The integrand has branch cuts at 
						\begin{equation}
							0 \le \Re z \le 1 \quad \text{ and } \quad \Im z = \frac{ (2n+1) \pi }{ r (B-A) } , \ n \in \mathbb{Z} ,
						\end{equation}
						and the integration cycle $\Gamma$ lies on their left.\par
						Using the general solution \eqref{eq:genericrho} for the eigenvalue density, we get 
						\begin{align}
							\left\langle \mathcal{W}_{\mathsf{A}_{K}} \right\rangle = e^{r A K } r (B-A) & \oint_{\Gamma} \frac{ dz}{2 \pi i } ~ e^{ z N \kappa r (B-A)}  \left[  1+ e^{-z r (B-A)  } \right]^{N c_A}   \label{eq:integralWAKres} \\
								& \times \left[  1+ e^{(1-z) r \left( B - A \right) } \right]^{N c_B} ~ \prod_{\alpha=1} ^{F} \left[  1+ e^{ - r \left( m_{\alpha} +A + z (B-A) \right)} \right]^{N c_{\alpha}} . \notag
						\end{align}
						There are two ways to obtain the Wilson loop vev from \eqref{eq:integralWAKres}. We can go back to the complex $w$-plane, compute the residue, take the logarithm and then retain only the leading order in $\frac{1}{r}$. The alternative approach consists in approximating the integrand in \eqref{eq:integralWAKres} at large $r$ first, obtaining the exponential of a piecewise linear function of $z$. Then, we use the fact that we are working at large $N$, thus the leading contribution to the integral will come from a neighbourhood of the local extrema of the integrand. Direct inspection easily shows that these extrema are to be looked for in the region $0 \le \Re z \le 1 $ and $\Im z = 2 \pi n$, $n \in \mathbb{Z}$. Due to the presence of the branch cut, however, only those points with $\Im z=0$ should be retained. We have checked in the explicit examples to be discussed below that the results computed in the two ways agree.\par
						Regardless of the details of each specific phase of any theory, the upshot is that $\log \left\langle \mathcal{W}_{\mathsf{A}_{K}} \right\rangle $ grows linearly in $N$ and $r$, meaning that it follows a perimeter law, and is of the general form 
						\begin{equation}
							\log \left\langle \mathcal{W}_{\mathsf{A}_{K}} \right\rangle \approx c_1 r K +  c_2 r N 
						\end{equation}
						at leading order in both $N$ and $r$, with $c_1$ and $c_2$ simple functions of $A$, $B$ and the masses $\left\{ m_{\alpha} \right\}$.

						\subsection{Hypermultiplets in other representations}
						\label{sec:adjhyper}

							So far the spotlight has been on theories with fundamental hypermultiplets. We now turn our attention to other types of matter content and analyze the large $N$ limit of $SU(N)$ theories with hypermultiplets in the adjoint, symmetric or rank-two antisymmetric representation.\par

					\subsubsection{Adjoint hypermultiplet}
					\label{sec:pureadjoint}
							We consider \YM~theory with a massive adjoint hypermultiplet \cite{Minahan:2013jwa}. This model has enhanced $\mathcal{N}=2$ supersymmetry at $m=0$. The SPE in the large $N$ decompactification limit is 
							\begin{align}
								\frac{2}{\lambda} \phi  = & - \int_A ^B \dd \psi \rho (\psi) \left( \phi - \psi \right)^2 \text{sign} \left( \phi - \psi \right) \\
								 & + \int_A ^B \dd  \psi \rho (\psi) \left[ \frac{1}{2} \left( \phi - \psi +m \right)^2 \text{sign} \left( \phi - \psi +m \right) + \frac{1}{2} \left( \phi - \psi -m \right)^2 \text{sign} \left( \phi - \psi -m \right)   \right]  . \notag 
							\end{align}
							The Lagrange multiplier $\tilde{\xi}$ has been omitted because the solution turns out to be automatically balanced, with $c_A=c_B$ and $A=-B$.\par
							Without loss of generality we assume $m>0$, and also take $\lambda>0$ for concreteness, being the case $\lambda<0$ completely analogous. It is not hard to check that the eigenvalue density is given by 
							\begin{equation}
								\rho (\phi) = \begin{cases}  \frac{1}{2} \delta \left(  \phi + \frac{1}{\lambda} \right) + \frac{1}{2} \delta \left(  \phi - \frac{1}{\lambda} \right)  & m > \frac{2}{\lambda} \\ \frac{1}{2} \delta \left(  \phi -m + \frac{1}{\lambda} \right) + \frac{1}{2} \delta \left(  \phi +m - \frac{1}{\lambda} \right)  & \frac{1}{\lambda} \le  m \le \frac{2}{\lambda} . \end{cases} 
							\end{equation}
							At $m= \lambda^{-1}$ nothing special happens, but $-B$ and $B$ cross and we should rename the endpoints of the interval. The free energy in this limit is 
							\begin{equation}
							\label{eq:FEadj}
								\mathcal{F}_{\mathbb{S}^5} = \begin{cases} \frac{5}{3} \frac{1}{\lambda^3} - \frac{m}{\lambda^2} - \frac{m^3}{6} &  m > \frac{2}{\lambda} \\ - \frac{1}{3} \left(  m - \frac{1}{\lambda} \right)^3 - \frac{m^3}{6} & 0 \le  m \le \frac{2}{\lambda} , \end{cases}
							\end{equation}
							which implies that $\frac{\partial^2 \mathcal{F}_{\mathbb{S}^5} }{\partial m^2 }$ is discontinuous. The model shows a second order phase transition. We ought to stress that the large $N$ limit we take differs form that in \cite{Minahan:2014hwa}, and hence the transition we find is different in nature. Besides, taking $m \to 0$ first in \eqref{eq:FEadj}, we are left with a third order phase transition at $\frac{1}{\lambda} \to 0$, which corresponds to pass through a $6d$ $\mathcal{N}=(2,0)$ superconformal point. This transition reflects a flop transition in the dual Calabi--Yau geometry (see appendix \ref{sec:geometric}).\par
							The free energy in \eqref{eq:FEadj} is a monotonically increasing function of $\frac{1}{\lambda}$, thus satisfying the $\mathcal{F}$-theorem \eqref{eq:Ftheo}, discussed in subsection \ref{sec:Ftheorem}, all along the RG flow from the SCFT to the deep IR.\par
							The vev of a Wilson loop in the fundamental representation in this model is 
							\begin{equation}
								\langle \mathcal{W}_{\mathsf{F}} \rangle = \begin{cases} \cosh \left( \frac{2 \pi r }{\lambda}  \right) & m > \frac{2}{\lambda} \\  \cosh \left( 2 \pi r \left(  m - \frac{1}{\lambda} \right) \right)  & 0 \le  m \le \frac{2}{\lambda}   \end{cases}
							\end{equation}
							with discontinuous derivative, in agreement with the result for $\mathcal{F}_{\mathbb{S}^5}$.

					\subsubsection{Antisymmetric or symmetric hypermultiplets}
							$5d$ $SU(N)$ gauge theories with $n_{\mathsf{A}} \in \left\{ 0,1,2 \right\}$ hypermultiplets in the rank-two antisymmetric representation or $n_{\mathsf{S}} \in \left\{ 0, 1 \right\}$ hypermultiplets in the symmetric representation descend from SCFTs \cite{Bhardwaj:2020gyu}. The free energies of the theories with $n_{\mathsf{S}} =1$ or $n_{\mathsf{A}} =1$ differ by terms that are sub-leading at large $N$ and therefore have identical phase diagram. The case $n_{\mathsf{A}}=2$ does not admit a large \CS~level nor a large number of additional fundamental hypermultiplets. The phase structure of $SU(N)$ theories with (anti-)symmetric matter is derived in subsection \ref{sec:SuNantisym}.\par

		\section{Phases of ${U(N)}$ theories}
		\label{sec:UN}
			In this section, the large $N$ limit \eqref{eq:tHooft}-\eqref{eq:Veneziano} of $U(N)$ gauge theories with $N_f$ fundamental flavours is studied. For unitary group, we set $\tilde{\xi}=0$.\par
\begin{figure}[htb]
						\centering
						\includegraphics[width=0.4\textwidth]{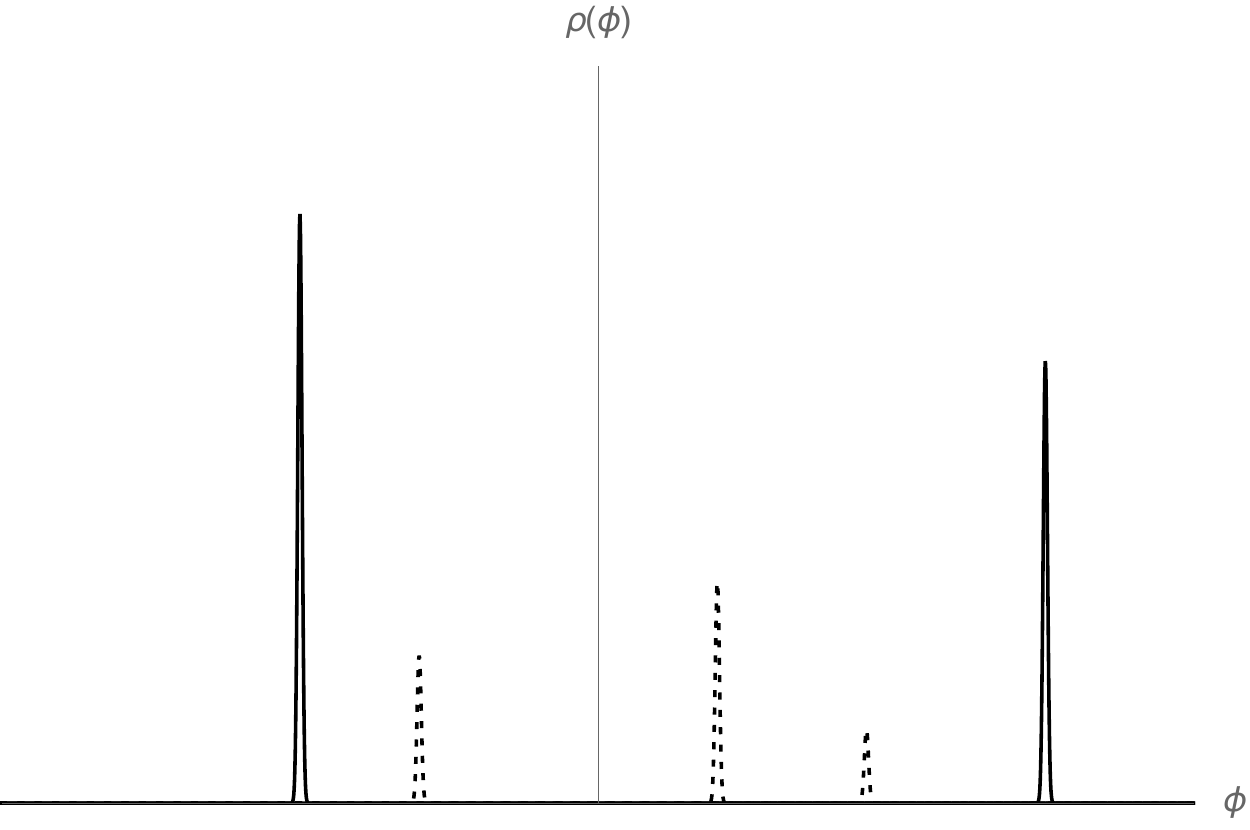}
						\caption{Eigenvalue density at large $N$, in a phase in which three out of the $F$ mass parameters fall inside $\text{supp} \rho$. The solid lines are the eigenvalues at the endpoints, the dashed lines are the eigenvalues at $-m_{\alpha}$. The range of the vertical axis is $\left[ 0,\frac{1}{2} \right]$ for a better visualization.}
						\label{fig:eigenvaluesGenschem}
				\end{figure}

			Before delving into the detailed analysis, it is instructive to analyze the solution. When $-m_{\alpha} \in [A,B]$ for a subset $\mathscr{F} \subset \left\{ 1,2, \dots, F \right\}$ of the $F$ mass scales, we see from \eqref{eq:genericrho} that the eigenvalue density is a sum of $\delta$-functions supported at $-m_{\alpha}$, for $\alpha \in \mathscr{F}$, as well as at the endpoints of $\text{supp} \rho $. The situation is schematically represented in figure \ref{fig:eigenvaluesGenschem}. The eigenvalues are clustered at $\vert \mathscr{F} \vert +2$ points, breaking the $U(N)$ group
			\begin{equation}
				U(N) \to U \left( c_A N \right) \times  U \left( c_B N \right) \times \prod_{\alpha \in \mathscr{F}} U \left( \frac{n_{\alpha} }{2}\right) 
			\end{equation}
			with each factor rotating the eigenvalues placed at the support of the corresponding $\delta$-function.
						
				\par
			Moving a mass, the corresponding $\delta$-function will eventually cross the boundary of $\text{supp} \rho$ and drop out. When $-m_{\alpha}$ hits $A$ or $B$, the corresponding coefficient $c_A $ or $c_B$ jumps by $\frac{\zeta_{\alpha} }{2}$ in order to preserve the total number $N$ of eigenvalues.
						
			\subsection{Pure gauge theory}
			\label{sec:UNPure}
				
				We start our analysis with the pure Yang--Mills-Chern--Simons theory without matter, thus setting $n_{\alpha}=0$. This theory lives in the IR of all the other theories with charged hypermultiplets, and is reached giving large masses to the matter fields and integrating them out. The presence of a \CS~level $k$ is therefore necessary, because it is generated dynamically along the RG flow as the effect of integrating out hypermultiplets.\par
\begin{figure}[ht]
					\centering
						\includegraphics[width=0.4\textwidth]{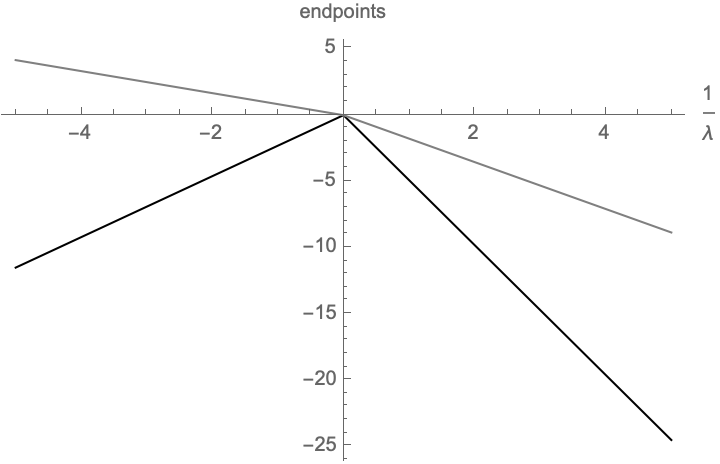}%
						\hspace{0.05\textwidth}%
						\includegraphics[width=0.4\textwidth]{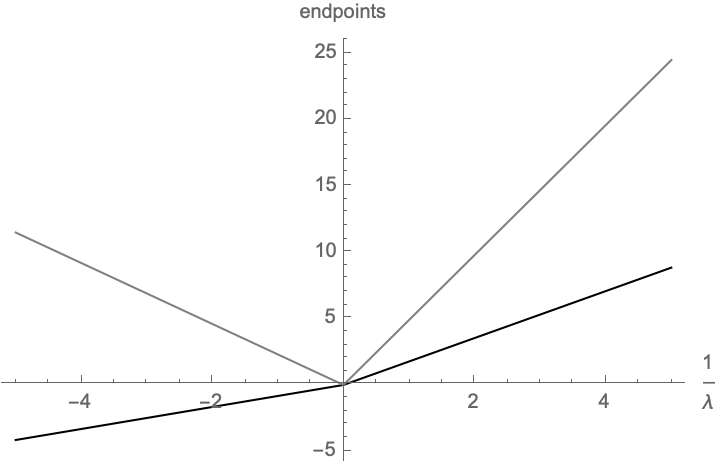}
						\caption{Plot of $A$ (black) and $B$ (gray) in the pure gauge theory. Left: $t=1.3$. Right: $t=-1.3$.}
					\label{fig:ABCSSYMUn}
					\end{figure}

				The SPE in pure \YM-\CS~theory is 
				\begin{equation}
				\label{eq:SPECSYMUN}
					- \int_A ^{B}  \dd  \psi \rho (\psi)~ \left( \phi - \psi \right)^2 \text{sign} \left( \phi - \psi \right)  =  \frac{1}{t} \phi^2 + \frac{2}{\lambda} \phi 
				\end{equation}
				and is solved by the ansatz $\rho (\phi) =  c_A \delta \left( \phi - A  \right) + c_B \delta \left( \phi - B \right)$. Following the steps described in subsection \ref{sec:largeNlimit}, we find 
				\begin{subequations}
				\begin{align}
						c_A & = \frac{1}{2} - \frac{1}{2t} , \qquad A = - \frac{t}{\lambda} \left( 1 \pm \sqrt{ \frac{t+1}{t-1} } \right) , \\
						c_B & = \frac{1}{2} + \frac{1}{2t} , \qquad B = - \frac{t}{\lambda} \left( 1 \pm \sqrt{ \frac{t-1}{t+1} } \right) ,
				\end{align}
				\end{subequations}
					with the same choice of sign of the square root in $A$ and $B$. Notice that we have derived the equations assuming $A<B$, and we must retain the solution which respects this hypothesis, depending on $\text{sign} (\lambda)$. We plot them in figure \ref{fig:ABCSSYMUn}.\par
					Removing the \YM~term sending $\vert \lambda \vert \to \infty$, the eigenvalues are attracted to the origin and, for a pure \CS~theory without any mass deformation, the saddle point configuration reduces to the trivial one. 
					
				\subsection{One mass scale}
				\label{sec:F1}
					We now consider a single mass scale, $F=1$. In other words, the theory has $N_f$ hypermultiplets all of equal mass $m$, and thus a single Veneziano parameter $\zeta$ as defined in \eqref{eq:Veneziano}.\par
					At very large values of the mass, the hypermultiplets can be integrated out to obtain an effective theory with Chern--Simons level $k - \frac{N_f}{2}  \text{sign} (m)$. Therefore, as $m$ is increased from $- \infty$ up to $+ \infty$, the effective description interpolates between two different pure Chern--Simons theories. At large gauge coupling, $\lambda \to \pm \infty$, there is no mass scale other than $m$, thus we expect a phase transition at $m=0$. Nevertheless, a finite $\lambda^{-1}$ sets a scale under which the hypermultiplet cannot be integrated out. We now show how this picture is realized.\par 
					The SPE reads 
					\begin{align}
						- \int^{B} _{A} \dd  \psi \rho (\psi ) \left( \phi - \psi \right)^2 \text{sign} \left( \phi - \psi \right) = \frac{1}{t} \phi^2 + \frac{2}{\lambda} \phi  - \frac{\zeta}{2} \left( \phi + m \right)^2 \text{sign} \left( \phi + m \right) .  \label{eq:SPEF1}
					\end{align}
					For clarity, we focus first on the limit $\vert \lambda \vert \to \infty$, in which the \YM~contribution drops out of the computations, and come back to the more general setting below.
					
					\subsubsection{Infinite Yang--Mills~'t Hooft coupling}
					\label{sec:UNF1YMinfty}
					
					We start increasing $m$ from $- \infty$, which gives $ \text{sign} \left( \phi + m \right) <0$. This inequality characterizes the first phase of the theory, which extends as long as $B<-m$. The explicit expressions of the solutions are reported in appendix \ref{app:rhoUN}, equation \eqref{eq:rhoUNF1YMinf}.\par
					The solution we have found holds as long as $B<-m$. Increasing $m$ from large negative values, $-m$ will descend and eventually hit $\text{supp} \rho $ at $B$. From the explicit form of $B$ in \eqref{eq:rhoUNF1YMinf} we find that the inequality $B<-m$ breaks down at $m=0$.\par
					We may assume the existence of an intermediate phase in which $m \in \text{supp} \rho$, but then the solution to \eqref{eq:SPEF1} would only be consistent with $A=0=B$. Therefore we pass to a new phase, for which $A>-m$ and hence $\text{sign} (\phi+m)>0$. The solution is found exactly as before, and is also recovered from the ones at $m<0$ flipping the sign of the Veneziano parameter, $\zeta \mapsto -\zeta$.\par
					We notice an important aspect: $\rho (\phi)$ is supported on the real line only for $t< -\left( 1-\frac{\zeta}{2} \right)^{-1}$ when $m<0$, and only for  $t>\left( 1-\frac{\zeta}{2} \right)^{-1}$ when $m>0$. We also find real solutions in the region $0<t<\left( 1-\frac{\zeta}{2} \right)^{-1}$ when $m<0$, and with opposite sign when $m>0$, which however fall out of the window \eqref{eq:tzetale1}. These solutions should not be discarded in principle, because the matrix model could still be defined at large $N$ if $\text{supp} \rho$ lies entirely on the positive real axis when $0<t<\left( 1-\frac{\zeta}{2} \right)^{-1}$, or on the negative real axis for negative $t$. However, evaluating $A$ and $B$ in that range, we find that the solutions do not satisfy the convergence condition, and therefore are inconsistent with the matrix model we have started with.\par
					We use $\rho (\phi)$ to evaluate the free energy $\mathcal{F}_{\mathbb{S}^5}$ \eqref{eq:freeenergy}. In the large $N$ and large $r$ limit and at infinite $\vert \lambda \vert $, $\mathcal{F}_{\mathbb{S}^5}$ is given by 
					\begin{equation}
						\mathcal{F}_{\mathbb{S}^5}  \left( t, \lvert \lambda \rvert \to  \infty, \zeta, m \right) = \frac{c_A c_B}{3} \vert B-A \vert^3 + \frac{1}{3t} \left( c_A A^3 + c_B B^3 \right) - \frac{\zeta}{6} \left( c_A \vert A+m \vert^3 + c_B \vert B + m \vert^3 \right) ,
					\end{equation}
					with $(c_A,c_B,A,B)$ functions of the gauge theoretical parameters as given in \eqref{eq:rhoUNF1YMinf}. The phase transition is third order, as proven by direct calculations, but it can also be predicted looking at the formula for $\mathcal{F}_{\mathbb{S}^5}$. It is a cubic function of $\vert m \vert ^3$, because $A$ and $B$ are linear functions of $m$: the expressions up to the second derivative will automatically vanish at $m=0$, determining the order of the phase transition.\par
\begin{figure}[htb]
					\centering
					\includegraphics[width=0.45\textwidth]{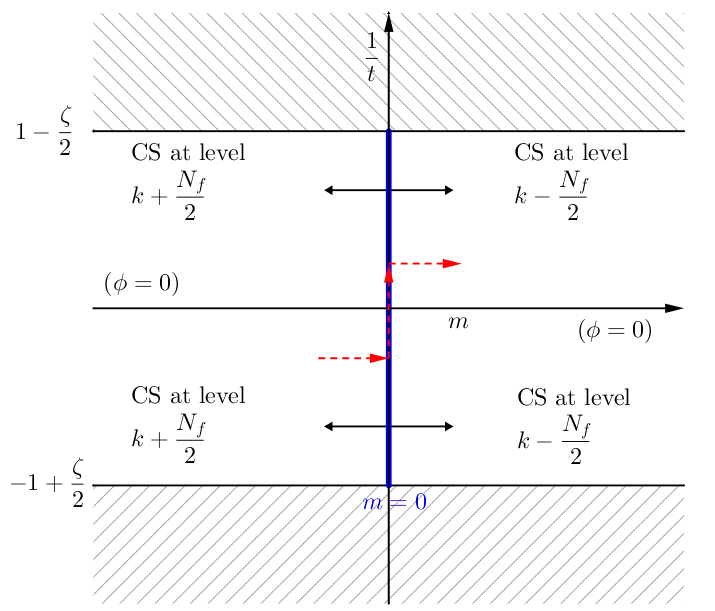}%
					\caption{Phases diagram of the theory with $N_f$ hypermultiplets all of mass $m$ at $\lambda \to \pm \infty$, in the $\left( m, \frac{1}{t} \right)$-plane. In the shaded region the matrix model is ill-defined. Crossing the blue wall $m_{\text{cr}}=0$, the theory undergoes a third order phase transition, indicated by the solid, black arrows in the picture. The dashed, red arrows indicate a phase transition between the two non-trivial regions.}
					\label{fig:sumF1Lambdainfty}
					\end{figure}

					To summarize, at infinite Yang--Mills 't Hooft coupling there are two phases separated by the critical surface $m_{\text{cr}}=0$. The result is schematically presented in figure \ref{fig:sumF1Lambdainfty}.\par
					
					\par

					\subsubsection{Finite \YM~'t Hooft coupling}
					
					We now come back to the more general setting with $\vert \lambda \vert < \infty $, hence turning on an additional massive deformation. We start again increasing $m$ from $- \infty$. The first phase is as in the large $\lambda$ limit studied above, but now for $\lambda^{-1} \ne 0$ the inequality $B<-m$ breaks down at a critical mass $m_{\text{cr},1}<0$. On the other hand, we could equivalently start decreasing $m$ from $+ \infty$, and see that the theory is in a phase equivalent to the second phase above. However, also in this case, the inequality $A>-m$ only holds for $m> m_{\text{cr},2}>0$. \par
					We see that the theory develops an intermediate phase 
					\begin{equation}
					\label{eq:middelphaseNf1}
						m_{\text{cr},1} < m < m_{\text{cr},2} ,
					\end{equation}
					in which the mass of the hypermultiplets is comparable to scale of the problem, determined by $\lambda^{-1}$. The matter fields cannot be integrated out and enter the IR dynamics. The deformation by $\lambda^{-1}$ has moved the two critical parameters away from $m_{\text{cr}}=0$.\par
\begin{figure}[bht]
					\centering
					\includegraphics[width=0.4\textwidth]{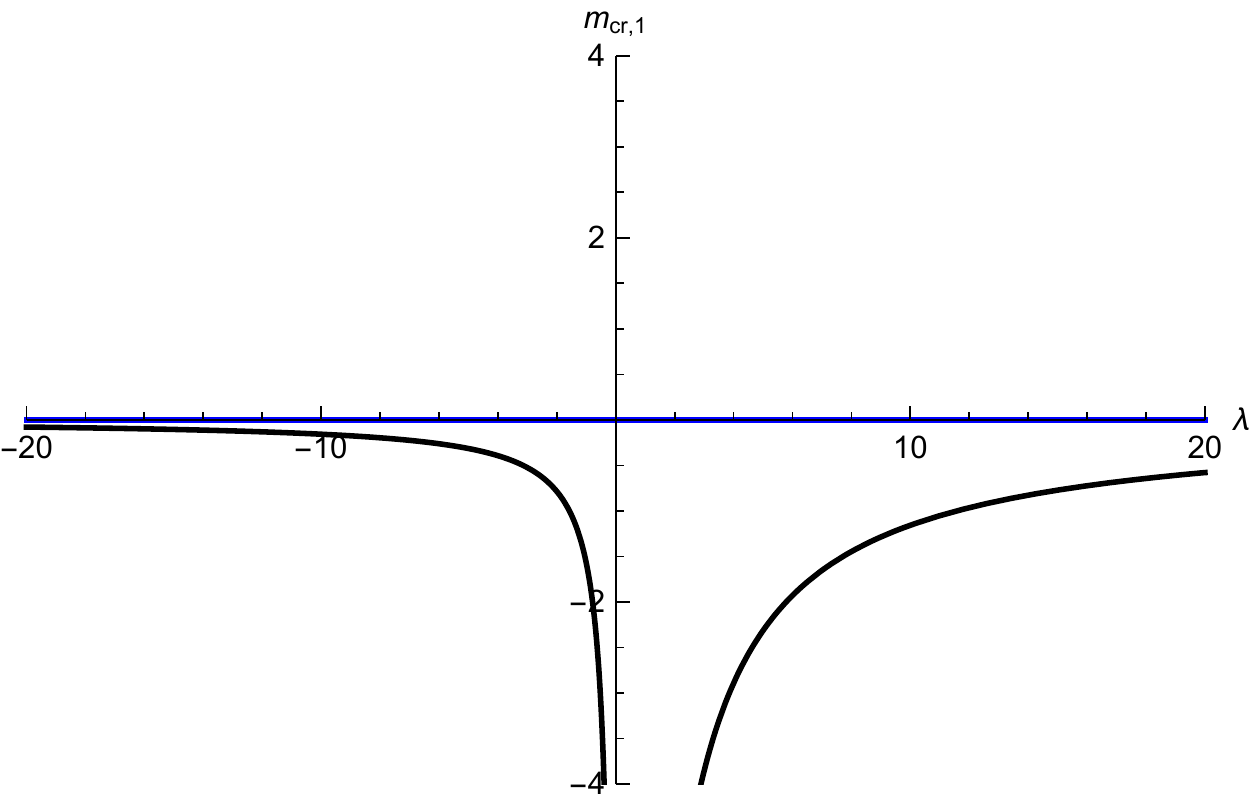}%
					\hspace{0.05\textwidth}
					\includegraphics[width=0.4\textwidth]{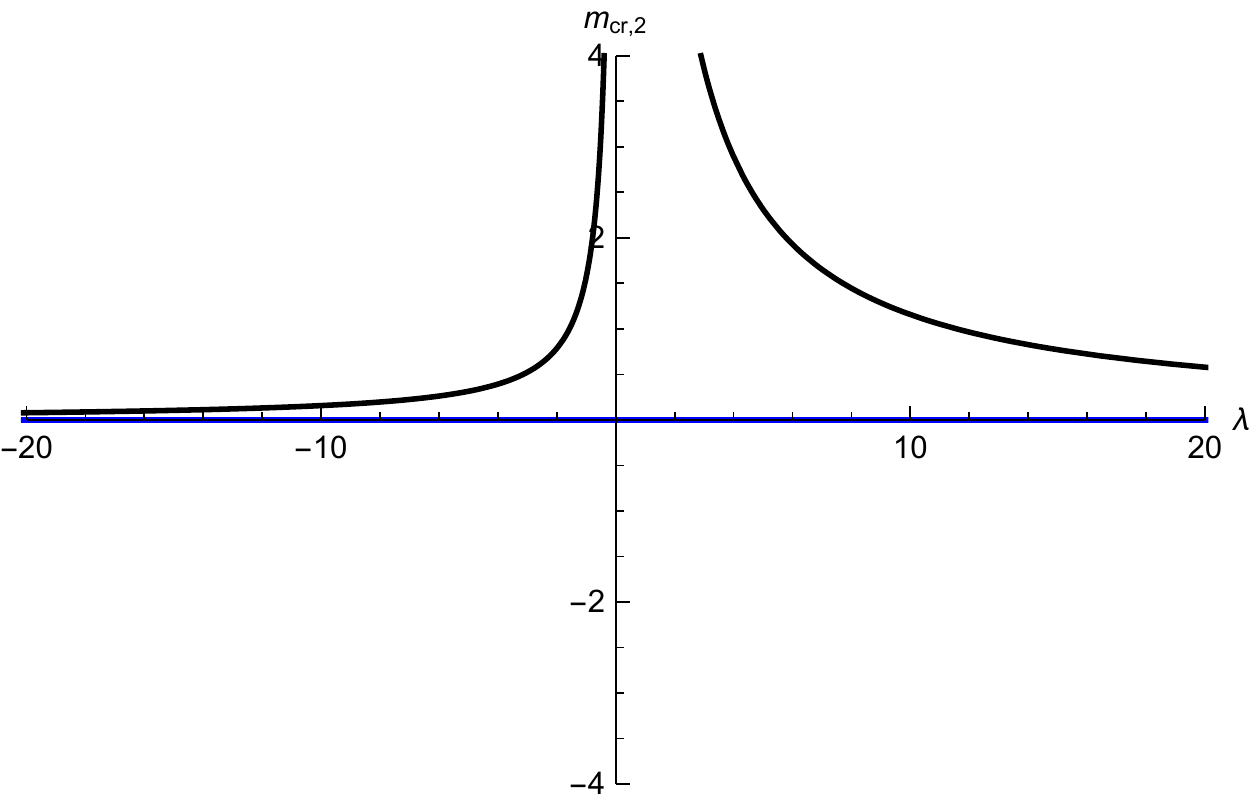}
					\caption{Critical surfaces plotted as functions of $\lambda$. Left: $m_{\text{cr},1}(t, \lambda, \zeta)$ at $\zeta= \frac{1}{2}$ and $t=-5$. Right: $m_{\text{cr},2}(t, \lambda, \zeta)$ at $\zeta= \frac{1}{2}$ and $t=5$. The blue horizontal line is the asymptote $m=0$.}
					\label{fig:mcrF1}
					\end{figure}

					The explicit form of the eigenvalue density $\rho (\phi)$ is given in \eqref{eq:rhoUNF1gen} in appendix \ref{app:rhoUN}. 
					As for large $\lambda$, we find that the first and third phases are non-trivial only for negative $t$ and for positive $t$, respectively. Imposing the condition 
					\begin{equation}
						B^{(\mathrm{I})} (t, \lambda, \zeta, m) = - m ,
					\end{equation}
					with $B$ computed under the assumption $B<-m$, we find the first critical surface $m=m_{\text{cr},1} (t, \lambda, \zeta)$. A direct computation using \eqref{eq:rhoUNF1gen} gives:
					\begin{equation}
						m_{\text{cr},1} = 	 \frac{t}{\lambda} \left( 1+ \sqrt{\frac{ (\zeta -2)t+2 }{\lambda^2  ((\zeta -2)t-2)} }  \right)  	
					\label{eq:mcr1F1}
					\end{equation}
					as plotted in the left panel of figure \ref{fig:mcrF1}. This solution vanishes for $\lambda \to \pm \infty$, in agreement with the discussion at infinite \YM~'t Hooft coupling.\par
					The second transition point is likewise determined decreasing $m$ from large positive values until the inequality $A>-m$ breaks down, see the right panel of figure \ref{fig:mcrF1}. Explicitly, this second critical surface is 
					\begin{equation}
						m_{\text{cr},2} (t, \lambda, \zeta ) = 	\frac{t}{\lambda} \left( 1- \sqrt{ \frac{ (\zeta -2)t- 2 }{ \lambda^2  ((\zeta -2)t+2) } } \right) 	.
					\label{eq:mcr2F1}
					\end{equation}
					
					\par
					
					We now pass to the intermediate phase \eqref{eq:middelphaseNf1}. The solution is well behaved and non-trivial in the whole allowed $(\zeta,t)$-region \eqref{eq:tzetale1}, and $A$ and $B$ do not depend explicitly on $m$, as we already know from the general solution \eqref{eq:genericrho}. This region is characterized by $-m \in [A,B]$, and therefore we can as well extract the critical values $m_{\text{cr},1}$ and $m_{\text{cr},2}$ from 
					\begin{equation}
						- m_{\text{cr},1} (t , \lambda, \zeta ) = B^{(\mathrm{II})} (t, \lambda, \zeta) , \qquad - m_{\text{cr},1} (t , \lambda, \zeta ) = A^{(\mathrm{II})} (t, \lambda, \zeta) ,
					\end{equation}
					where $^{(\mathrm{II})}$ means the quantity evaluated in the intermediate phase. The solutions \eqref{eq:mcr1F1}-\eqref{eq:mcr2F1} are correctly reproduced.\par
					\medskip
					We compute the free energy $\mathcal{F}_{\mathbb{S}^5}$ at finite $\lambda$. In the first and last phase, it has the form
					\begin{align}
						\mathcal{F}_{\mathbb{S}^5}  \left( t, \lambda , \zeta, m \right) &= \frac{c_A c_B}{3} \vert B-A \vert^3 + \frac{1}{3t} \left( c_A A^3 + c_B B^3 \right)  + \frac{1}{\lambda} \left(  c_A A^2 + c_B B^2   \right) \label{eq:FreeEnergyF1phI}\\
						& - \frac{\zeta}{6} \left( c_A \vert A+m \vert^3 + c_B \vert B + m \vert^3 \right) \notag
					\end{align}
					while in the middle phase we obtain 
					\begin{equation}
						\mathcal{F}_{\mathbb{S}^5}  \left( t, \lambda, \zeta, m \right) = \frac{ c_A c_B}{3} \vert B-A \vert^3 + \frac{1}{3t} \left( c_A A^3 + c_B B^3 - \frac{\zeta}{2} m^3 \right) + \frac{1}{\lambda} \left(  c_A A^2 + c_B B^2 + \frac{\zeta}{2} m^2 \right) . \label{eq:FreeEnergyF1phII}
					\end{equation}
					In these expressions, $\left(c_A, c_B, A,B  \right)$ are explicitly known functions of the gauge theoretical parameters $(t, \lambda,\zeta, m)$, given in \eqref{eq:rhoUNF1gen}.\par
\begin{figure}[bht]
					\centering
					\includegraphics[width=0.6\textwidth]{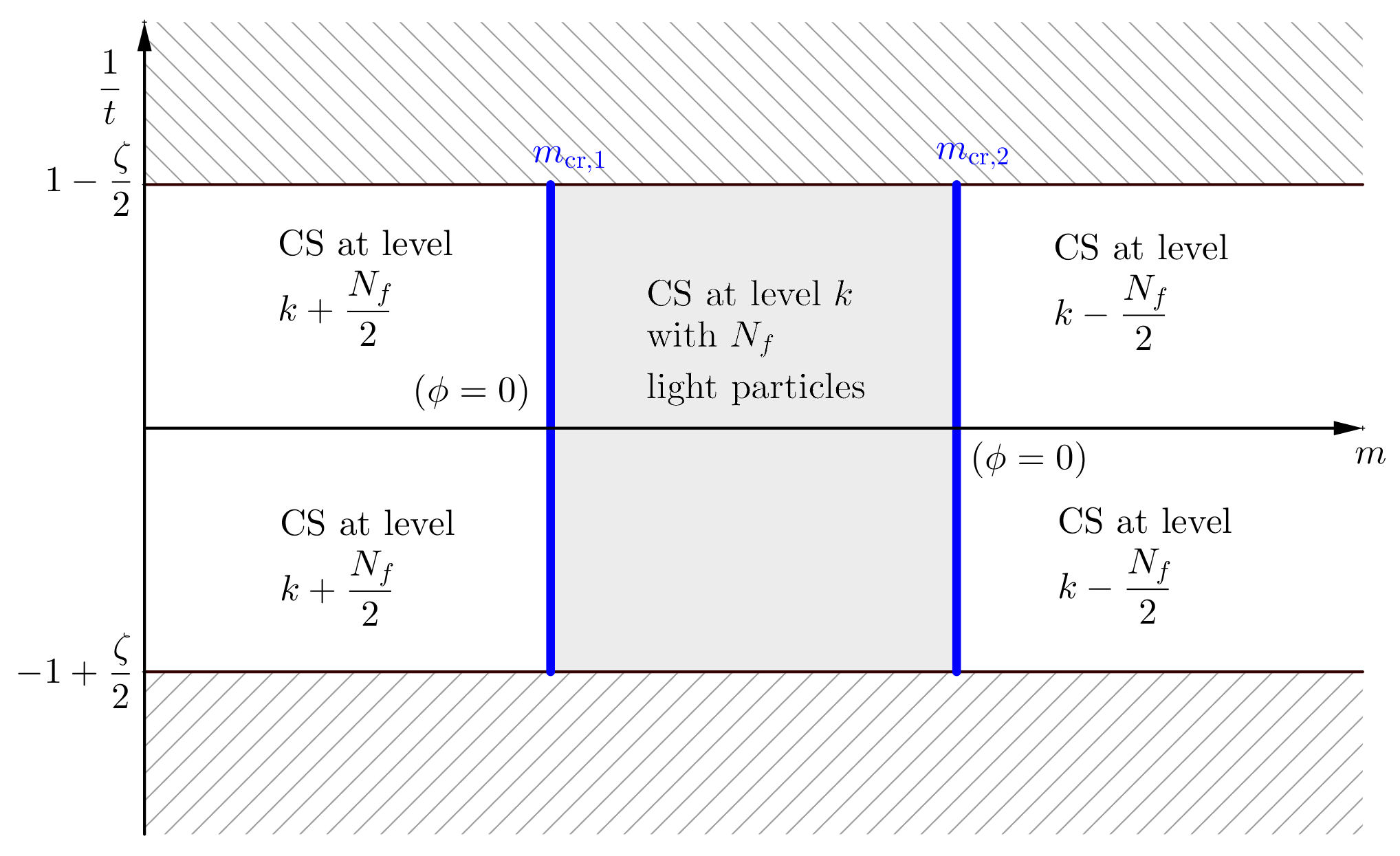}%
					\caption{Phase diagram of the theory with a single mass scale, plotted in the $\left( m, \frac{1}{t} \right)$-plane.}
					\label{fig:summaryF1}
					\end{figure}

					Taking the derivative of $\mathcal{F}_{\mathbb{S}^5}$ with respect to $m$, we find at one critical point 
					\begin{equation}
						 \left. \frac{ \partial  \mathcal{F}_{\mathbb{S}^5}  }{ \partial m  }  \right\rvert_{m \uparrow m_{\text{cr},1}}  =  \frac{ 2 t \zeta }{ \lambda^2 \left( 2+t (2 - \zeta )  \right) }=  \left. \frac{ \partial  \mathcal{F}_{\mathbb{S}^5}  }{ \partial m  }  \right\rvert_{m \downarrow m_{\text{cr},1}} 
					\end{equation}
					and a closely related expression at the other critical point. The second derivative however is discontinuous, thus we find a pair of second order phase transitions. We summarize the result in figure \ref{fig:summaryF1}.\par
					
					\par

					\subsubsection{Large Chern--Simons~'t Hooft coupling}
					\label{sec:F1larget}
					
						We now consider the limit of large \CS~'t Hooft coupling, $t \to \pm \infty $, that realizes the large $N$ limit at fixed \CS~level $k$.\par
						We start noting that, if $\lambda>0$ and $\zeta \le 2$ (possibly analytically continued to negative values), the effective action $S_{\text{eff}} (\phi)$ is non-negative definite, and the large $N$ limit describes trivial dynamics. On the contrary, for $\lambda<0$, $S_{\text{eff}} (\phi)$ admits a non-trivial saddle point configuration.\par
						The solution is given in \eqref{eq:rhoUNF1tinfty}. The critical values are
						\begin{subequations}
						\begin{align}
							m_{\text{cr},1} \left( \lvert t \rvert \to  \infty , \lambda, \zeta \right) & = \left[ \lambda \left( 1 - \frac{\zeta}{2} \right) \right]^{-1} , \\
							m_{\text{cr},2} \left( \lvert t \rvert \to  \infty , \lambda, \zeta \right) & = - \left[ \lambda \left( 1 - \frac{\zeta}{2} \right) \right]^{-1} .
						\end{align}
						\end{subequations}
						Recall that $\lambda<0$, so $m_{\text{cr},1}<0$ and $m_{\text{cr},2}= -m_{\text{cr},1} >0$.\par
						In the intermediate phase we find an even $\rho (\phi)$ with symmetric support $A=-B$: 
						\begin{equation}
							\rho (\phi) = \frac{2 - \zeta}{4} \left[ \delta \left( \phi + B \right) +  \delta \left( \phi - B \right) \right] +  \frac{\zeta}{2} \delta (\phi+m), \qquad B= - \left[ \lambda \left( 1 - \frac{\zeta}{2} \right) \right]^{-1} .
						\end{equation}
						The phase transitions are still second order.\par
						In the intermediate phase, the saddle point configuration clusters the eigenvalues in three peaks, around $A,B$ and $-m$. Approaching a critical value, the peak at $-m$ moves towards $A$ or $B$, and eventually the two sets of eigenvalues coalesce. The phase transition is thus a signal of the partial restoration of symmetry 
						\begin{equation}
							U \left(  \frac{N}{2} - \frac{N_f}{4} \right)^2  \times  U \left( \frac{N_f}{2}  \right)   \longrightarrow     U \left( \frac{N}{2} - \frac{N_f}{4}  \right) \times U \left(  \frac{N}{2} + \frac{N_f}{4}   \right) 
						\end{equation}
						in going from the second to the first or third phase. Note that this is a symmetry enhancement because $\frac{N_f}{2} \le N$.

				\subsection{Two opposite mass scales}
				\label{sec:F2s}
					We proceed in our analysis breaking the degeneracy in the masses of the hypermultiplets, setting $F=2$ distinct mass scales. We start with a symmetric setting, in which $n_1$ out of the $N_f < 2N$ fundamental hypermultiplets have mass $m$ and the others have mass $-m$. We work in the Veneziano limit \eqref{eq:Veneziano} and assume $\zeta_1 = \zeta_2 \equiv \zeta$ in this symmetric setting. The case with vanishing Chern--Simons level has been addressed in \cite{Nedelin}, finding two phases separated by a third order transition.\par
					The SPE is 
					\begin{align}
						- \int_{A} ^{B} \dd  \psi \rho (\psi) \left( \phi - \psi \right)^2 \text{sign} \left(  \phi - \psi \right) &= \frac{1}{t} \phi^2 + \frac{2}{\lambda} \phi    \label{eq:SPEF2}  \\
							& - \frac{\zeta}{2} \left[ \left( \phi +m \right)^2 \text{sign} \left(  \phi +m \right) +  \left( \phi -m \right)^2 \text{sign} \left(  \phi -m \right) \right] . \notag 
					\end{align}

					\subsubsection{Infinite \YM~'t Hooft coupling}
					\label{sec:F2YMinfty}
						We first consider the limit $\vert \lambda \vert \to \infty$. The solution to the SPE \eqref{eq:SPEF2} is given in \eqref{eq:rhoUNF2symYMinftpos}-\eqref{eq:rhoUNF2symYMinftneg}. We again find two phases, with a phase transition at $m=0$, as anticipated from general arguments. As discussed in section \ref{sec:largeNlimit}, we find a pair of solutions for $A$ and $B$ in each phase. One solution, reported in \eqref{eq:rhoUNF2symYMinftpos}, is consistent with $t> \left( 1- \zeta \right)^{-1}$ and the other, reported in \eqref{eq:rhoUNF2symYMinftneg}, is consistent with $t<- \left( 1- \zeta \right)^{-1}$. Crossing from one phase to the other, the solutions are mapped consistently.\par
\begin{figure}[htb]
						\centering
							\includegraphics[width=0.45\textwidth]{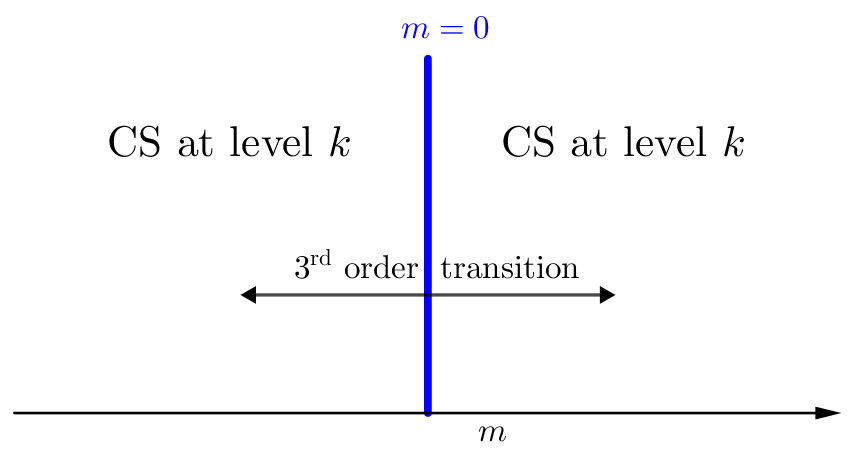}
							\caption{Phase diagram of the theory with two opposite mass scales at $\vert \lambda \vert \to \infty$: two effective \CS~theories are separated by a third order phase transition at $m=0$.}
							\label{fig:3rdorderm0}
						\end{figure}

						The free energy and its first and second derivatives are continuous at $m=0$ in this limit, but the third derivative is not. The situation is summarized in figure \ref{fig:3rdorderm0}.\par
						
						\par

					\subsubsection{Finite \YM~'t Hooft coupling}
						Reintroducing the mass deformation leading to a \YM~term brings in a new mass scale, and consequently an intermediate phase when $m$ is small compared to $\lambda^{-1}$. The phases corresponding to large positive or negative mass are found as for infinite \YM~coupling. The solution is given in \eqref{eq:rhoUNF2symtpos}.\par
						The asymmetry of $\text{supp} \rho$, that is $A \ne -B$, implies that we find different solutions for the critical value $m_{\text{cr},1}$, and the physically realized is the first one for which any of the two inequalities $A>m$ and $B<-m$ breaks down. We find that one scenario is realized for $t \lambda >0$ and the other for $t \lambda <0 $: 
						\begin{equation}
							m_{\text{cr},1} \left( t, \lambda, \zeta \right)  = \begin{cases} \frac{2 t}{\lambda  \left(\sqrt{t^2-1}+(2 \zeta -1) t-1\right)} & t \lambda >0 \\ \frac{t \left(\sqrt{t^2-1}+(2 \zeta -1) t+1\right)}{\lambda +\lambda  t (2 (\zeta +(\zeta -1) \zeta  t)-1)} &  t \lambda <0 . \end{cases}
						\end{equation}
						Beyond this first critical point, the system is in a new phase, in which the singularity at $\phi=m$ or at $\phi =-m$ enters in the interval $[A,B]$, while the other singularity falls out of the interval, see \eqref{eq:rhoUNF2symtpos} for the explicit solution. The effective theory in this new phase is equivalent to the $F=1$ theory with renormalized \CS~coupling. We find consistent solutions for $\lambda<0$. This second phase holds until the second singularity at $\pm m$ (depending on the sign of $t$) reaches $[A,B]$. For negative $t$, this means that a new phase transition takes place at $B=-m$, with $B$ computed in the second phase. This equation yields two solutions, but only one is consistent with $m_{\text{cr},1}$. The analogous reasoning applies to the other situation with positive $t$. We find a second phase transition at 
						\begin{equation}
							m_{\text{cr},2} \left( t, \lambda, \zeta \right)  = \begin{cases}  \frac{(\zeta -1) t^2+\sqrt{t^2 \left((\zeta -1)^2 t^2-1\right)}+t}{\lambda +(\zeta -1) \lambda  t} & t \lambda >0  \\ -  \frac{\sqrt{t^2 \left((\zeta -1)^2 t^2-1\right)}}{(\zeta -1) \lambda t- \lambda } - \frac{t}{\lambda} & t \lambda <0 ,  \end{cases}
						\end{equation}
						beyond which both $m$ and $-m$ belong to $[A,B]$. The solution is given in \eqref{eq:rhoUNF2symtpos}. In this third phase $\rho (\phi)$ has the same form for both positive and negative $t$, and holds for $\lambda^{-1} \in \mathbb{R}$. Increasing $m$ further, the system goes through the same phases in the converse direction, with the role of $m$ and $-m$ swapped. Such behaviour is expected in this especially symmetric case, due to the $\mathbb{Z}_2$ invariance under exchange of the masses, $m_1 \leftrightarrow m_2$.\par
\begin{figure}[htb]
						\centering
						\includegraphics[width=0.9\textwidth]{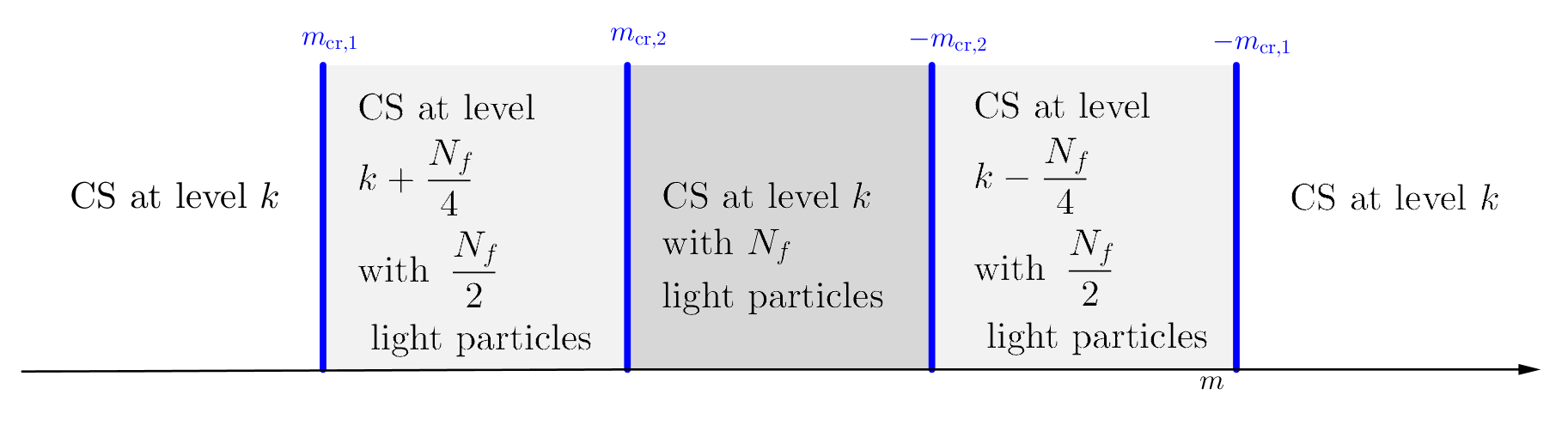}
						\caption{Phase diagram of the theory with two opposite mass scales.}
						\label{fig:F2Symphases}
						\end{figure}

						We summarize the phase structure in figure \ref{fig:F2Symphases}.\par
						\par
						
						The free energy $\mathcal{F}_{\mathbb{S}^5}$ is evaluated using the eigenvalue density $\rho (\phi)$ in each phase, as in \eqref{eq:freeenergygeneral}. Taking derivatives of the resulting expression we find a third order phase transition for both signs of $t \lambda$. This extends the result of \cite{Nedelin} to a more general setting.\par

					\subsubsection{Large \CS~'t Hooft coupling}
					\label{sec:UNF2symlarget}
						Consider two symmetric masses and $\vert t \vert \to \infty$. As in the one-mass setting, a non-trivial saddle point configuration requires $\lambda<0$. We find a symmetric $\rho (\phi)$ supported on $[-B,B]$, with 
						\begin{equation}
							 B= \begin{cases}  -\frac{1}{\lambda} - m \zeta  & m<m_{\text{cr}} \\ - \frac{1}{\lambda \left(1-\zeta \right)}   & m>m_{\text{cr}} .  \end{cases}
						\end{equation}
						The intermediate phases disappear in this limit, because 
						\begin{equation}
							\lim_{t \to \pm \infty } m_{\text{cr},2} \left( t, \lambda, \zeta \right) = \lim_{t \to \pm \infty } m_{\text{cr},1} \left( t, \lambda, \zeta \right) = \frac{1}{\lambda \left( 1- \zeta \right) } .
						\end{equation}
						The results of \cite{Nedelin} are then recovered.\footnote{The dictionary between \cite{Nedelin} and the present work is: $\Lambda \vert_{\text{there}} = - \lambda^{-1} \vert_{\text{here}}$ and $\frac{1}{2} \zeta \vert_{\text{there}} = \zeta \vert_{\text{here}}$.}\par
\begin{figure}[htb]
						\centering
						\includegraphics[width=0.35\textwidth]{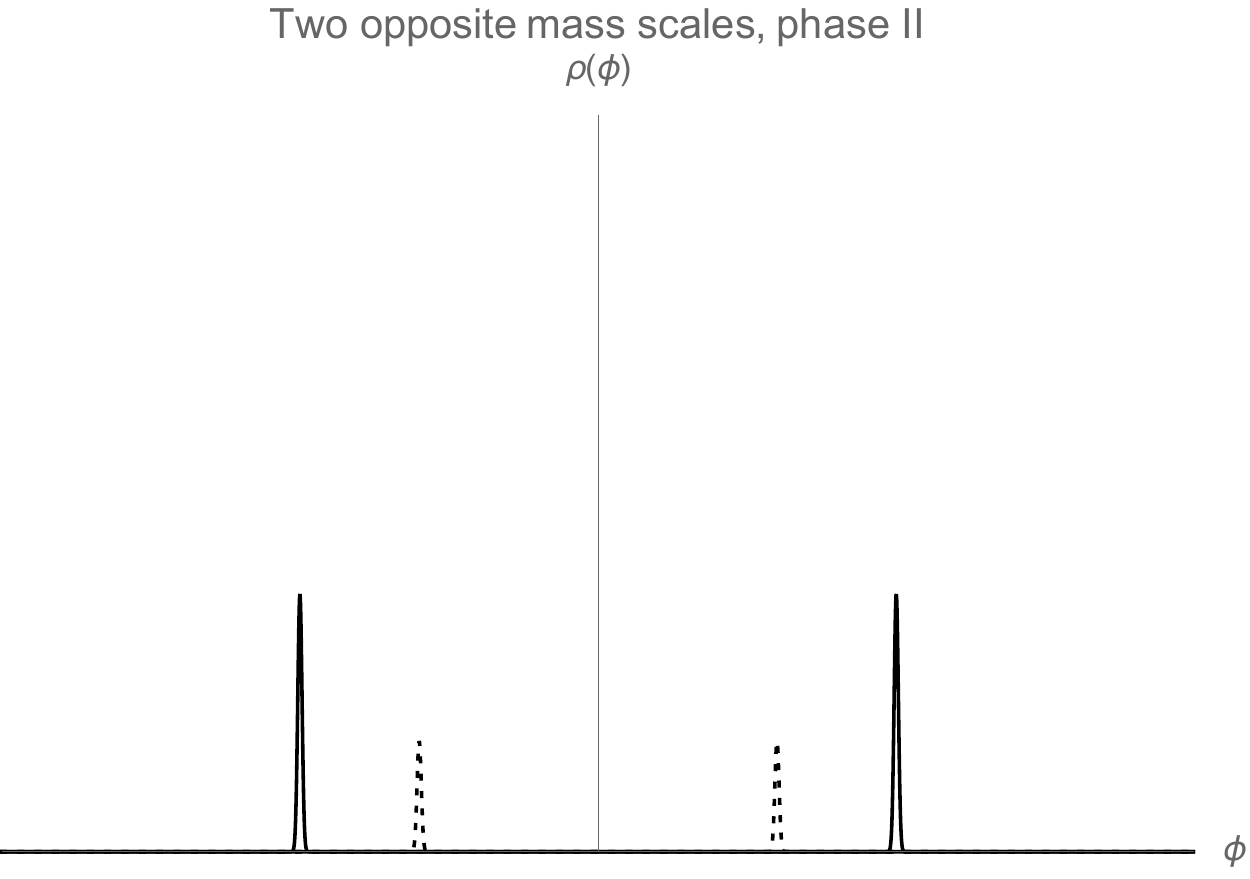}
						\hspace{0.05\textwidth}
						\includegraphics[width=0.35\textwidth]{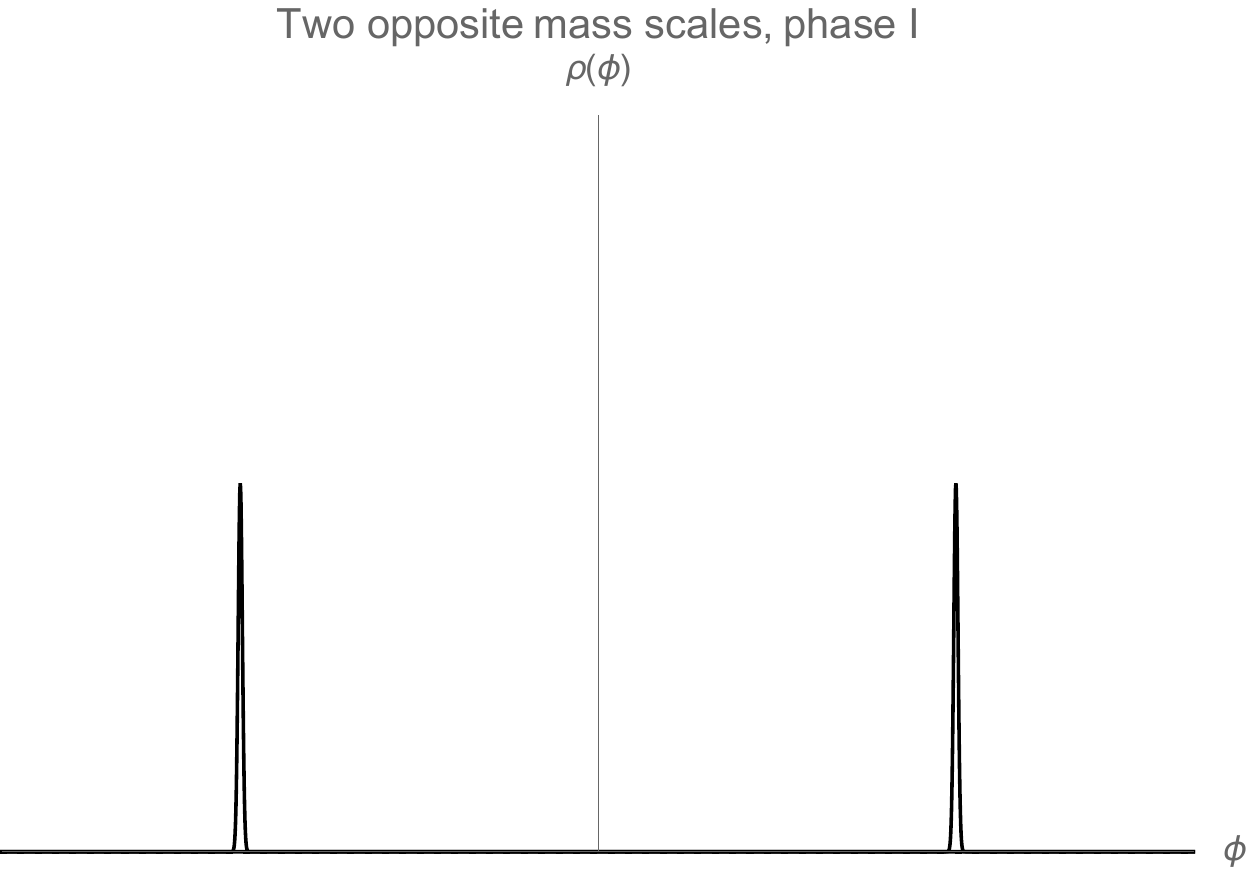}
						\caption{Schematic representation of the clustering of eigenvalues at large $N$ in the theory with two opposite mass scales. Left: Eigenvalue density in the intermediate phase, with $-m,m \in [-B,B]$ (dashed lines). Right: Eigenvalue density when $-m,m \notin [-B,B]$. The range of the vertical axis is $[0,1]$.}
						\label{fig:eigenvaluesF2tinf}
						\end{figure}

						This model has a $\mathbb{Z}_2$ symmetry. In the intermediate phase, $\rho (\phi)$ presents four clusters of eigenvalues, placed at $\pm B$ and $\pm m$. Approaching the critical locus, the eigenvalues at $m$ and the ones at $-B$ coalesce, and simultaneously the eigenvalues at $-m$ and the ones at $B$ coalesce, see figure \ref{fig:eigenvaluesF2tinf}. The phase transition is thus a signal of the symmetry enhancement 
						\begin{equation}
							U \left( \frac{N}{2} - \frac{N_f}{4}  \right)^2 \times U \left( \frac{N_f}{4}   \right)^2  \longrightarrow  U \left(  \frac{N}{2} \right)^2 .
						\label{eq:F2largetRestor}
						\end{equation}
						The $\mathbb{Z}_2$ symmetry is manifest on both sides of the arrow in \eqref{eq:F2largetRestor}.
						\par

					\subsection{Two mass scales}
					
						Consider now a generic assignment of masses $m_1$ and $m_2$, and two Veneziano parameters $\zeta_1$ and $\zeta_2$. The SPE is 
						\begin{align}
						- \int_{A} ^{B} \dd  \psi \rho (\psi) \left( \phi - \psi \right)^2 \text{sign} \left(  \phi - \psi \right)= \frac{1}{t} \phi^2 + \frac{2}{\lambda} \phi  & - \frac{\zeta_1}{2} \left( \phi +m_1 \right)^2 \text{sign} \left(  \phi +m_1 \right)  \label{eq:SPEnosymF2}  \\
							& - \frac{\zeta_2}{2} \left( \phi +m_2 \right)^2 \text{sign} \left(  \phi +m_2 \right) . \notag 
						\end{align}
						\par
						When $\vert m_1 \vert $ and $\vert m_2 \vert$ are both large, we find the usual solution $\rho (\phi)$ with two $\delta$-function singularities at the endpoints. The first phase transition takes place when one of the masses hits $\text{supp} \rho$. We focus first on infinite \YM~'t Hooft coupling limit and reintroduce the corresponding deformation later.

						\subsubsection{Infinite \YM~'t Hooft coupling}
						\label{sec:UNF2YMinfty}
							The complete solution to the SPE \eqref{eq:SPEnosymF2} in the limit $\frac{1}{\lambda} \to 0$ is given in \eqref{eq:rhoUNF2gen}.\par
							In order to effectively have a single free mass modulus, throughout the present subsection we impose this constraint
							\begin{equation}
							\label{eq:SUNf2const}
								\zeta_1 m_1 + \zeta_2 m_2 =0 .
							\end{equation}\par
							The first phase, as usual, arises when the masses fall out of $\text{supp} \rho$, and the solution $\rho (\phi)$ is a sum of $\delta$-functions at the endpoints $A$ and $B$ of the support, reported in equation \eqref{eq:rhoUNF2gen}. At this point, two possible scenarios disclose: moving the values of the masses with the constraint \eqref{eq:SUNf2const}, either the singularity at $\phi=-m_2$ hits $\text{supp} \rho $ from below, or the singularity at $\phi=-m_1$ hits $\text{supp} \rho $ from above. A simple computation imposing \eqref{eq:SUNf2const} shows that these two scenarios are realized simultaneously: the mass parameter hit the endpoints at $m_2=0=m_1$ for all values of $t$ consistent with \eqref{eq:tzetale1}. Crossing the critical point, the eigenvalue density in the new phase is obtained via the formal substitution $(\zeta_1, \zeta_2) \leftrightarrow (-\zeta_1, - \zeta_2)$, see \eqref{eq:rhoUNF2gen}.\par
							The free energy $\mathcal{F}_{\mathbb{S}^5}$ is, as usual, a cubic function of $A$, $B$ and $m_1$, $m_2$, and all of them vanish at the critical point. Taking derivatives, we find a third order phase transition. We conclude that the picture is equivalent to the symmetric case studied in subsection \ref{sec:F2YMinfty} and summarized in figure \ref{fig:3rdorderm0}.

						\subsubsection{Infinite \YM~'t Hooft coupling revisited}
						\label{sec:F2genlambdainftym1m2}
							Let us consider the situation in which the massive deformation leading to a \YM~term is removed, $\vert \lambda \vert \to \infty$, but dropping the constraint \eqref{eq:SUNf2const}. In this way, we have two real mass parameters to play with.\par
							The explicit solution for $\rho (\phi)$ is found by the standard calculation, and is reported in \eqref{eq:rhoUNF2YMinftyI}. Having two real mass moduli, we can either increase $m_1$ keeping $m_2$ fixed, or decrease $m_2$ keeping $m_1$ fixed, or any linear combination of the two. In the former case, the system undergoes a phase transition when the singularity at $-m_1$ hits $B$ from above, whilst in the latter case a phase transition takes place when the singularity at $-m_2$ hits $A$ from below.\par
							We move $m_1$ and study the new phase, characterized by $-m_1 \in [A,B]$. The explicit solution for $\rho (\phi)$ is reported in \eqref{eq:rhoUNF2YMinftyIIm1}. The critical surface is determined imposing $-m_1=B^{(\mathrm{II})}$, with $B$ evaluated in the second phase. If we compute the critical point from the first phase, we get two solutions, and the physical one is the lowest value, that is, the first value for which $-m_1>B$ does not hold as $m_1$ is increased from $- \infty$. The critical value obtained in this way matches the value of $B$ in the second phase, as required by consistency.\par
							From the second phase, we can either keep $-m_1 \in [A,B]$ and decrease $m_2$ until $-m_2$ reaches $A$, or increase $m_1$ further until $-m_1 <A$. Let us first focus on the former choice. We notice that, decreasing $m_2$ the support $[A,B]$ of $\rho (\phi)$ shrinks. Therefore, to keep $-m_1 \in [A,B]$ we should in fact decrease $m_1$ at the same time as we decrease $m_2$. This procedure leads to a third order phase transition at $m_2=0=m_1$.\par
							Increasing $m_1$ further with $m_2$ fixed at a large positive value, the singularity at $-m_1$ eventually reaches the lower boundary of $\text{supp} \rho$. This triggers a new phase transition. The critical values for both transitions encountered at fixed $m_2$ and moving $m_1$ are linear functions of $m_2$, see \eqref{eq:rhoUNF2YMinftyIIm1}. The rest of the phase diagram is described analogously.\par
							Evaluating the free energy in each phase, we find third order phase transitions.

						\subsubsection{Finite \YM~'t Hooft coupling}
							We come back to the general setting reintroducing a \YM~term.\par
							In the middle phase, when both $-m_1$ and $-m_2$ fall inside the support $[A,B]$ of the eigenvalue density, we find  
							\begin{subequations}
							\begin{align}
								c_A & = \frac{2-\zeta_1 - \zeta_2}{4} - \frac{1}{2t} , \qquad A= - \frac{t}{\lambda}  \left( 1 \pm \sqrt{  \frac{c_B}{c_A}  }  \right) , \\
								c_B & = \frac{2-\zeta_1 - \zeta_2}{4} + \frac{1}{2t} , \qquad B= - \frac{t}{\lambda}  \left( 1 \pm \sqrt{  \frac{c_A}{c_B}  }  \right) .
							\end{align}
							\end{subequations}
							In the latter expression the sign must be chosen in consistency with our starting assumption $A<B$, and is the same in both formulas.\par
							We henceforth focus on $\lambda<0$ for concreteness. In this case, $A<0$ and $B>0$ and are explicitly given in \eqref{eq:rhoUNF2genmiddle}. Let us assume we increase $m_2$ to positive values: the $\delta$-function singularity at $\phi=-m_2$ moves toward the endpoint $A$ and eventually hits the boundary of $\text{supp} \rho$ at 
							\begin{equation}
							\label{eq:F2m2cr1}
								m_{2 \text{cr},1} = \frac{t}{\lambda} \left(1- \sqrt{1-\frac{4}{\left(\zeta _1+\zeta _2\right) t-2 t+2}}\right)  .
							\end{equation}\par
							Increasing $m_2$ beyond this point, the gauge theory enters in a new phase in which the solution, that holds for positive $t$, is reported in \eqref{eq:rhoUNF2genintA}. Notice that $B$ soon becomes negative in this phase as we increase $m_2$, meaning that we should increase $m_1$ at the same time so that $-m_1 \in [A,B]$. Then, we can either increase $m_1$ further or decrease it, driving the system toward a new phase. The explicit results are reported in equations \eqref{eq:rhoUNF2genextA}-\eqref{eq:rhoUNF2genextB}.\par
							We can equivalently begin keeping $-m_2 \in [A,B]$ and moving $m_1$. If we increase $m_1>0$, we recover the setting just analyzed, upon relabelling $\zeta_1 \leftrightarrow \zeta_2$. Decreasing $m_1<0$, instead, the singularity at $\phi=-m_1$ is moved toward $B$, and eventually the theory undergoes a phase transition at 
							\begin{equation}
							\label{eq:F2m1cr1}
								m_{1 \text{cr},1} = - \frac{t}{\lambda} \left(\sqrt{\frac{4}{\left(\zeta _1+\zeta _2\right) t-2(t+1)}+1}-1\right)  .
							\end{equation}
							The solution in the new phase, characterized by $-m_2 \in [A,B]$ and $B<-m_1$, is given in \eqref{eq:rhoUNF2genintB}, and holds for negative $t$. The description of the other phases is obtained in a completely analogous fashion.\par
\begin{figure}[ht]
							\centering
								\includegraphics[width=0.45\textwidth]{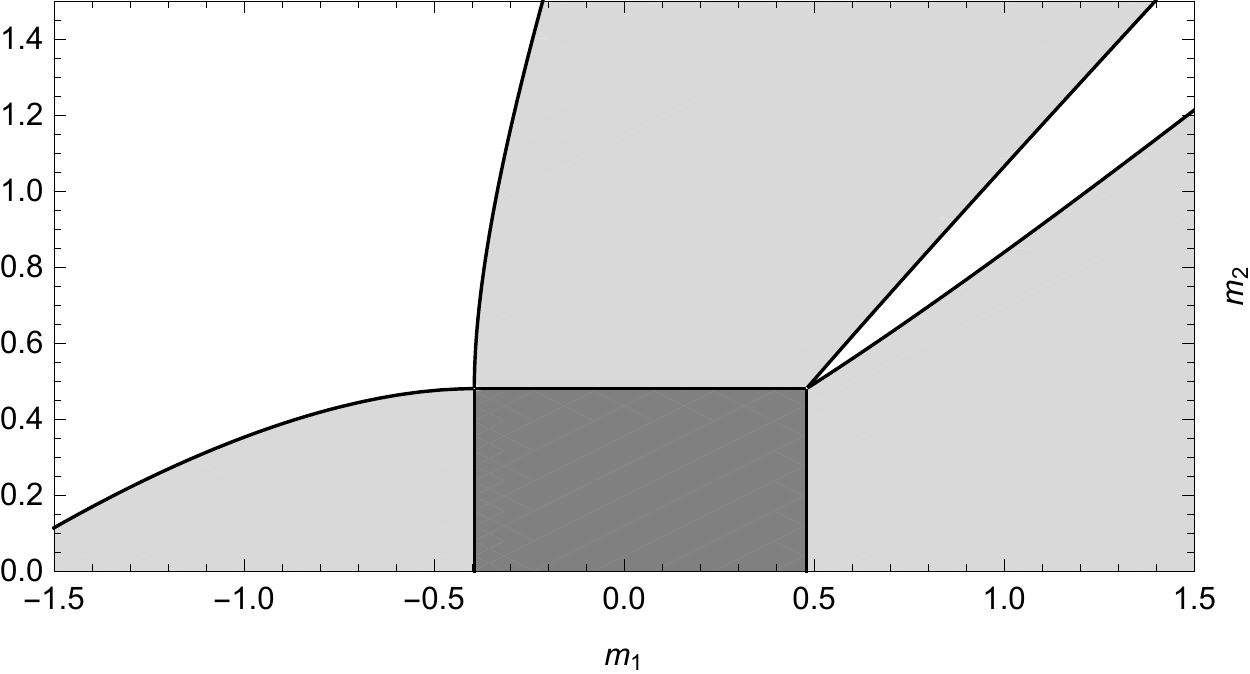}
								\caption{Phase diagram in the half-plane $\left\{ m_1 \in \mathbb{R} , m_2 \ge 0 \right\}$. The plot is at $(t, \lambda, \zeta_1, \zeta_2)=  \left( 22,-10, \frac{1}{3}, \frac{6}{5} \right)$. In the darker shaded region both singularities lie in the support of the eigenvalue density. In the lighter shaded regions one singularity lies inside and the other lies outside the support. In the white region none of the singularities lies inside the support.}
								\label{fig:F2phasem1m2}
							\end{figure}
\begin{figure}[htb]
							\centering
								\includegraphics[width=0.75\textwidth]{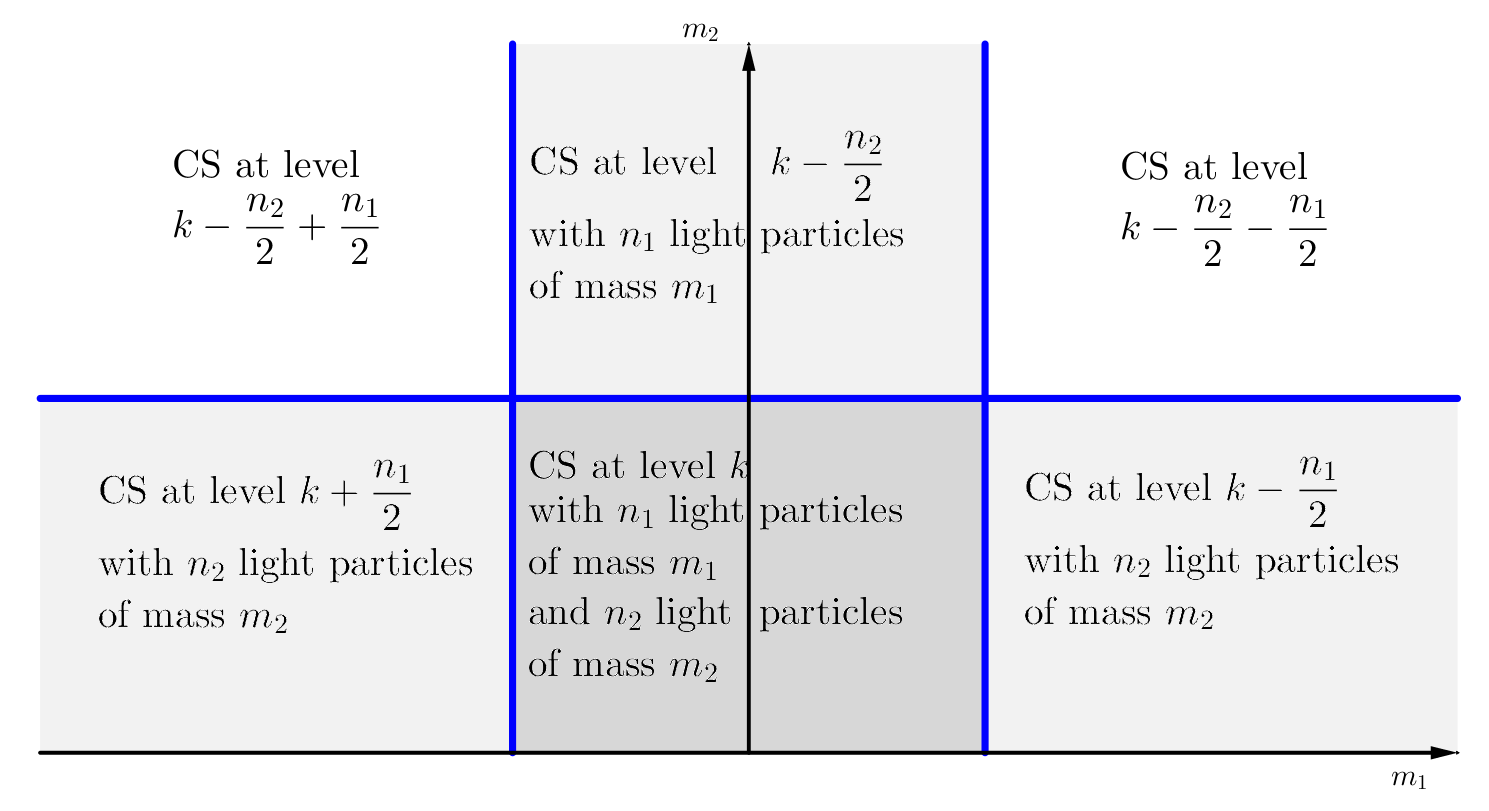}
								\caption{Phase diagram of the theory with two mass scales. Note that the solid blue walls, representing the critical surfaces, are not straight lines in the $(m_1,m_2)$-plane, cf. figure \ref{fig:F2phasem1m2}.}
								\label{fig:F2PhaseStrqual}
							\end{figure}

							We plot the phase structure of the $F=2$ theory in the $(m_1,m_2)$-plane in figure \ref{fig:F2phasem1m2}. A more qualitative description of the phases is in figure \ref{fig:F2PhaseStrqual}.\par
							
							\par
							
						\subsubsection{Limiting cases}
							The present framework with $F=2$ encompasses the previously studied theories as special cases. Setting $\zeta_2=0$ we expect to recover the $F=1$ theory of subsection \ref{sec:F1}, whilst setting $\zeta_2 = \zeta_1$ we should recover the symmetric framework of subsection \ref{sec:F2s}.\par
							In the first mentioned limiting case we check that, sending $\zeta_2 \downarrow 0 $, all the expressions in appendix \ref{app:rhoUN} for the $F=2$ theory reduce to $F=1$. An alternative approach is to take $\vert m_2 \vert  \to \infty$ and integrate out the massive hypermultiplets. Then we obtain the $F=1$ theory with a renormalized \CS~'t Hooft coupling $t\vert_{F=1} = t\vert_{F=2} - \frac{n_2}{2}$. This result may be visualized comparing the upper strip of figure \ref{fig:F2PhaseStrqual} with the phase diagram of the $F=1$ theory in figure \ref{fig:summaryF1}.\par
							To recover the $F=2$ symmetric case with two opposite masses, set $m_1 =-m_2$. Moving along the diagonal in the $(m_1 \le 0, m_2 \ge 0)$-quadrant of figure \ref{fig:F2phasem1m2} reproduces the phases on the right half $(m_2 \ge 0)$ of figure \ref{fig:F2Symphases}.

					\subsection{Three or more mass scales}
					\label{sec:UNF3}
						The generic solution for $F$ real mass scales is \eqref{eq:genericrho}, and the procedure is a direct extension of what we have presented so far. The explicit determination of the phase diagram requires a detailed case by case study, with each mass $m_{\alpha}$ moved independently. The upshot is that, for generic $\lambda$ and $\left\{ m_{\alpha} \right\}$, a second order phase transition takes place whenever one of the singularities drops in or out of $[A,B]$.\par
						We have followed a bottom-up approach in our presentation, starting with a pure gauge theory in subsection \ref{sec:UNPure} and increasing $F$. We might have adopted a top-down approach as well, following the RG flow. Indeed, starting with a given $F$, the other theories with lower $F^{\prime}<F$ are phases of the original theory, reached giving large mass to $(F- F^{\prime})$ families of hypermultiplets.

	\subsubsection{Infinite \YM~'t Hooft coupling and constrained masses}
	\label{sec:UNF3YMinfty}
	
		While, as we have shown, the phase transitions are generically second order for $U(N)$, there is a selected sub-class of theories for which we find third order transitions. These are \CS~theories at infinite \YM~coupling and with a single modulus controlling the theory, 
		\begin{equation}
		\label{eq:malphaxm}
			m_{\alpha} = x_{\alpha} m , \qquad x_{\alpha} \in \R, \ \alpha =1, \dots, F .
		\end{equation}
		Without loss of generality, we impose 
		\begin{equation}
			\sum_{\alpha=1} ^{F} x_{\alpha} ^2 =1 ,
		\end{equation}
		as any scaling of all the $x_{\alpha}$ together can be absorbed in a redefinition of $m$. The definition \eqref{eq:malphaxm} is a change to polar coordinates in $\mathbb{R}^F$ for each sign of $m$, with $ \vert m \vert $ parametrizing the radial direction. Note, however, that the theories we consider allow $m \in \R$.\par
		For very large $m$, which we take positive for concreteness, the singularities fall out of $\text{supp} \rho$, either above or below depending on the sign of $x_{\alpha}$. We get 
		\begin{equation}
			c_A = - \frac{1}{2t} + \frac{1}{2} + \sum_{\alpha} \frac{\zeta_{\alpha}}{4} \text{sign} ( x_{\alpha} )  , \quad c_B= \frac{1}{2t} + \frac{1}{2} - \sum_{\alpha} \frac{\zeta_{\alpha}}{4} \text{sign} ( x_{\alpha} ) ,
		\end{equation}
		while $A$ and $B$ are linear functions of $m$, 
		\begin{equation}
			A = m A_0, \ B = m B_0, \qquad \text{$A_0$ and $B_0$ independent of $m$} .
		\end{equation}
		This implies that, moving $m$, all the singularities reach the boundary of $\text{supp} \rho$ simultaneously at $m=0$. Recall that the free energy $\mathcal{F}_{\mathbb{S}^5}$ is a cubic function of $A$, $B$ and $\left\{ m_{\alpha} \right\}$. It follows that the free energy is continuous and vanishing at the critical locus, up to its second derivative. We establish that the phase transition is third order.\par
		In the geometric picture sketched in appendix \ref{sec:geometric}, the rewriting \eqref{eq:malphaxm} corresponds to take all the K\"{a}hler parameters that are dual to non-compact divisors in a resolution of the Calabi--Yau threefold $X$ to be proportional to a single parameter $m$. Then, sending $\lambda^{-1} \to 0$ first, corresponds to keep the volume $\mathrm{vol} \left( \mathbb{P}^1 _{0} \right)$ of a certain curve $\mathbb{P}^1 _0$ finite while by the number of exceptional divisors fibered over it grows to infinity (see appendix \ref{sec:geometric} for notation and definitions). After that, we decrease the K\"{a}hler parameter $m$ controlling the volumes of the non-compact exceptional divisors, until it vanishes. Then, the gauge theory undergoes a third order phase transition, which agrees with the expected geometric flop transition.\par
		From the explicit results in subsection \ref{sec:F2genlambdainftym1m2}, the present picture with the associated third order transition is expected to hold even dropping the constraint \eqref{eq:malphaxm}.

						\subsection{Wilson loops}
						\label{sec:UNWL}
	
							We have argued in subsection \ref{sec:WLlimit} that Wilson loops are always continuous but generically not differentiable. It is worthwhile to focus on the special instances in which the partition function undergoes a third order transition, and analyze the behaviour of the Wilson loops.
	
							\subsubsection{Fundamental Wilson loop: Two opposite mass scales}
							We consider the theory without \CS~term and with two opposite mass scales, with equal Veneziano parameters, discussed in \cite{Nedelin} and revisited in subsection \ref{sec:UNF2symlarget}. The vev of a Wilson loop in the fundamental representation is 
							\begin{equation}
								\langle \mathcal{W}_{\mathsf{F}} \rangle = \begin{cases} \cosh \left( 2 \pi r \left( \zeta m - \frac{1}{\lambda} \right) \right) & \vert m \vert > \frac{1}{\lambda (1- \zeta)}  \\  (1 - \zeta ) \cosh \left( \frac{2 \pi r }{ \lambda (1-\zeta) } \right) + \zeta \cosh \left( 2 \pi r m \right)   & \vert m \vert < \frac{1}{\lambda (1- \zeta)} . \end{cases}
							\end{equation}
							Taking the logarithm and differentiating, we find 
							\begin{subequations}
							\begin{align}
								\left. \frac{\partial \ }{\partial m} \log \langle \mathcal{W}_{\mathsf{F}} \rangle  \right\rvert_{m \downarrow  \frac{1}{\lambda (1- \zeta)} } -  \left. \frac{\partial \ }{\partial m} \log \langle \mathcal{W}_{\mathsf{F}} \rangle  \right\rvert_{m \uparrow  \frac{1}{\lambda (1- \zeta)} } & = 0 \\
								\left. \frac{\partial^2  \ }{\partial m^2 } \log \langle \mathcal{W}_{\mathsf{F}} \rangle  \right\rvert_{m \downarrow  \frac{1}{\lambda (1- \zeta)} } -  \left. \frac{\partial^2  \ }{\partial m^2 } \log \langle \mathcal{W}_{\mathsf{F}} \rangle  \right\rvert_{m \uparrow  \frac{1}{\lambda (1- \zeta)} } & = \zeta (1- \zeta)
							\end{align}
							\end{subequations}
							meaning that the Wilson loop vev experiences a second order non-analyticity, one order less than the free energy. Note that the second and higher derivatives vanish as $\zeta \to 1$, because in that case there exists a single phase valid for all $m$. To conclude, we mention that, as we work in the decompactification limit, the functions $\cosh (2 \pi r x)$ should be replaced by $e^{\vert 2 \pi x \vert }$ in all the expression above. Using this substitution before taking the derivatives does not alter the conclusion.

						\subsubsection{Fundamental Wilson loop: Infinite \YM~coupling}
							The other situation in which the phase transition is third order is for theories without \YM~'t Hooft coupling, $\vert \lambda \vert \to \infty$. We consider the $F=1$ theory of section \ref{sec:F1} as an explicit example. The endpoints $A=m A_0$ and $B=m B_0$ are linear functions of $m$, and 
							\begin{equation}
								\left. \frac{\partial \ }{\partial m}  \log \langle \mathcal{W}_{\mathsf{F}} \rangle  \right\rvert_{m \to 0} = 2 \pi r \frac{ c_A A_0 + c_B B_0 }{ c_A + c_B} = 2 \pi r  \left( c_A A_0 + c_B B_0  \right) .
							\end{equation}
							This does not vanish unless $c_A A_0 +c_B B_0 =0$, and therefore the Wilson loop vev has a first order discontinuity. This may indicate an inconsistency in the strong coupling limit of non-balanced $U(N)$ theories, or at least an ambiguity in the order of strong coupling and large $N$ limits.

						\subsubsection{Antisymmetric Wilson loop: Pure gauge theory}
							We now apply the framework presented in section \ref{sec:WLlimit} to compute the expectation value of half-BPS Wilson loops in antisymmetric representations of large rank.\par
							We begin with the pure gauge theory analyzed in section \ref{sec:UNPure}. The theory has no mass scales other than the inverse \YM~'t Hooft coupling $\lambda^{-1}$, and presents a single phase. Specializing the argument of section \ref{sec:WLlimit} to such theory without hypermultiplets, we have to evaluate 
							\begin{equation}
								\left\langle \mathcal{W}_{\mathsf{A}_K} \right\rangle = e^{rAK} \oint \frac{d \tilde{w}}{2 \pi i } ~ \tilde{w}^{K-1} \left[  1+ \frac{1}{\tilde{w}} \right]^{ \lfloor N c_A \rfloor }  \left[  1+ \frac{e^{r (B-A)} }{\tilde{w}} \right]^{ \lfloor N c_B \rfloor } 
							\end{equation}
							keeping the leading contribution at large radius. We observe that such contribution will differ depending on $c_B> \kappa $ or $c_B < \kappa $, where the scaling parameter $\kappa= \frac{K}{N}$ has been introduced in \eqref{eq:defkappaWL}. We find 
							\begin{equation}
								\log \left\langle \mathcal{W}_{\mathsf{A}_K} \right\rangle = \begin{cases} r  B K  &  K \le  \frac{N + k }{2}  \\  r \left[  A K +  \left(B-A \right)  \left( \frac{N + k }{2} \right) \right] &  K >  \frac{N + k }{2}  \end{cases}
							\end{equation}
							with $k$ the \CS~level. The inequalities are understood at large $N$. Besides, strictly speaking this solution only holds as long as $\kappa \le c_A + c_B$, but recalling the normalization $c_A+c_B=1$ and that $0 \le \kappa \le 1$ by definition, this latter requirement is always satisfied. We stress that the Wilson loop is a continuous but not differentiable function of $\kappa$.\par
							This result holds for all gauge theories with massive matter, in the phases in which all the masses fall outside of $\text{supp} \rho$, up to a renormalization of the \CS~coupling.

						\subsubsection{Antisymmetric Wilson loop: One mass scale}
							We discuss the antisymmetric Wilson loop in the $U(N)$ theory with a single mass scale $m$. In the first and last phase, with $-m>B$ and $-m<A$ respectively, the solution is analogous to the one for the pure gauge theory, upon replacement $\frac{N+k}{2} \mapsto \frac{ N + k }{2} + \frac{N_f}{4}$ if $-m>B$, and $\frac{N+k}{2} \mapsto \frac{ N + k }{2} - \frac{N_f}{4}$ if $-m<A$.\par
							In the intermediate phase, characterized by $A<-m<B$, we must take into account two possibilities, namely $A< -m < B-A$ and $B-A <-m <B$. The final results in the two sub-cases are 
							\begin{subequations}
							\begin{align}
								\left.  \log \left\langle \mathcal{W}_{\mathsf{A}_K} \right\rangle^{(\mathrm{II})} \right\rvert_{-m < B-A}  &  = \begin{cases}  {\scriptstyle r B K  } &  {\scriptstyle 0 \le K  < \frac{N + k }{2} - \frac{ N_f}{4}  }  \\  {\scriptstyle r \left[ \left(  A - m \right)  \left( K - \frac{N + k }{2} + \frac{ N_f}{4} \right)  +  B  \left( \frac{N + k }{2} - \frac{ N_f}{4} \right) \right] } &    {\scriptstyle \frac{N + k }{2} - \frac{ N_f}{4}   \le K  < \frac{N + k }{2} + \frac{ N_f}{4}   }  \\  {\scriptstyle  r \left[  A  \left( K - \frac{N + k }{2} + \frac{ N_f}{4} \right)   + B  \left( \frac{N + k }{2} - \frac{ N_f}{4} \right)    - m  \frac{N_f}{2} \right]  } &    {\scriptstyle  K \ge  \frac{N + k }{2} + \frac{ N_f}{4}  }  , \end{cases}   \\
								\left.  \log \left\langle \mathcal{W}_{\mathsf{A}_K} \right\rangle^{(\mathrm{II})} \right\rvert_{-m > B-A}  &  = \begin{cases}  {\scriptstyle r \left( A -m  \right) K } &   {\scriptstyle 0 \le K  < \frac{ N_f}{2}  }  \\  {\scriptstyle r \left[  \left( A -m  \right) \frac{ N_f}{2} +  B  \left( K  - \frac{ N_f}{2} \right)   \right] } &  {\scriptstyle \frac{N_f}{2} \le K < \frac{N +k}{2} + \frac{ N_f}{4}  } \\  {\scriptstyle r \left[   A  \left( K - \frac{N + k }{2} + \frac{ N_f}{4} \right)   + B  \left( \frac{N + k }{2} - \frac{ N_f}{4} \right)    - m  \frac{N_f}{2} \right] }  &    {\scriptstyle K \ge  \frac{N + k }{2} + \frac{ N_f}{4}  }  , \end{cases}
							\end{align}
							\end{subequations}
							where the superscript $^{(\mathrm{II})}$ means that we have computed the Wilson loop vev in the intermediate phase. Condition \eqref{eq:tzetale1} guarantees that the inequalities are always well posed.\par
							The Wilson loop is continuous but not differentiable function of $m$ at both critical loci. It is also a continuous but not differentiable function of the scaling parameter $\kappa$ in every phase.\par
							The study of the expectation value of a Wilson loop in a large antisymmetric representation $\mathsf{A}_K$ for any number of mass scales can be addressed by the method presented here, specializing the argument of section \ref{sec:WLlimit} to a given $F$ and analyzing the various sub-cases in each phase.

				\section{Phases of ${SU(N)}$ theories}
				\label{sec:SuN}
				
				$SU(N)$ gauge theories descend from (twisted) compactifications of $6d$ SCFTs on a circle of radius $\beta$, with the \YM~deformation $h \propto \beta^{-1}$.\par
				The analysis is closely related to that of the $U(N)$ theory, but we now reintroduce the Lagrange multiplier $\tilde{\xi}$. At the end, we will set it to its physical value, determined by the requirement 
					\begin{equation}
					\label{eq:condSun}
						\int_{A} ^{B} \phi \rho (\phi)  \dd \phi = 0 .
					\end{equation}\par
				The details for finding the explicit solution $\rho (\phi)$ in each $SU(N)$ theory are exactly as in the corresponding $U(N)$ theory analyzed in section \ref{sec:UN}, except that the endpoints $A$, $B$ will carry an additional dependence on $\tilde{\xi}$. Notice that $c_A$ and $ c_B$ are the same in the $U(N)$ and $SU(N)$ theory, and for this reason we will not explicitly discuss them throughout this section.\par
				The generic solution $\rho (\phi)$ is again given by \eqref{eq:genericrho}, and the constraint \eqref{eq:condSun} reads 
				\begin{equation}
				\label{eq:condxi0}
					c_A A + c_B B - \sum_{\alpha=1}^{F} c_{\alpha} m_{\alpha} =0.
				\end{equation}
				This is meant as an equation fixing $\tilde{\xi}$ through the dependence of $A$ and $B$ on it.\par
				\medskip
				A remark is in order to clarify the role of the multiplier $\tilde{\xi}$. A power counting in the integral representation \eqref{eq:Zloc} of the partition function suggests that the difference between $SU(N)$ and $U(N)$ is sub-leading in a $\frac{1}{N}$ expansion. Indeed, only the ratio $\tilde{\xi} \propto \frac{\xi}{ N} $ enters the SPE \eqref{eq:LargeNspe}, showing that ungauging an Abelian factor $U(1) \subset U(N)$ gives a next-to-leading order correction. In this work we do not go beyond the leading order at large $N$, and adopt the approach of \cite{Minahan:2014hwa} scaling the Lagrange multiplier in a 't Hooft-like way, with $\tilde{\xi}$ fixed at large $N$, to keep track of the tracelessness condition at large $N$. Stated more formally, we work in the direct limit Lie algebra $\mathfrak{u} (\infty)$ and restrict to the traceless subspace.\par
				From the geometric engineering viewpoint, the 't Hooft limit \eqref{eq:tHooft} blows up the volume of a curve $\mathbb{P}_0 ^1$ belonging to the base $\mathscr{B}$ of the elliptic fibration $\widetilde{X} \to \mathscr{B}$. Imposing the same scaling for $\tilde{\xi}$ corresponds to scale the metric on the base $\mathscr{B}$ in such a way that the volume of a different curve $\mathbb{P}^1 _{\ast} \subset \mathscr{B}$, transverse to $\mathbb{P}_0 ^1$, grows linearly with $\mathrm{vol}\left( \mathbb{P}_0 ^1 \right)$. Essentially, this procedure amounts to keep track of the difference between ALF and ALE metrics on the Calabi--Yau threefold.\footnote{The author thanks M. Del Zotto for this remark.}

						\subsection{Pure gauge theory}
						\label{sec:SunPure}
						The first theory we consider is the pure \YM-\CS~gauge theory without matter. We compute $A$ and $B$ when the multiplier $\tilde{\xi}$ is taken into account in the SPE, and then impose \eqref{eq:condxi0} and solve for $\tilde{\xi}$. This gives 
						\begin{equation}
							\tilde{\xi} = - \frac{t}{\lambda^2 (t^2 -1)} .
						\end{equation}
						Plugging this value back in $A$ and $B$ we obtain 
						\begin{equation}
							A= - \frac{t}{\lambda} \mp \frac{ t^2}{\lambda (t-1)} , \qquad B= - \frac{t}{\lambda} \pm \frac{ t^2}{\lambda (t+1)} , 
						\end{equation}
						with signs chosen consistently depending on $\text{sign}(\lambda)$. There is no crucial difference between $SU(N)$ and $U(N)$ pure gauge theories, except for the details in determining $A$ and $B$. In the present case, the tracelessness condition implies $A<0$ and $B>0$ $\forall \lambda^{-1} \in \R$.

						\subsection{One mass scale}
						The first example which includes matter is the $F=1$ theory. As in subsection \ref{sec:F1} we discuss first the $\vert \lambda \vert \to \infty $ case, and then reintroduce a finite \YM~term.

							\subsubsection{Infinite Yang--Mills~'t Hooft coupling}
\begin{figure}[htb]
								\centering
								\includegraphics[width=0.5\textwidth]{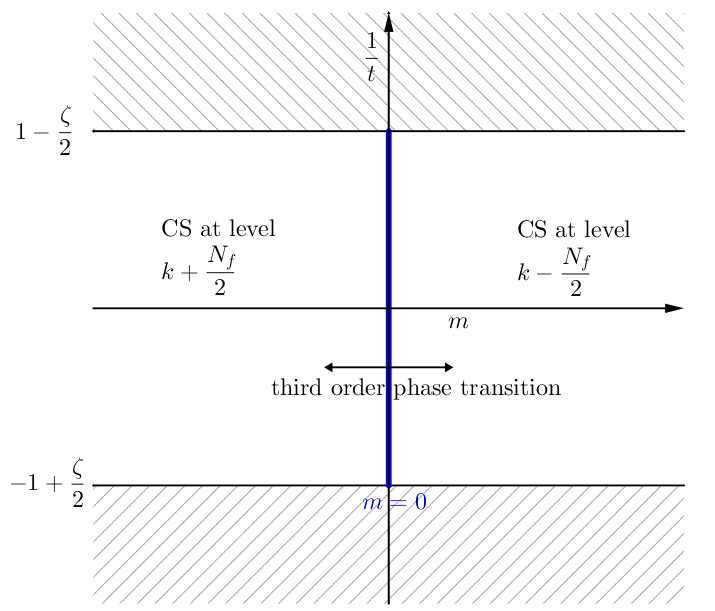}%
								\caption{Phase diagram of the $SU(N)$ theory with $N_f$ hypermultiplets all of mass $m$ at $\lambda \to \pm \infty$. Across the blue wall $m_{\text{cr}}=0$ the theory undergoes a third order phase transition.}
								\label{fig:sumSunF1Lambdainf}
							\end{figure}

							We begin with the analysis of the $SU(N)$ theory with all hypermultiplets of equal mass and no \YM~term. Solving for $A$, $B$ and imposing \eqref{eq:condxi0} we find 
							\begin{equation}
								\tilde{\xi} = \begin{cases} -\frac{\zeta  m^2 (t (\zeta -2 t)+2)}{\left(\zeta ^2-4\right) t^2+4 \zeta  t+4} & m<0 \\ 
									-\frac{\zeta  m^2 (t (\zeta +2 t)-2)}{\left(\zeta ^2-4\right) t^2-4 \zeta  t+4}  & m>0 . \end{cases}
							\end{equation}
							Plugging this back into $A$ and $B$ in both phases gives the explicit solution reported in appendix \ref{app:rhoSuN}, equation \eqref{eq:rhoSuNF1YMinf}. These expressions are much simpler than the ones obtained in the $U(N)$ theory. We find a third order phase transition at $m=0$ but, in contrast to the $U(N)$ theory, the solution is non-trivial for all values of $\frac{1}{t}$ in the window \eqref{eq:tzetale1} on both sides of the critical wall. The phase structure is represented in figure \ref{fig:sumSunF1Lambdainf}.\par
		
							\par

							\subsubsection{Finite \YM~'t Hooft coupling}
							At finite \YM~'t Hooft coupling $\vert \lambda \vert < \infty $, we solve the SPE and impose the constraint \eqref{eq:condxi0}, which gives 
							\begin{equation}
								\tilde{\xi} = \begin{cases}  \frac{\zeta  \lambda ^2 \left(-m^2\right) (t (\zeta -2 t)+2)+2 \zeta  \lambda  m t (\zeta t+2)+2 t (\zeta  t+2)}{\lambda ^2 ((\zeta -2) t+2) ((\zeta +2) t+2)} & m< m_{\text{cr},1} \\
									\frac{t (- m \zeta  \lambda (\zeta  \lambda  m-2 (\zeta -2) t)-4)}{\lambda ^2 \left((\zeta -2)^2 t^2-4\right)} & m_{\text{cr},1} < m< m_{\text{cr},2} \\
									\frac{\zeta  \lambda ^2 \left(-m^2\right) (t (\zeta +2 t)-2)+2 \zeta  \lambda  m t (\zeta  t-2)-2 t (\zeta  t-2)}{\lambda ^2 ((\zeta -2) t-2) ((\zeta +2) t-2)}	 & m> m_{\text{cr},2} .  \end{cases}
							\end{equation}
							The endpoints of the support, reported in \eqref{eq:rhoSuNF1full}, are uniquely determined and take an especially simple form. The two critical surfaces are 
							\begin{subequations}
							\begin{align}
								m_{\text{cr},1} (t, \lambda , \zeta ) & = \left[  \lambda \left( 1 + \frac{1}{t} \right) \right]^{-1} \\ 
								m_{\text{cr},2} (t, \lambda , \zeta ) & = \left[ - \lambda \left( 1 - \frac{1}{t} \right) \right]^{-1} 
							\end{align}
							\end{subequations}
							which, remarkably, are independent of $\zeta$. These solutions have been obtained under the assumption $A<B$, which is self-consistent only for $\lambda<0$, in agreement with the analysis of the $U(N)$ theory. A major difference with the $U(N)$ theory is that the condition \eqref{eq:condxi0} has introduced an explicit dependence on $m$ in $A$ and $B$ in the intermediate phase. This dependence is necessary to balance the average $\int_A ^B \phi \rho(\phi) \dd  \phi$ as the $\delta$-function at $-m$ is moved inside the interval $[A,B]$.\par
							The free energy is explicitly given by
							\begin{subequations}
							\begin{align}
								\left. \mathcal{F}_{\mathbb{S}^5}  \right.\rvert_{m<m_{\text{cr},1}} & = \frac{2 \zeta  \lambda ^3 m^3+2 \zeta ^2 \lambda ^3 m^3 t-t^2 (\zeta  \lambda  m (\lambda m (3 \zeta +2 \lambda  m)+6)+4)}{3 \lambda ^3 ((\zeta -2) t+2) ((\zeta +2) t+2)} \label{eq:FSuNF1phase1} \\ 
								\left. \mathcal{F}_{\mathbb{S}^5} \right.\rvert_{ m_{\text{cr},1}<m<m_{\text{cr},2} } & = \frac{2 \zeta  \lambda ^3 m^3-(\zeta -2) t^3 \left(3 \zeta  \lambda ^2 m^2+2\right)+2 \zeta  \lambda  m t^2 \left((\zeta -1) \lambda ^2 m^2+3\right)-6 \zeta  \lambda ^2 m^2 t}{3 \lambda ^3 \left((\zeta -2)^2 t^3-4 t\right)} 
							\end{align}
							\end{subequations}
							in the first and second phase respectively, and by \eqref{eq:FSuNF1phase1} with $\zeta \mapsto - \zeta$ in the third phase. 
							Taking derivatives, we obtain 
							\begin{subequations}
							\begin{align}
								\left. \frac{ \partial \mathcal{F}_{\mathbb{S}^5} }{ \partial m } \right\rvert_{m \uparrow m_{\text{cr},1}} & =  -\frac{2 \zeta  t^3}{\lambda ^2 (t+1)^2 ((\zeta -2) t+2)} = \left. \frac{ \partial \mathcal{F}_{\mathbb{S}^5} }{ \partial m } \right\rvert_{m \downarrow m_{\text{cr},1}}  \\
								\left. \frac{ \partial^2 \mathcal{F}_{\mathbb{S}^5} }{ \partial m^2 } \right\rvert_{m \uparrow m_{\text{cr},1}} & = -\frac{2 \zeta  (t-1) t}{\lambda  (t+1) ((\zeta -2) t+2)} = \left. \frac{ \partial^2 \mathcal{F}_{\mathbb{S}^5} }{ \partial m^2 } \right\rvert_{m \downarrow m_{\text{cr},1}}
							\end{align}
							\end{subequations}
							yielding a third order phase transition. We conclude that the phase diagram of the $SU(N)$ theory is similar to that of the corresponding $U(N)$ theory, but with different critical loci and the order of the transitions is increased from second to third.\footnote{A change from second to third order phase transition upon removal of a centre of mass ``gauge'' degree of freedom was observed in \cite{Santilli:2020ueh} in a different context. Likewise, in the Gross--Witten--Wadia model \cite{Gross:1980he,Wadia:1980cp,Wadia:2012fr} the centre $U(1) \subset U(N)$ decouples, so the third order transition agrees with the predictions of the present section.}

							\subsubsection{Large Chern--Simons~'t Hooft coupling}
		
							For $\lambda<0$ and $\vert t \vert \to \infty $ the solution is given in \eqref{eq:rhoSuNF1CSinf}. The system has three phases, separated by third order transitions at the critical curves  
							\begin{equation}
								m_{\text{cr},1} = \lambda^{-1} = - m_{\text{cr},2}  .
							\end{equation}
							In contrast to subsection \ref{sec:F1larget}, the support of the eigenvalue density does not collapse sending $ \vert \lambda \vert \to \infty$.

						\subsection{Two opposite mass scales}
						\label{sec:SuNF2opposite}
						We consider the $SU(N)$ theory with $F=2$ opposite mass scales $(-m,m)$ and equal number of hypermultiplets per mass, $\zeta_1 = \zeta_2 \equiv \zeta$. This is the traceless counterpart of the analysis carried out in subsection \ref{sec:F2s}. The SPE reads 
						\begin{align}
							- \int_{A} ^{B} \dd  \psi \rho (\psi) \left( \phi - \psi \right)^2 \text{sign} \left(  \phi - \psi \right) &= \frac{1}{t} \phi^2 + \frac{2}{\lambda} \phi  + \tilde{\xi}  \label{eq:SPESuNF2}  \\
								& - \frac{\zeta}{2} \left[ \left( \phi +m \right)^2 \text{sign} \left(  \phi +m \right) +  \left( \phi -m \right)^2 \text{sign} \left(  \phi -m \right) \right] . \notag 
						\end{align}

							\subsubsection{Infinite \YM~'t Hooft coupling}
							To begin with, we remove the \YM~term sending $\vert \lambda \vert \to \infty$, thus the unique mass scale remaining in the problem in $m$. Solving \eqref{eq:SPESuNF2} for $A$ and $B$ as functions of $\tilde{\xi}$ and imposing \eqref{eq:condxi0} we get 
							\begin{equation}
							\label{eq:tildeiSuNF2s}
								\tilde{\xi} =  -\frac{4 \zeta ^2 m^2 t}{t^2-1} 
							\end{equation}
							in both phases. $A$ and $B$ are given in \eqref{eq:rhoSuNF2sYMinf}. We find a third order phase transition at $m=0$, consistent with the general arguments presented so far.

							\subsubsection{Finite \YM~'t Hooft coupling}
							Without loss of generality we restrict the analysis to $m>0$, thanks to the $\mathbb{Z}_2$ symmetry exchanging the two sets of hypermultiplets.\par
							Reintroducing a finite \YM~term, $\vert \lambda \vert < \infty$, we solve \eqref{eq:SPESuNF2} in analogy with the $U(N)$ theory of section \ref{sec:F2s}. The solution in each phase is reported in equation \eqref{eq:rhoSuNF2sfull}. The Lagrange multiplier $\tilde{\xi}$ takes the values 
							\begin{equation}
								\tilde{\xi} = \begin{cases}   -\frac{t \left(\frac{1}{\lambda }-\zeta  m\right)^2}{t^2-1} & m > m_{\text{cr},1} \\ 
								\frac{\zeta  \lambda ^2 \left(-m^2\right)+\zeta  (t-(\zeta -1) \lambda  m t)^2+t (2-2 \zeta  \lambda  m)}{2 \lambda ^2 (t+1) ((\zeta -1) t+1)} & m_{\text{cr},2} < m < m_{\text{cr},1} \text{ and } t \lambda <0 \\
								\frac{\zeta  \lambda ^2 m^2-\zeta  (t-(\zeta -1) \lambda  m t)^2+t (2-2 \zeta  \lambda  m)}{2 \lambda ^2 (t-1) ((\zeta -1) t-1)}   & m_{\text{cr},2} < m < m_{\text{cr},1} \text{ and } t \lambda >0 \\
								-\frac{t}{\lambda ^2 \left((\zeta -1)^2 t^2-1\right)}   & m< m_{\text{cr},2}  .  \end{cases}
							\end{equation}\par
							The analysis is carried out as for the corresponding unitary theory. The critical loci are 
							\begin{subequations}
							\begin{align}
								m_{\text{cr},1} (t, \lambda, \zeta ) & = \begin{cases} \frac{t}{\lambda  (\zeta  t-t+1)} & t \lambda <0 \\  \frac{t}{\lambda  (\zeta  t-t-1)} & t \lambda >0 ,  \end{cases} \\ 
								m_{\text{cr},2} (t, \lambda, \zeta ) & = \begin{cases} -\frac{t}{\lambda -\zeta  \lambda  t+\lambda  t} & t \lambda <0 \\ \frac{t}{\lambda +(\zeta -1) \lambda  t}   & t \lambda >0 . \end{cases}
							\end{align}
							\end{subequations}\par
							The phase diagram is qualitatively analogous to that of the corresponding $U(N)$ theory, but the expressions for $A$ and $B$, as well as the critical loci, are much simpler, as shown in \eqref{eq:rhoSuNF2sfull}. The free energy is directly evaluated in each phase, giving 
							\begin{subequations}
							\begin{align}
								\left. \mathcal{F}_{\mathbb{S}^5}  \right\rvert_{m>m_{\text{cr},1}} & =  { \scriptstyle \frac{\zeta  \lambda ^3 m^3-\zeta  \lambda  m t^2 \left(\lambda  m \left(\left(\zeta ^2+1\right) \lambda  m-3 \zeta \right)+3\right)+t^2}{3 \lambda ^3 \left(t^2-1\right)}  }  \\
								\left. \mathcal{F}_{\mathbb{S}^5}  \right\rvert_{ m < m_{\text{cr},1}} & =   \begin{cases}  { \scriptstyle  \frac{1}{6 (t+1)} \left(  \frac{3 \zeta  m^2 (\zeta  t+t+1)}{\lambda }-\frac{\zeta  m^3 (\zeta  t+t+1)^2}{t}-\frac{3 \zeta  m t}{\lambda ^2}+\frac{(\zeta -2) t^2}{\lambda ^3 ((\zeta -1) t+1)}   \right)  } &   { \scriptstyle  t \lambda < 0  } \\   
									{ \scriptstyle \frac{3 \zeta  \lambda ^2 m^2 t ((\zeta -1) t-1) (\zeta  t+t-1)-\zeta  \lambda ^3 m^3 ((\zeta -1) t-1) (\zeta  t+t-1)^2+3 \zeta  \lambda  m t^2 (-\zeta  t+t+1)+(\zeta -2) t^3}{6 \lambda ^3 (t-1) t ((\zeta -1) t-1)} } &  { \scriptstyle t \lambda > 0  }  \end{cases}  \\
								\left. \mathcal{F}_{\mathbb{S}^5}  \right\rvert_{m<m_{\text{cr},2}} & = { \scriptstyle \frac{\zeta  \lambda ^2 m^2 (3-4 \zeta  \lambda  m)-(\zeta -1)^3 t^4 \left((\zeta -1) \zeta  \lambda ^2 m^2 (4 \zeta  \lambda  m-3)+1\right)+(\zeta -1) t^2 \left(2 (\zeta -1) \zeta  \lambda ^2 m^2 (4 \zeta  \lambda  m-3)+5\right)}{3 \lambda ^3 \left((\zeta -1)^2 t^2-1\right)^2}  } .
							\end{align}
							\end{subequations}
							Differentiating, a third order phase transition is found.

						\subsection{Wilson loops}
						\label{sec:SuNWL}
						From subsection \ref{sec:WLlimit}, the vev of a half-BPS Wilson loop in the fundamental representation is 
						\begin{equation}
							\left.\langle \mathcal{W}_{\mathsf{F}} \right\rangle = c_A e^{2 \pi r A } + c_B e^{2 \pi r B} + \sum_{\alpha=1} ^{F} c_{\alpha} e^{- 2 \pi r m_{\alpha}} 
						\end{equation}
						and its continuity at the critical surfaces follows from the continuity of $A$ and $B$ and the associated jump of $c_A$ or $c_B$ by $\frac{\zeta}{2}$.\par
						When the gauge group is $SU(N)$, we can exploit the additional constraint \eqref{eq:condxi0} to prove that the derivative of $\left.\langle \mathcal{W}_{\mathsf{F}} \right\rangle$ is continuous, too. We show this for the phase with $m_{\alpha} \notin [A,B]$ $\forall \alpha=1, \dots, F$, being the extension to any other phase straightforward.\par
						Differentiate equation \eqref{eq:condxi0} together with \eqref{eq:system3eqsUNb} with respect to a given $m$ on both sides of the critical wall, and use the resulting expressions to get rid of the derivatives $\frac{\partial A}{\partial m}$ and $\frac{\partial B}{\partial m}$ in the formula for the derivative of the Wilson loop vev. This gives 
						\begin{subequations}
						\begin{align}
							\frac{1}{2 \pi r}  \left.  \frac{ \partial \ }{ \partial m } \left.\langle \mathcal{W}_{\mathsf{F}} \right\rangle  \right\rvert_{m<m_{\text{cr}}} & = \frac{ \zeta }{4}  \tilde{s} \left( e^{2 \pi r B} -  e^{2 \pi r A} \right) \\
							\frac{1}{2 \pi r}  \left.  \frac{ \partial \ }{ \partial m } \left.\langle \mathcal{W}_{\mathsf{F}} \right\rangle  \right\rvert_{m>m_{\text{cr}}} & = \frac{ \zeta }{4} \left( e^{2 \pi r B} +  e^{2 \pi r A }  \right) - \frac{ \zeta}{2} e^{- 2 \pi r m} 
						\end{align}
						\end{subequations}
						where we have followed the notation of subsection \ref{sec:largeNlimit} and introduced the auxiliary variable $\tilde{s}$, which is $-1$ if $-m>B$ and $+1$ if $-m<A$ in the first phase. Sending $m \to m_{\text{cr}}$, either $m=A$ or $m=B$ at the critical point, whence the continuity of $ \frac{ \partial \ }{ \partial m } \left.\langle \mathcal{W}_{\mathsf{F}} \right\rangle $ follows.\par
						We conclude that, in the $SU(N)$ theory, the vevs of Wilson loops in the fundamental representation are always differentiable. This is consistent with the explicit calculations, yielding third order phase transitions.\par
						A similar reasoning can be applied to the expectation value of Wilson loops in large antisymmetric representations. The computations to determine $\langle \mathcal{W}_{\mathsf{A}_K} \rangle$ follow closely those in section \ref{sec:UNWL}. The additional constraints on the partial derivatives of $A$ and $B$ in the $SU(N)$ theory allow to show that the first derivative of the vev with respect to the mass is a continuous function.

						\subsection{Hypermultiplets in the symmetric representation}
						\label{sec:SuNantisym}
						In this subsection we analyze the phase structure of $SU(N)$ theories with hypermultiplets in the symmetric representation.\par

						\subsubsection{Only symmetric hypermultiplet}
						\label{sec:SuNpuresym}
						Let us start with the simpler case of only a symmetric hypermultiplet of mass $m$, without \CS~term. The SPE of this model is: 
						\begin{align}
							\frac{2}{\lambda} \phi  = & - \int_A ^B \dd  \psi \rho (\psi) \left( \phi - \psi \right)^2 \text{sign} \left( \phi - \psi \right) \label{eq:SPESuN1anti}\\
							 & + \frac{1}{4} \int_A ^B \dd  \psi \rho (\psi) \left[  \left( \phi + \psi +m \right)^2 \text{sign} \left( \phi + \psi +m \right) +  \left( \phi + \psi -m \right)^2 \text{sign} \left( \phi - \psi -m \right)   \right]  . \notag 
						\end{align}
						We can take $m \ge 0$ without loss of generality. The symmetric ansatz 
						\begin{equation}
						\label{eq:symmansatzna}
							\rho (\phi) = \frac{1}{2} \delta ( \phi + B) + \frac{1}{2} \delta ( \phi - B) 
						\end{equation}
						solves \eqref{eq:SPESuN1anti} with 
						\begin{equation}
							B= \begin{cases}  \frac{m}{2} - \frac{1}{\lambda}  & m \ge - \frac{2}{\lambda} \\ - \frac{2}{\lambda} & m  \le - \frac{2}{\lambda} . \end{cases}
						\end{equation}
						As we are taking $m \ge 0$, the phase transition takes place at negative values of the \YM~'t Hooft coupling. The Wilson loop in the fundamental representation acquires a vev 
						\begin{equation}
							\left\langle \mathcal{W}_{\mathsf{F}} \right\rangle = \begin{cases} \cosh \left( 2 \pi r \left(  \frac{m}{2} - \frac{1}{\lambda} \right) \right) & m \ge - \frac{2}{\lambda} \\ \cosh \left( \frac{4 \pi r }{\lambda} \right) & m  \le - \frac{2}{\lambda} . \end{cases}
						\end{equation}
						We find a second order phase transition. As for the model with a single adjoint hypermultiplet of subsection \ref{sec:adjhyper}, at $m=0 $ we find a third order transition at the superconformal point $\frac{1}{\lambda} \to 0$, mirroring a flop transition in the dual Calabi--Yau geometry (see appendix \ref{sec:geometric}).\par
						We can easily obtain a solution for the theory analytically continued to any real $n_{\mathsf{S}} < 2$. In that case the symmetric ansatz \eqref{eq:symmansatzna} solves the SPE with 
						\begin{equation}
							B= \begin{cases}  \frac{n_{\mathsf{S}}}{2} m - \frac{1}{\lambda}  & m \ge - \frac{2}{\lambda (2 - n_{\mathsf{S}} )  } \\ - \frac{2}{\lambda (2 - n_{\mathsf{S}} ) } & m  \le - \frac{2}{\lambda (2 - n_{\mathsf{S}} ) } . \end{cases}
						\end{equation}
						The features of the phase diagram extend to this case.

						\subsubsection{Symmetric and fundamental hypermultiplets}
						We now consider $SU(N)$ theory with a massless symmetric hypermultiplet and two families of fundamental hypermultiplets with opposite masses $\pm m$, with equal Veneziano parameters $\zeta_1 = \zeta_2 \equiv \zeta$. That is, we introduce a massless symmetric or rank-two antisymmetric hypermultiplet in the model of subsection \ref{sec:SuNF2opposite}. In absence of a \CS~term, the Veneziano parameter is constrained by $\zeta \le \frac{1}{2}$. The SPE is 
						\begin{multline}
							\int_{A} ^{B} \dd \psi \rho (\psi)  \left[ - \left( \phi - \psi \right)^2 \text{sign} \left( \phi - \psi \right) + \frac{1}{2} \left( \phi + \psi \right)^2 \text{sign} \left( \phi + \psi \right)  \right] = \\ \frac{2}{\lambda } \phi + \tilde{\xi} - \frac{\zeta}{2} \left[  \left( \phi +m \right)^2 \text{sign} \left( \phi +m \right) +  \left( \phi -m \right)^2 \text{sign} \left( \phi -m \right)  \right] .
						\label{eq:SuNsymSPE}
						\end{multline}
						It is solved by a simple extension of the method in subsection \ref{sec:largeNlimit}, with $\tilde{\xi}=0$. Let us begin with the case $\lambda <0$. Then, starting with the phase in which $m \notin \text{supp} \rho$, we find that the symmetric ansatz \eqref{eq:symmansatzna} solves \eqref{eq:SuNsymSPE} with 
						\begin{equation}
							B = 2 \zeta m - \frac{2}{\lambda} , \qquad m > B .
						\end{equation}
						This phase holds for $m > m_{\text{cr}}$, with 
						\begin{equation}
							m_{\text{cr}} \left( \lambda <0, \zeta \right) = - \frac{2}{\lambda (1- 2 \zeta )} ,
						\end{equation}
						which is positive. Beyond the critical point we find the solution 
						\begin{equation}
							\rho (\phi) = \frac{1-2 \zeta}{2} \left[  \delta ( \phi + B)  + \delta ( \phi - B)  \right] + \zeta \left[  \delta ( \phi + m)  + \delta ( \phi - m)  \right] , \qquad B= - \frac{2}{\lambda (1- 2 \zeta )} .
						\end{equation}\par
						For $\lambda>0$ the solution in the first phase is identical, but $B$ becomes negative at $m= \frac{1}{\zeta \lambda}$, thus we should rename $B $ and $-B$. The phase transition then takes place at 
						\begin{equation}
							m_{\text{cr}} \left( \lambda >0, \zeta \right) = \frac{2}{\lambda (1+ 2 \zeta )} >0 ,
						\end{equation}
						which also equals $B^{(\mathrm{II})}$ computed in the second phase. Note that, consistently with the derivation for $\lambda>0$, 
						\begin{equation}
							 \frac{1}{\zeta \lambda} > \frac{2}{\lambda (1+2 \zeta)} .
						\end{equation}\par
						Computing the free energy, we find a third order phase transition.\par
						The Wilson loop in the fundamental representation acquires a vev 
						\begin{equation}
							\left\langle \mathcal{W}_{\mathsf{F}} \right\rangle = \begin{cases} \cosh \left( 2 \pi r \left(  2 \zeta m  - \frac{2}{\lambda } \right) \right) & m \ge m_{\text{cr}} \left( \lambda <0, \zeta \right) \\ (1 - 2 \zeta ) \cosh \left( \frac{4 \pi r }{\lambda (1- 2 \zeta)} \right)  + 2 \zeta \cosh (2 \pi r m) & m  \le m_{\text{cr}} \left( \lambda <0, \zeta \right) , \end{cases}
						\end{equation}
						whose derivative is a continuous but not differentiable function of $m$. This confirms that the phase transition is third order. The case of positive $\lambda$ is analogous.

						\subsubsection{Spontaneous one-form symmetry breaking}
						\label{sec:1formSSB}
						$5d$ $\mN=1$ gauge theories with simple gauge group have a one-form symmetry associated to the centre of the group \cite{Morrison:2020ool,Albertini:2020mdx}. It is $\mathbb{Z}_N$ for $SU(N)$ and $\mathbb{Z}_2$ for $USp (2N)$. This symmetry is compatible with matter in the adjoint or rank-two antisymmetric representation, whilst fundamental hypermultiplets break it explicitly \cite{BenettiGenolini:2020doj}.\par
						It is argued in \cite{BenettiGenolini:2020doj} that, for theories with adjoint or antisymmetric matter, the one-form symmetry is spontaneously broken. This expectation is confirmed by our results, in the regime considered, as signalled by $\left\langle \mathcal{W}_{\mathsf{F}} \right\rangle$ following a perimeter law. Moreover we observe that transitions between two phases with spontaneously broken one-form symmetry are always second order (cf. subsections \ref{sec:pureadjoint} and \ref{sec:SuNpuresym}). Instead, whenever the one-form symmetry is absent from the beginning, the phase transitions are third order.\par
						\medskip
						We note a subtlety concerning the pure $SU(N)$ gauge theory at \CS~level $k$ of subsection \ref{sec:SunPure}. The one-form symmetry should be restored at $k=0$, however, directly taking the limit $t \to \infty$ in $\left\langle \mathcal{W}_{\mathsf{F}} \right\rangle$ is problematic. Instead, we write the result for large but finite $N$ and tune $k \to 0$ first. The Wilson loop vev is then damped as 
						\begin{equation}
							\left\langle \mathcal{W}_{\mathsf{F}} \right\rangle \approx  e^{- \frac{4 \pi r \lvert h \rvert }{k}} \left[ \frac{N-k}{2N} e^{\frac{2 \pi r}{\lambda} \left( 1+ \frac{k}{N} \right)} +  \frac{N+k}{2N} e^{\frac{2 \pi r}{\lambda} \frac{k}{N} } \right] 
						\end{equation}
						for one sign of $h$, and a similar expression for the other sign. The term in backet is finite in the $k \to 0$ limit with fixed $N$, thus we find agreement with \cite{Morrison:2020ool,Albertini:2020mdx}.\par

			\section{Phases of ${USp}(2N)$, $SO(2N)$ and $SO(2N+1)$ theories}
			\label{sec:SpN}
				In this section we study the large $N$ phase structure of gauge theories with the other classical gauge groups: ${USp} (2N)$, $SO (2N)$ and $SO (2N+1)$.\footnote{For orthogonal groups the hypermultiplets are taken in the vector representation.} In the large $N$ setup of section \ref{sec:CBloc}, the difference between $SO(2N)$ or $SO(2N+1)$ and $USp(2N)$ is sub-leading, thus it suffices to study the compact symplectic gauge group $USp(2N)$. The eigenvalues of the $\mathfrak{usp} (2N)$-valued adjoint scalar $\phi$ are 
				\begin{equation}
					(\phi_1 , \dots, \phi_N , - \phi_1 , \dots, - \phi_N) ,
				\end{equation}
				and we can take $\phi_a \ge 0$ for all $a=1, \dots, N$ without loss of generality. These groups do not admit a \CS~term but have a $\mathbb{Z}_2$-valued theta parameter \cite{Intriligator:1997pq}, which we set to zero.\par
				Repeating the argument of section \ref{sec:CBloc} for $USp(2N)$ we arrive at the SPE:
				\begin{multline}
					- \int_A ^{B}  \dd  \psi \rho (\psi) \left[  \left( \phi - \psi \right)^2 \text{sign} \left( \phi - \psi \right)  + \left( \phi+ \psi \right)^2   \text{sign} \left( \phi + \psi \right) \right] \\
					= \frac{4}{\lambda} \phi - \sum_{\alpha=1} ^{F} \frac{\zeta_{\alpha}}{2} \left[ \left( \phi + m_{\alpha} \right)^2 \text{sign} \left( \phi + m_{\alpha} \right) + \left( \phi - m_{\alpha} \right)^2 \text{sign} \left( \phi - m_{\alpha} \right) \right] .
				\label{eq:SPESpNlargeN}
				\end{multline}
				The eigenvalue density is assumed to be supported on a single interval $[A,B]$ on the positive real axis. The contributions to the vector multiplet one-loop determinant from each pair of opposite eigenvalues are sub-leading at large $N$ and do not appear in the SPE.\par
				Convergence of the localized partition function in the large $N$ limit requires 
				\begin{equation}
				\label{eq:condzetale2SpN}
					\sum_{\alpha=1}^{F} \zeta_{\alpha} \le 2 ,
				\end{equation}
				which matches the condition for the gauge theory to sit in the IR of a SCFT \cite{Jefferson:2017ahm}. Taking three derivatives of \eqref{eq:SPESpNlargeN} we find the solution 
				\begin{equation}
				\label{eq:rhogenSpN}
					\rho (\phi) = c_A \delta \left( \phi - A \right) + c_B \delta \left( \phi - B \right) + \sum_{\alpha=1} ^{F} \left[  c_{\alpha} ^{-} \delta \left( \phi + m_{\alpha} \right) + c_{\alpha} ^{+} \delta \left( \phi - m_{\alpha} \right) \right] ,
				\end{equation}
				where the coefficients $c_{\alpha} ^{\pm}$ are 
				\begin{equation}
					c_{\alpha} ^{\pm } = \begin{cases}  \frac{\zeta_{\alpha}}{4}  & \pm m_{\alpha} \in [A,B] \\ 0 & \text{otherwise} \end{cases} 
				\end{equation}
				for all $\alpha=1, \dots, F$. Note that, as we are taking $0<A<B$, at most one between $c_{\alpha}^{-}$ and $c_{\alpha} ^{+}$ is non-zero. The normalization condition \eqref{eq:rhonorm} imposes 
				\begin{equation}
				\label{eq:cAcBfromnormSpN}
					c_A + c_B + \sum_{\alpha=1} ^{F} \left( c_{\alpha} ^{-} + c_{\alpha} ^{+} \right) = 1  .
				\end{equation}
				To lighten the notation, let us define $\tilde{c}_{\alpha} = c_{\alpha} ^{-} + c_{\alpha} ^{+}  $ and 
				\begin{equation}
					\tilde{s}_{\alpha} ^{-} = \begin{cases} -1 & -m_{\alpha} >B \\ 0 & A \le -m_{\alpha} \le B \\ +1 & -m_{\alpha} <A ,  \end{cases}  \qquad \tilde{s}_{\alpha} ^{+} = \begin{cases} -1 & m_{\alpha} >B \\ 0 & A \le m_{\alpha} \le B \\ +1 & m_{\alpha} <A .  \end{cases}
				\end{equation}
				Plugging \eqref{eq:rhogenSpN} back into the SPE \eqref{eq:SPESpNlargeN} we find the three additional conditions 
				\begin{subequations}
				\label{eq:SpNsystem3eqs}
				\begin{align}
					2 c_A + \sum_{\alpha=1} ^{F} \tilde{c}_{\alpha} \left( \tilde{s}_{\alpha} ^{-}  + \tilde{s}_{\alpha} ^{+}  \right) & = \sum_{\alpha=1} ^{F} \frac{\zeta_{\alpha}}{2} \left( \tilde{s}_{\alpha} ^{-}  + \tilde{s}_{\alpha} ^{+}  \right) \label{eq:SpNsystem3eqsa} \\
					2 c_B B + \sum_{\alpha=1} ^{F} \tilde{c}_{\alpha} \left( \tilde{s}_{\alpha} ^{-}  - \tilde{s}_{\alpha} ^{+}  \right)  m_{\alpha}  & =  - \frac{2}{\lambda}   +   \sum_{\alpha=1} ^{F} \frac{\zeta_{\alpha}}{2} \left( \tilde{s}_{\alpha} ^{-}  - \tilde{s}_{\alpha} ^{+}  \right) m_{\alpha} \label{eq:SpNsystem3eqsb} \\
					2 c_A A^2 + \sum_{\alpha=1} ^{F} \tilde{c}_{\alpha} \left( \tilde{s}_{\alpha} ^{-}  + \tilde{s}_{\alpha} ^{+}  \right) m_{\alpha} ^2  & =    \sum_{\alpha=1} ^{F} \frac{\zeta_{\alpha}}{2} \left( \tilde{s}_{\alpha} ^{-}  + \tilde{s}_{\alpha} ^{+}  \right) m_{\alpha} ^2 . \label{eq:SpNsystem3eqsc}
				\end{align}
				\end{subequations}
				These three equations together with \eqref{eq:cAcBfromnormSpN} determine $c_A$, $c_B$ and the endpoints $A$ and $B$. Inspection of the possible values of $\tilde{s}_{\alpha} ^{-}  + \tilde{s}_{\alpha} ^{+}$ shows that the assumption of a $\delta$-function supported at $\phi=A$ is pleonastic, because either $c_A=0$ or the point $A$ is merged with the singularities at $\phi=m_{\alpha}$ (or at $\phi=-m_{\alpha}$, depending on the sign of the mass). We therefore obtain the solution 
				\begin{equation}
				\label{eq:rhofinalSpN}
					\rho (\phi) =  c_B \delta \left( \phi - B \right) + \sum_{\alpha=1} ^{F} c_{\alpha}   \delta \left( \phi - \vert m_{\alpha} \vert \right) ,
				\end{equation}
				with coefficients 
				\begin{equation}
					c_B= 1 - \sum_{\alpha=1} ^{F} c_{\alpha} , \qquad c_{\alpha} = \begin{cases} \frac{\zeta_{\alpha}}{2} & -B < m_{\alpha} < B \\ 0 & \text{otherwise} \end{cases}
				\end{equation}
				and endpoint 
				\begin{equation}
					B= \left(  1 - \sum_{\alpha=1} ^{F} c_{\alpha} \right)^{-1}  \left[ - \frac{1}{\lambda} +  \sum_{\alpha=1} ^{F} \left( \frac{\zeta_{\alpha}}{2} - c_{\alpha} \right) m_{\alpha} \right] .
				\end{equation}
				In particular, $B$ is independent of $m_{\alpha}$ when $-B < m_{\alpha} < B$. The full eigenvalue density $\rho_{ USp(2N) } (\phi)$, that accounts for all the $2N$ eigenvalues, is symmetric under $\phi \mapsto - \phi $, and reads 
				\begin{equation}
					\rho_{ USp(2N) } (\phi) = \frac{  \rho (\phi) + \rho (- \phi)  }{2} .
				\end{equation}
				\par
				The free energy computed using the solution \eqref{eq:rhofinalSpN} is 
				\begin{align}
					\mathcal{F}_{\mathbb{S}^5} &  = \frac{4}{3} c_B ^2 B^3 + \frac{c_B}{3} \sum_{\alpha=1} ^{F} c_{\alpha} \left( \vert B - m_{\alpha} \vert^3  +  \vert B + m_{\alpha} \vert^3  \right) \\
						& + \frac{1}{6} \sum_{\alpha=1} ^{F} c_{\alpha} \sum_{\alpha^{\prime}=1} ^{F} c_{\alpha^{\prime}} \left(  \vert m_{\alpha } - m_{\alpha^{\prime} } \vert^3  +   \vert m_{\alpha } + m_{\alpha^{\prime} } \vert^3  \right)  + \frac{2}{\lambda} \left( c_B B^2 +   \sum_{\alpha=1} ^{F} c_{\alpha} m_{\alpha}^2 \right) \notag \\
						& - \sum_{\alpha=1} ^{F} \frac{\zeta_{\alpha}}{6} \left[  c_B \left(  \vert B - m_{\alpha} \vert^3  +  \vert B + m_{\alpha} \vert^3  \right) + \sum_{\alpha^{\prime}=1} ^{F} c_{\alpha^{\prime}} \left(  \vert m_{\alpha } - m_{\alpha^{\prime} } \vert^3  +   \vert m_{\alpha } + m_{\alpha^{\prime} } \vert^3  \right) \right] \notag .
				\end{align}
				Let us study the phase structure. We focus for clarity on the first phase transition, assuming that all masses are larger than $B$ and a mass, which we take to be $m_1$, is decreased until it eventually crosses $B$. On one side of the wall we use $c_{\alpha}=0$ for all $\alpha=1, \dots, F$ and find 
				\begin{equation}
					\mathcal{F}_{\mathbb{S}^5} ^{(\mathrm{I})} = \frac{4}{3} \left(B^{(\mathrm{I})}\right) ^3 + \frac{2}{\lambda} \left( B^{(\mathrm{I})}\right) ^2  - \sum_{\alpha=1} ^{F} \frac{\zeta_{\alpha}}{6} \left(  \vert B^{(\mathrm{I})}  - m_{\alpha} \vert^3 + \vert B^{(\mathrm{I})} + m_{\alpha} \vert^3 \right) , 
				\end{equation}
				while on the other side we use $c_1= \frac{\zeta_1}{2}$ and $c_{\alpha}=0$ for $\alpha=2, \dots, F$ and get 
				\begin{align}
					\mathcal{F}_{\mathbb{S}^5}  ^{(\mathrm{II})} & = \frac{4}{3} \left(  1 - \frac{\zeta_1}{2} \right)^2 \left( B^{(\mathrm{II})}\right) ^3 + \frac{2}{\lambda} \left( \left( 1 - \frac{\zeta_1}{2} \right) \left( B^{(\mathrm{II})}\right) ^2 + \frac{\zeta_1}{2} m_1 ^2 \right) \\
					& - \frac{\zeta_1 ^2}{24} \left( \vert B^{(\mathrm{II})} -m_1 \vert^3 +  \vert B^{(\mathrm{II})} + m_1 \vert^3 \right) - \sum_{\alpha=2} ^{F} \frac{\zeta_{\alpha}}{6} \left(  \vert B^{(\mathrm{II})} - m_{\alpha} \vert^3 + \vert B^{(\mathrm{II})} + m_{\alpha} \vert^3 \right)  ,  \notag 
				\end{align}
				where the superscripts ${}^{(\mathrm{I})}$ and ${}^{(\mathrm{II})}$ indicate the quantity evaluated in the corresponding phase. At the critical point, $B=m_1$ by definition, which guarantees the continuity of $\mathcal{F}_{\mathbb{S}^5}$. Taking derivatives and using 
				\begin{equation}
					\frac{ \partial B^{(\mathrm{I})}}{\partial m_1} = \frac{\zeta_1}{2}, \qquad \frac{\partial B^{(\mathrm{II})}}{\partial m_1} =0 , 
				\end{equation}
				we find that the first and second derivatives of the free energy are continuous, and the phase transition is third order.\par
				We notice that the calculations of the free energy are akin to those in \cite{Nedelin}. On a computational level, this stems from the symmetric form of the eigenvalues together with the lack of a \CS~term. This is an incarnation of the fact that the $SU(N)$ and $USp (2N)$ gauge theories are UV-completed into the same SCFT, up to a shift in $N_f$ that is invisible in the Veneziano limit.\par
				\medskip
				The argument extends to theories with hypermultiplets in the adjoint or rank-two antisymmetric representation. The phase diagram is easily recovered from subsections \ref{sec:adjhyper} and \ref{sec:SuNantisym}.\footnote{$USp(2N)$ theories have been investigated from different angles in \cite{Minahan:2020ifb,Bergman:2013koa,Hwang:2014uwa,Bourget:2020gzi,Bergman:2020myx,Li:2021rqr}.}

				\section{Quiver theories}
				\label{sec:quiverGT}

			The aim of the present section is to analyze the large $N$ limit of various $5d$ $\mN=1$ quiver gauge theories. Let the gauge group be
			\begin{equation}
				U \left( N_1 \right) \times U \left( N_2 \right) \times \cdots \times U \left( N_L \right) ,
			\end{equation}
			corresponding to a quiver with $L$ nodes, that we label by $j=1, \dots, L$. These models have been recently constructed in \cite{c}.

		\subsection{Short quivers}
		\label{sec:quivershort}
		
		\subsubsection{Homogeneous $U(N) \times U(N)$ quiver}
\begin{figure}[htb]
					\centering
					\begin{tikzpicture}[auto,square/.style={regular polygon,regular polygon sides=4}]
						\node[circle,draw] (gauge1) at (1,0) {$N$};
						\node[circle,draw] (gauge2) at (-1,0) {$N$};
						\node[square,draw] (fl1) at (3,0) { \hspace{8pt} };
						\node[square,draw] (fl2) at (-3,0) { \hspace{8pt} };
						\node[draw=none] (aux1) at (3,0) {$N$};
						\node[draw=none] (aux2) at (-3,0) {$N$};
						\draw[-](gauge1)--(gauge2);
						\draw[-](gauge1)--(fl1);
						\draw[-](gauge2)--(fl2);
					\end{tikzpicture}
					\caption{Homogeneous $U(N) \times U(N)$ quiver.}
					\label{fig:homoUN2}
					\end{figure}
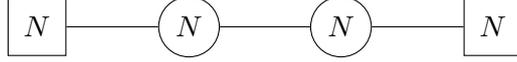

			The first model we discuss is the $U(N) \times U(N)$ homogeneous quiver, with a bi-fundamental hypermultiplet and $N$ fundamental flavours at each gauge node, represented in figure \ref{fig:homoUN2}. A \CS~term is forbidden by  \eqref{eq:condNnf}. For simplicity we assume that all the fundamental hypermultiplets at each node have equal masses, that we denote $m_1$ and $m_2$.
			
			\par
			The large $N$ SPE for the first node reads 
			\begin{align}
				- \int \dd  \psi \rho_1 (\psi)~ \left( \phi - \psi \right)^2 \text{sign} \left( \phi - \psi \right)  =  \frac{2}{\lambda_1} \phi & - \frac{1}{2}   \left( \phi +m_1 \right)^2 \text{sign} \left( \phi +m_1 \right)  \\
					& - \frac{1}{2} \int \dd  \psi \rho_2 (\psi)~ \left( \phi - \psi \right)^2 \text{sign} \left( \phi - \psi \right) \notag
			\end{align}
			and the SPE for the second node has the labels $1$ and $2$ exchanged. Here $\rho_1 (\phi)$ and $\rho_2 (\phi)$ are the eigenvalue densities corresponding to the first and second node respectively, and likewise for the \YM~couplings $\lambda_1$ and $\lambda_2$. In the SPE for the $j^{\mathrm{th}}$ node $\phi \in \text{supp} \rho_j$ is assumed, $j \in \left\{ 1,2 \right\}$.\par
			When both $\vert m_1 \vert$ and $\vert m_2 \vert$ are large, the solution is 
			\begin{subequations}
			\label{eq:homog2PhI}
			\begin{align}
				\rho_1 (\phi) & = \frac{1}{16} \delta \left( \phi -  A_2 \right) +  \frac{7}{16} \delta \left( \phi -  B_2 \right) + \frac{1}{8} \delta \left( \phi -  A_1 \right) + \frac{3}{8} \delta \left( \phi -  B_1 \right)  \\
				\rho_2 (\phi) & = \frac{1}{8} \delta \left( \phi -  A_2 \right) + \frac{7}{8}  \delta \left( \phi -  B_2 \right)  
			\end{align}
			\end{subequations}
			under the assumption that $\lambda_1 ^{-1}$, $\lambda_2 ^{-1}$ and $m_1$, $m_2$ are such that $B_2<B_1$, or with the roles of first and second node swapped otherwise. We do not spell the details explicitly, because they are exactly as in section \ref{sec:UN}, and only sketch the argument.\par
			From phase \eqref{eq:homog2PhI} the system can access two other phases, by either decreasing $\vert m_2 \vert$ keeping $\vert m_1 \vert $ large or the converse. In the first case, both $\rho_1 (\phi)$ and $\rho_2 (\phi)$ develop an additional $\delta$-function singularity at $\phi=-m_2$, $m_2 \in \left[ A_2, B_2 \right]$, and we get 
			\begin{subequations}
			\begin{align}
				\rho_1 (\phi) & = \frac{1}{8} \delta \left( \phi -  A_2 \right) +  \frac{1}{8} \delta \left( \phi -  B_2 \right) + \frac{1}{8} \delta \left( \phi -  A_1 \right) + \frac{3}{8} \delta \left( \phi -  B_1 \right) + \frac{1}{4} \delta \left( \phi + m_2 \right) \\
				\rho_2 (\phi) & = \frac{1}{4} \delta \left( \phi -  A_2 \right) + \frac{1}{4}  \delta \left( \phi -  B_2 \right)  + \frac{1}{2} \delta \left( \phi + m_2 \right) .
			\end{align}
			\end{subequations}
			In the second case, we find a solution in the region $B_2 < -m_1 < B_1$ or $A_1 < -m_1 < A_2$, with 
			\begin{subequations}
			\begin{align}
				\rho_1 (\phi) & = \frac{1}{8} \delta \left( \phi -  A_2 \right) +  \frac{1}{8} \delta \left( \phi -  B_2 \right) + \frac{1}{8} \delta \left( \phi -  A_1 \right) + \frac{3}{8} \delta \left( \phi -  B_1 \right) + \frac{1}{4} \delta \left( \phi + m_1 \right) \\
				\rho_2 (\phi) & = \frac{1}{4} \delta \left( \phi -  A_2 \right) + \frac{3}{4}  \delta \left( \phi -  B_2 \right)  .
			\end{align}
			\end{subequations}
			However, when $-m_1 \in [A_2, B_2]$, the eigenvalue density at the second node will develop a $\delta$-function singularity at $\phi=-m_1$ as well, and we obtain 
			\begin{subequations}
			\begin{align}
				\rho_1 (\phi) & = \frac{3}{32} \delta \left( \phi -  A_2 \right) +  \frac{5}{32} \delta \left( \phi -  B_2 \right) + \frac{1}{8} \delta \left( \phi -  A_1 \right) + \frac{1}{8} \delta \left( \phi -  B_1 \right) + \frac{1}{2} \delta \left( \phi + m_1 \right) \\
				\rho_2 (\phi) & = \frac{3}{16} \delta \left( \phi -  A_2 \right) + \frac{5}{16}  \delta \left( \phi -  B_2 \right)  + \frac{1}{2} \delta \left( \phi + m_1 \right).
			\end{align}
			\end{subequations}
			The rest of the phase structure is directly obtained from the cases discussed. The phase diagram of homogeneous quivers with more nodes is likewise derived by iteration of the ideas presented.\par

		\subsubsection{Circular $U(N) \times U(N)$ quiver}
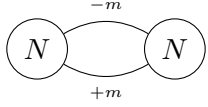
\begin{figure}[htb]
				\centering
			\begin{tikzpicture}[auto,node distance=1cm]
				\node[circle,draw] (gauge1) {$N$};
				\node[circle,draw] (gauge2) [left = of gauge1]{$N$};
				\path (gauge1) edge [bend left] node {${\scriptscriptstyle +m }$} (gauge2);
				\path (gauge2) edge [bend left] node {${\scriptscriptstyle -m }$} (gauge1);
			\end{tikzpicture}
			\caption{Circular $U(N)\times U(N)$ quiver.}
			\label{fig:ABJMlike}
			\end{figure}

			The next example we consider is the $U(N) \times U(N)$ circular quiver \cite{Minahan:2014hwa}, drawn in figure \ref{fig:ABJMlike}.\par
				 \par
			There are precisely $N_f=2N$ hypermultiplets in the fundamental representation of each $U(N)$, then by \eqref{eq:condNnf} the \CS~levels must vanish. We assign masses $\pm m$ to the hypermultiplets.\par
			The SPEs for this quiver at large $N$ and in the decompactification limit are:
			\begin{align}
				- \int \dd  \psi \rho_1 (\psi)~ \left( \phi - \psi \right)^2 \text{sign} \left( \phi - \psi \right)  = & - \frac{1}{2} \int \dd  \psi \rho_2 (\psi) \left( \phi - \psi +m  \right)^2 \text{sign} \left( \phi - \psi +m  \right) \\
				& - \frac{1}{2} \int \dd  \psi \rho_2 (\psi)  \left( \phi - \psi -m  \right)^2 \text{sign} \left( \phi - \psi -m  \right) +  \frac{2}{\lambda_1} \phi   \notag  
			\end{align}
			for the first node, and the same upon swapping the labels $1$ and $2$ for the second node. This system of two equations is solved generalizing the procedure for the single node theory with an adjoint hypermultiplet, studied in subsection \ref{sec:adjhyper}.\par
			In analogy with subsection \ref{sec:adjhyper}, we get a solution 
			\begin{subequations}
			\begin{align}
				\rho_1 (\phi) &= \frac{1}{2} \left[ \delta \left(  \phi + B_1 \right) + \delta \left(  \phi - B_1 \right)  \right] \\
				\rho_2 (\phi) &= \frac{1}{2} \left[ \delta \left(  \phi + B_2 \right) + \delta \left(  \phi - B_2 \right) \right]
			\end{align}
			\end{subequations}
			where we take $m \ge 0$ without loss of generality, and with
			\begin{equation}
				B_j= - \frac{1}{\lambda_j} +  m  , \qquad j \in \left\{  1,2 \right\} .
			\end{equation}
			This phase holds under the assumption $m \le \min \left\{ \frac{2}{\lambda_1}, \frac{2}{\lambda_2}  \right\} $. Beyond this point, the system is in a new phase, in which the eigenvalue density at one node sees the singularities from the eigenvalues of the other node. For instance, let us assume $\frac{1}{\lambda_1} < \frac{1}{\lambda_2}$. Then, when $m> \frac{2}{\lambda_1}$ we find 
			\begin{subequations}
			\begin{align}
				\rho_1 (\phi) &= \frac{1}{4} \left[ \delta \left(  \phi + B_1 \right) + \delta \left(  \phi - B_1 \right) +  \delta \left(  \phi + \widetilde{m}_1 \right) +  \delta \left(  \phi - \widetilde{m}_1 \right)   \right] \\
				\rho_2 (\phi) &= \frac{1}{2} \left[ \delta \left(  \phi + B_2 \right) + \delta \left(  \phi - B_2 \right) \right]
			\end{align}
			\end{subequations}
			where 
			\begin{equation}
				B_1 = \frac{1}{\lambda_1} , \qquad \widetilde{m}_1 = 2m - \frac{3}{ \lambda_1 } , \qquad B_2 = 2m - \frac{2}{\lambda_1} - \frac{1}{\lambda_2} .
			\end{equation}
			In obtaining the explicit value of $B_2$ we have used $\widetilde{m}_1 >0$. The next phase transition takes place at $m= 2 B_2$, that is, this phase extends in the region 
			\begin{equation}
				 \frac{1}{\lambda_1}  < m < \frac{2}{3 \lambda_1} + \frac{4}{3 \lambda_2} < \frac{2}{\lambda_2} ,
			\end{equation}
			with the last inequality showing the consistency with our previous assumption. When $m$ crosses the second critical value, the system enters in a third phase.

		\subsection{Long quivers}
		\label{sec:quivers}
		
			The goal of the present subsection is to study the large $N$ limit of unitary quiver gauge theories, when the number $L$ of nodes is large \cite{Uhlemann,UhlemannWL,Coccia:2020cku,Coccia:2020wtk}. Although here we follow the philosophy of \cite{Uhlemann}, there are a few major differences. The first is that we are interested in massive deformations of the theory away from the superconformal point. The second aspect is that we work in the decompactification limit.\par
			We write the ranks $\left\{ N_j \right\}$ of the gauge nodes as $N_j = N \nu_j$ and then take the limits $N \to \infty$ and $\frac{1}{r} \to 0 $. We have not managed to obtain the most general solution for a long quiver, and we limit ourselves to discuss various examples in this section.\par
			\medskip
				We are interested in the large $L$ limit, so that the discrete index $j$ is replaced by the continuous one 
				\begin{equation}
					z= \frac{j}{L} , \qquad 0 \le z \le 1 ,
				\end{equation}
				and all the quantities that depend on $j$ become functions of $z$. In particular the structure of the gauge group is encoded in a rank density function $\nu (z)$, defined through 
				\begin{equation}
					N \nu (z) = N \nu_{zL} = N_{zL} . 
				\end{equation}
				The eigenvalue densities $\rho_j (\phi)$ at each node $j$ are collected into the function $\rho (z, \varphi)$ of $0 \le z \le 1$, with $\varphi = \frac{\phi}{L}$ . Since every $\rho_j (\phi)$ must be normalized by $N_j$, the corresponding normalization condition for $\rho (z, \varphi)$ is 
				\begin{equation}
				\label{eq:normrhonuz}
					\int \rho (z,\varphi) \dd  \varphi = \nu (z) , \qquad \forall \ 0 \le z \le 1 .
				\end{equation}
				Note that we have introduced the scaled variable $\varphi = \frac{\phi}{L}$ taking into account the scaling of the eigenvalues $\phi$ with $L$ \cite{Uhlemann}. The number of flavours at each node is 
				\begin{equation}
					N_{f,j} = N_{j+1} + N_{j-1} + n_j , \qquad \forall \ j = 1, \dots, L
				\end{equation}
				with $N_{L+1}=0=N_0$ by convention. In the multiple limit we are considering, this condition becomes 
				\begin{equation}
				\label{eq:Nfcondquiver}
					\frac{ N_f (z)}{N} = \frac{1}{L^2} \frac{\partial^2 \nu }{\partial z^2} + 2 \nu (z) + \zeta (z) .
				\end{equation}
				Assuming the rank density $\nu (z)$ is of class $C^2 ([0,1])$, this condition implies that no \CS~term nor fundamental matter is allowed in the interior of the quivers, thus 
				\begin{equation}
				\label{eq:tzetalong}
					\frac{1}{t(z)} =  \frac{\delta (z)}{t (0)} + \frac{\delta (z-1)}{t (1)} , \qquad \zeta (z) = \zeta (0) \delta (z) +\zeta (1) \delta (z-1) .
				\end{equation}
				At this point, we adapt the discussion of section \ref{sec:CBloc} to long quivers. We fix $0<z<1$ and get the saddle point equation away from the head and tail of the quiver. The contributions of the vector and bi-fundamental hypermultiplets to the SPE, after simplifications that hold in the limit $L \to \infty$, are respectively 
				\begin{subequations}
					\begin{align}
					& L^3 \int \dd  \psi \rho (z, \varphi) \left( \varphi - \psi \right)^2 \text{sign} \left( \varphi - \psi \right) , \\
					- & \int \dd  \psi \left[ L^3 \rho (z, \psi) + \frac{L}{2} \frac{ \partial^2 \ }{ \partial z^2}  \rho (z, \psi)  \right]\left( \varphi - \psi \right)^2 \text{sign} \left( \varphi - \psi \right) .
					\end{align}
				\end{subequations}
			In this multiple limit, the contribution from the vector multiplet is cancelled against part of the contribution from the bi-fundamental hypermultiplets, at each $0<z<1$. The only other contribution in the interior of the long quiver comes from the \YM~term. The SPE is 
			\begin{equation}
			\label{eq:SPElongquiver}
				\frac{1}{2} \int \dd  \psi  \frac{ \partial^2 \ }{ \partial z^2}  \rho (z, \psi) \left( \varphi - \psi \right)^2 \text{sign} \left( \varphi - \psi \right) = \frac{2}{\lambda (z)} \varphi 
			\end{equation}
			which must hold for every $0<z<1$.\par
				Let us remark that the non-trivial scaling of the eigenvalues $\phi$ with $L$ is a consequence of the vanishing mass of the bi-fundamental hypermultiplets. This point is explicitly addressed in subsection \ref{sec:quivermassivebi}.\par
				We bring $\partial^2 _z$ out of the integral and solve the differential equation in $z$ first and then the integral equation. We obtain 
			\begin{equation}
				 \rho (z, \varphi ) =  \frac{\nu (z)}{2} \left[  \delta \left( \varphi + B (z) \right)  +  \delta \left( \varphi - B (z) \right)  \right]  ,
			\end{equation}
			where the coefficients are fixed by the normalization \eqref{eq:normrhonuz}, and $B(z)$ is given by 
			\begin{equation}
				B (z) = \frac{2 g(z)}{ \nu (z) } ,
			\end{equation}
			with $g(z)$ a function satisfying 
			\begin{equation}
			\label{eq:diffeqglambda}
				\frac{\dd ^2 \ }{ \dd z^2} g (z) = \frac{1}{\lambda (z)} .
			\end{equation}
			The two boundary data to integrate \eqref{eq:diffeqglambda} are fixed by the boundary conditions for the quiver at $z=0$ and $z=1$. \par

			\subsubsection{Circular quivers}
				If we consider an affine $\widehat{A}$-type quiver, that is a circular quiver in which each node is connected to two neighbours, the variable $z$ becomes periodic and the absence of boundary terms implies that the solution we have obtained holds for all $0 \le z<1$ with periodic identification $\left\{ z=1 \right\} \equiv \left\{ z=0  \right\} $. The boundary conditions to obtain $g(z)$ from \eqref{eq:diffeqglambda} are the consistency conditions 
				\begin{equation}
					g(0)=g(1) , \qquad \dd  g \vert_{z=0} = - \dd  g \vert_{z=1} .
				\end{equation}
				So, for example, if all the gauge nodes have equal \YM~'t Hooft coupling $\lambda$, $g(z)= \frac{z(z-1)}{2 \lambda }$.\par
				For linear quivers, instead, we have to take into account the boundary conditions for $\rho (z, \varphi)$ obtained solving the theory at the edges of the quiver.

			\subsubsection{Homogeneous quivers}
			\label{sec:homoquiver}
\begin{figure}[htb]
					\centering
					\begin{tikzpicture}[auto,square/.style={regular polygon,regular polygon sides=4}]
						\node[circle,draw] (gauge1) at (3,0) {$N$};
						\node[draw=none] (gaugemid) at (1,0) {$\cdots$};
						\node[circle,draw] (gauge3) at (-1,0) {$N$};
						\node[circle,draw] (gauge4) at (-3,0) {$N$};
						\node[square,draw] (fl1) at (5,0) { \hspace{8pt} };
						\node[square,draw] (fl2) at (-5,0) { \hspace{8pt} };
						\node[draw=none] (aux1) at (5,0) {$N$};
						\node[draw=none] (aux2) at (-5,0) {$N$};
						\draw[-](gauge1)--(gaugemid);
						\draw[-](gaugemid)--(gauge3);
						\draw[-](gauge4)--(gauge3);
						\draw[-](gauge1)--(fl1);
						\draw[-](gauge4)--(fl2);
					\end{tikzpicture}
					\caption{Homogeneous linear quiver.}
					\label{fig:quiverhomo}
					\end{figure}
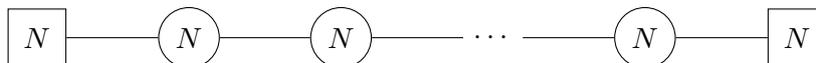

				Consider a constant function $\nu (z)=1$. We take as main example the quiver with all nodes of equal rank $N$, including two sets of $N$ fundamental hypermultiplets attached at the first and last node, as in figure \ref{fig:quiverhomo}. See \cite{c} for the M-theory derivation of such quiver.
					 \par
					
					In the interior of the quiver, the solution found above holds, while the solution at the edges is inherited from the discussion in section \ref{sec:UN}. If all the hypermultiplets are massless, we find $\rho (0, \varphi) = \delta (\varphi) = \rho (1, \varphi)$. Imposing the continuity at $z=0$ and $z=1$ we obtain the pair of conditions $g(0)=0=g(1)$, which serve as boundary conditions to integrate \eqref{eq:diffeqglambda}.\par
					We may also give two opposite mass scales to the hypermultiplets at the head and tail of the quiver, so that isolating each one of these two nodes we reproduce the theory of section \ref{sec:F2s}. We find 
					\begin{align}
						\rho (z, \varphi ) & = \left( \frac{1}{2} - \frac{\delta (z)}{4} - \frac{\delta (z-1)}{4} \right) \left[ \delta \left( \varphi + 2 g(z) \right) +  \delta \left( \varphi - 2 g(z) \right) \right] \\
						& + \left( \frac{\delta (z) }{4} +  \frac{\delta (z-1) }{4} \right) \left[  \delta \left( \varphi + \frac{m (z) }{L} \right) +  \delta \left( \varphi - \frac{m (z) }{L} \right) \right]  , \notag
					\end{align}
					and $g(z)$ is obtained integrating \eqref{eq:diffeqglambda} with boundary conditions $g(0)= \frac{m(0) }{2 L}$ and $g(1)= \frac{m(1) }{2 L}$. We notice that the masses must be scaled linearly with $L$, as we have done with $\phi$, to obtain a non-trivial dependence. This scaling stems from the necessity of putting the gauge and flavour part of the extended Coulomb branch $\mathscr{C} (X) = \mathscr{C}_{\text{gauge}} \times \mathscr{C}_{\text{flavour}} $ on equal footing.

				\subsubsection{Linear quivers}
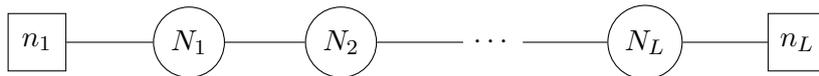
\begin{figure}[htb]
					\centering
					\begin{tikzpicture}[auto,square/.style={regular polygon,regular polygon sides=4}]
						\node[circle,draw] (gauge1) at (3,0) {$N_L$};
						\node[draw=none] (gaugemid) at (1,0) {$\cdots$};
						\node[circle,draw] (gauge3) at (-1,0) {$N_2$};
						\node[circle,draw] (gauge4) at (-3,0) {$N_1$};
						\node[square,draw] (fl1) at (5,0) { \hspace{8pt} };
						\node[square,draw] (fl2) at (-5,0) { \hspace{8pt} };
						\node[draw=none] (aux1) at (5,0) {$n_L$};
						\node[draw=none] (aux2) at (-5,0) {$n_1$};
						\draw[-](gauge1)--(gaugemid);
						\draw[-](gaugemid)--(gauge3);
						\draw[-](gauge4)--(gauge3);
						\draw[-](gauge1)--(fl1);
						\draw[-](gauge4)--(fl2);
					\end{tikzpicture}
					\caption{Admissible linear quiver when $\nu (z)$ is of class $C^2 ([0,1])$.}
					\label{fig:quiver1}
					\end{figure}

					The discussion for the homogeneous linear quiver is easily extended to more generic quivers when $\partial^2 _z \nu (z)$ is continuous and bounded, as represented in figure \ref{fig:quiver1}.
			
					 \par
						
						In this case we find the eigenvalue density 
						\begin{equation}
							\rho (z, \varphi) = \left(  \frac{\nu (z)}{2} - \frac{\zeta (z)}{4} \right) \left[ \delta \left(  \varphi + B(z) \right) + \delta \left(  \varphi - B(z) \right)  \right] + \frac{\zeta (z)}{2} \delta ( \varphi )  ,
						\end{equation}
						with $\zeta(z)$ as defined in \eqref{eq:tzetalong} and supported at the head and tail of the quiver. The continuity condition on $g(z)$ again fixes the eigenvalue density.

				\subsubsection{$T[SU(N)]$ quivers}
				\label{sec:TSUNquiver}
\begin{figure}[htb]
					\centering
					\begin{tikzpicture}[auto,square/.style={regular polygon,regular polygon sides=4}]
						\node[circle,draw] (gauge1) at (3,0) { \hspace{18pt} };
						\node[draw=none] (gaugemid) at (1,0) {$\cdots$};
						\node[circle,draw] (gauge3) at (-1,0) { \hspace{18pt} };
						\node[circle,draw] (gauge4) at (-3,0) { \hspace{18pt} };
						\node[square,draw] (fl1) at (5,0) { \hspace{8pt} };
						\node[draw=none] (aux1) at (5,0) {$N$};
						\node[draw=none] (aux3) at (-1,0) {$2$};
						\node[draw=none] (aux4) at (-3,0) {$1$};
						\node[draw=none] (aux2) at (3,0) {${\scriptstyle N-1}$};
						\draw[-](gauge1)--(gaugemid);
						\draw[-](gaugemid)--(gauge3);
						\draw[-](gauge4)--(gauge3);
						\draw[-](gauge1)--(fl1);
					\end{tikzpicture}
					\caption{The $T[SU(N)]$ ascending quiver.}
					\label{fig:TSUNquiver}
					\end{figure}
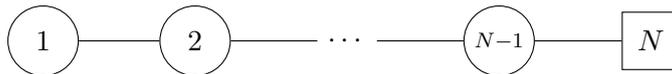

					Another class of examples is constituted by the so-called $T[SU(N)]$ quivers \cite{Hayashi:2014hfa,Eckhard:2020jyr}, represented in figure \ref{fig:TSUNquiver}. These quivers are other particular examples of the constriction in \cite{c}. They have $L=N-1$ gauge nodes, with unitary groups $U(j)$, $j=1, \dots, N-1$, and a set of $N_f=N$ fundamental hypermultiplets at the last node. Therefore the quiver is balanced, in the sense that there are precisely $2j$ fields in the fundamental representation of $U(j)$, for $j=1, \dots, N-1$. The large $N$ limit of the $3d$ counterpart of this theory has been addressed in \cite{Coccia:2020cku}.\par
					
					 \par
						
					In this example, $L$ depends on $N$ and therefore we cannot take the large $N$ limit first and the long quiver limit at the end. However, from the simple relation $L=N-1$, we get $\nu (z) = z$, $0 \le z \le 1$, and the large $N$ limit automatically enforces the long quiver limit.\par
					More generally, we can fix a positive integer $\nu$ and build the ascending quiver with gauge group
					\begin{equation}
						U(\nu) \times U(2 \nu) \times \cdots \times U((N-1)\nu)
					\end{equation}
					with $\nu N$ fundamental hypermultiplets at the last node. The rank density function is again a linear function, $\nu (z)= \nu z $, on the interval $z \in [0,1]$, and $\nu=1$ gives back the $T[SU(N)]$ theory. The dependence of the rank of each node on $j$ requires care: for a given massive deformation $h_j$ leading to a \YM~term for the $j^{\mathrm{th}}$ node, the corresponding 't Hooft parameter is $\lambda_j ^{-1} = \frac{h_j}{z \nu N}$, which would be ill-defined in the na\"{i}ve $z \to 0$ limit. This is of course a consequence of the rank not being large for the first nodes. This does not invalidate our procedure, but we should keep in mind that as $z \to 0$ from above, we must take $\lambda^{-1} (z) \to \infty $.\par
					We arrive at the eigenvalue density 
					\begin{equation}
							\rho (z, \varphi) = \left(  \frac{\nu z}{2} - \frac{\delta (z-1)}{4} \right) \left[ \delta \left(  \varphi + B(z) \right) + \delta \left(  \varphi - B(z) \right)  \right] + \frac{\delta (z-1)}{2} \delta ( \varphi )  
					\end{equation}
					for massless hypermultiplets. Using this density to compute the free energy, $\mathcal{F}_{\mathbb{S}^5}$ is uniquely determined by the profile of $\lambda (z)^{-1}$ on $z \in [0,1]$.

				\subsubsection{Gluing}
					The solutions for the homogeneous and $T[SU(N)]$ quivers of subsections \ref{sec:homoquiver} and \ref{sec:TSUNquiver} respectively are manifestly compatible with the gluing operation. That is, we can construct a linear quiver of length $L=L_1 + L_2$ by identifying and gauging the flavour nodes of two linear quivers of length, respectively, $L_1$ and $L_2$. The assignment of equal real masses to the two flavour nodes and the subsequent integration are translated into a continuity condition for the function $g(z)$ at the junction. For the homogeneous quiver, gluing the head and the tail is allowed and produces the circular $\widehat{A}_L$ quiver.\par

				\subsubsection{Remarks on special unitary quivers}
					For balanced quivers we have found eigenvalue distributions $\rho (z, \varphi) $ that are explicitly even in the large $N$ and decompactification limit. Therefore, if we replace the $U(N)$ nodes by $SU(N)$ nodes, we would get the same answer, as the $\delta$-function constraint on the eigenvalues would be automatically fulfilled. We remark that the argument will hold on a $\mathbb{S}^5$ of finite radius, since the saddle point equations will still be even in $\phi$, yielding an eigenvalue density with symmetric support.\par
					Further comments on special unitary groups as seen by the matrix model, at finite $N$, are collected in appendix \ref{app:specialunitary}.

				\subsubsection{Massive matter}
				\label{sec:quivermassivebi}
				
					We now revisit the discussion of long quivers when the bi-fundamental hypermultiplets are massive. Let us denote $\mu_{j}$ the mass of the bi-fundamental hypermultiplet between the $j^{\mathrm{th}}$ and the $(j+1)^{\mathrm{th}}$ node. In the large $L$ limit, the masses are encoded in a function $\mu (z)$, that we assume continuous on $0<z<1$.\par
					In the large $N$ and large $L$ limit, this theory does not require scaling the scalar $\phi$ with $L$, and we arrive at the SPE 
					\begin{align}
						 \int \dd \psi \rho (z, \psi) \left(  \phi - \psi \right)^2 \text{sign}  \left(  \phi - \psi \right)  & =   \frac{1}{2} \int \dd \psi \rho (z, \psi) \left(  \phi - \psi + \mu (z) \right)^2 \text{sign}  \left(  \phi - \psi + \mu (z) \right)  \label{eq:SPElongquiv} \\
						& + \frac{1}{2} \int \dd \psi \rho (z, \psi) \left(  \phi - \psi - \mu (z) \right)^2 \text{sign}  \left(  \phi - \psi - \mu (z) \right)  -  \frac{2 \phi}{\lambda (z) }  . \notag 
					\end{align}
					The solution to this equation is similar to that in subsection \ref{sec:adjhyper}, and a phase transition must take place as the product $\mu (z) \vert \lambda (z) \vert$ is increased from 0 to $\infty$.

				\subsection{Codimension-two defects}
				\label{sec:defects}

					This conclusive subsection is devoted to the study of the large $N$ and decompactification limit of a particular class of $5d$ theories with codimension-two defects \cite{Gaiotto:2014ina,Ashok:2017bld,Gutperle:2020rty}. These defects are described by an $\mathbb{S}^3 \subset \mathbb{S}^5$ hosting a $3d$ \CS~quiver preserving half of the supercharges.\par
\begin{figure}[htb]
					\centering
					\begin{tikzpicture}[auto,square/.style={regular polygon,regular polygon sides=4}]
						\node[circle,draw,fill=gray!10] (gauge1) at (3,0) {$N_L$};
						\node[draw=none] (gaugemid) at (1,0) {$\cdots$};
						\node[circle,draw,fill=gray!10] (gauge3) at (-1,0) {$N_2$};
						\node[circle,draw,fill=gray!10] (gauge4) at (-3,0) {$N_1$};
						\node[square,draw] (fl1) at (5,0) { \hspace{8pt} };
						\node[draw=none] (aux1) at (5,0) {$N$};
						\node[draw=none] (kL) at (3.5,0.5) {${\scriptscriptstyle (3d)}$};
						\node[draw=none] (k2) at (-0.5,0.5) {${\scriptscriptstyle (3d)}$};
						\node[draw=none] (k1) at (-2.5,0.5) {${\scriptscriptstyle (3d)}$};
						\draw[-](gauge1)--(gaugemid);
						\draw[-](gaugemid)--(gauge3);
						\draw[-](gauge4)--(gauge3);
						\draw[-](gauge1)--(fl1);
					\end{tikzpicture}
					\caption{A $3d$ \CS~theory on the defect, drawn with shaded nodes. The last node is the five-dimensional gauge group, seen as a flavour symmetry from the $3d$ viewpoint.}
					\label{fig:3d5dquiver}
					\end{figure}
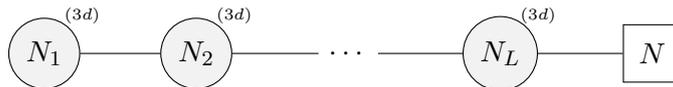

					We follow the construction in \cite{Ashok:2017bld}. Given a $SU(N)$ five-dimensional \CS~theory, the collection of integers 
					\begin{equation}
						\left\{ n_J \right\}_{J=1, \dots, L+1}, \quad \text{ with } \sum_{J=1} ^{L+1} n_J = N 
					\end{equation}
					specifies a defect. The $3d$ quiver has gauge group $U(N_1) \times \cdots \times U(N_L)$, with non-decreasing ranks 
					\begin{equation}
						N_j = \sum_{J=1} ^{j} n_J ,
					\end{equation}
					and the ambient space $SU(N)$ is coupled to the last node as a flavour symmetry. The generic $3d$ \CS~theory has four supercharges, but our methods are best suited for non-chiral theories. Therefore, we focus presently on the special case of $5d$ $\mN=2$ \YM, admitting defects that preserve eight supercharges. We add \CS~terms at each node, eventually obtaining a $3d$ $\mN=3$ quiver as in figure \ref{fig:3d5dquiver}. The study of more general defects is left for future work.
					 \par

					\subsubsection{Minimal partition}
						The first example is the defect with $L=1$, that is, we choose $n_1= \nu N $, $n_2 =(1- \nu) N$, for $0 \le \nu \le \frac{1}{2}$. The resulting system is a $3d$ $U(\nu N)_k$ \CS~theory coupled to $5d$ $SU(N)$ \YM. In the $\mN=1$ formalism, the latter corresponds to the theory with an adjoint hypermultiplet of subsection \ref{sec:pureadjoint}, setting $m=0$. Therefore, the $3d$ system sees two sets of $\frac{N}{2}$ massive hypermultiplets, of masses $ \pm \frac{1}{\lambda_{5d}}$. The SPE is 
						\begin{equation}
						\label{eq:3dsingleSPE}
							\int_{A_{3d}} ^{B_{3d}} \dd \psi \rho_{3d} (\psi ) ~ \text{sign} (\phi - \psi) = \frac{\phi}{t_{3d}} + \frac{1}{4 \nu } \left[  \text{sign}  \left( \phi +  \frac{1}{\lambda_{5d}} \right) +  \text{sign}  \left( \phi -  \frac{1}{\lambda_{5d}} \right) \right] ,
						\end{equation}
						where $t_{3d}$ is the $3d$ \CS~'t Hooft coupling, normalized with the rank $\nu N$ of the $3d$ system. Note that we have defined it intrinsically as the five-dimensional lift of a $3d$ \CS~term, following \cite{KZ}. The outcome agrees with the prescription in \cite{Ashok:2017bld}. Equation \eqref{eq:3dsingleSPE} has been solved in \cite{Barranco} (see also \cite{STWL}), to which we refer for the detailed phase diagram. Varying the $5d$ gauge coupling, the $3d$ system undergoes two third order phase transitions.

					\subsubsection{Long partition}
						The choice of codimension-two defect is encoded in the choice of a partition of $N$, up to shuffling. Therefore, in the large $N$ limit, a typical defect will be described by a typical partition of $N$, picked with uniform distribution. We thus expect $L \sim \sqrt{N}$ and $N_j \sim \sqrt{N}$ at large $N$. The $3d$ theory is then a long quiver, and can be analyzed adapting the procedure of \cite{Coccia:2020cku,Coccia:2020wtk}. As for $5d$ long quivers, the contribution from bi-fundamental hypermultiplets cancels part of the contribution from the vector multiplet in the long quiver limit. The SPE is 
						\begin{equation}
						\label{eq:3dlongSPE}
							- \frac{1}{2} \int \dd  \psi  \frac{ \partial^2 \ }{ \partial z^2}  \rho_{3d} (z, \psi) ~  \text{sign} \left( \varphi - \psi \right) = \frac{\varphi }{t_{3d} (z)}  + \frac{\delta (z-1)}{4}  \left[ \text{sign}  \left( \varphi +  \frac{1}{\tilde{\lambda}_{5d}} \right) +  \text{sign}  \left( \varphi -  \frac{1}{\tilde{\lambda}_{5d}} \right) \right] .
						\end{equation}
						In this expression we have scaled the scalar in the vector multiplet linearly with $L$, keeping $\varphi$ finite. This requires scaling the $5d$ 't Hooft coupling $\lambda_{5d}$ with $L$, keeping $\tilde{\lambda}_{5d}$ finite. If we do not enforce this scaling, the flavour node will reduce to massless hypermultiplets.\par
						The solution to \eqref{eq:3dlongSPE} is akin to subsection \ref{sec:quivers}. We get 
						\begin{equation}
							\rho_{3d} (z, \varphi) = g_{3d} (z) , \qquad \text{supp} \rho_{3d} = [-B_{3d} (z), B_{3d}(z)] , \qquad B_{3d}(z) = \frac{ \nu (z)}{2 g_{3d}(z)} ,
						\end{equation}
						where $g_{3d} (z)$ solves 
						\begin{equation}
							\frac{\dd^2 \ }{\dd z^2} g_{3d} (z) = - \frac{1}{t_{3d} (z)} , \qquad g_{3d} (1) =  \tilde{\lambda}_{5d}  \frac{\nu(1)}{2} .
						\end{equation}
						The solution is determined by the profile of $\frac{1}{t_{3d} (z)}$, with the $5d$ \YM~'t Hooft coupling serving as a boundary condition.\par

	\vspace{0.7cm}
	\paragraph*{Acknowledgements} The author thanks Miguel Tierz for many fruitful discussions. The work was supported by the Funda\c{c}\~{a}o para a Ci\^{e}ncia e a Tecnologia through the doctoral grant {SFRH/BD/129405/2017} and the project {PTDC/MAT-PUR/30234/2017}.
	\vspace{0.4cm}

		\begin{appendix}

			\section{Calabi--Yau varieties and localization}
			\label{app:locgeom}

				As presented in the introduction, the Coulomb branches of the $5d$ $\mathcal{N}=1$ gauge theories we are interested in can be built from resolutions of a singular Calabi--Yau threefold $X$. If $X$ is realized as an elliptic fibration, the theory has special unitary gauge group \cite{Intriligator:1997pq,Esole:2015xfa,a,FF1}, while if $X$ is realized as a $\mathbb{C}^{\ast}$-fibration the gauge group is unitary \cite{c}.\par
				Appendix \ref{sec:geometric} embeds the results of the main text in the geometric framework and explains how to extract matrix models from Calabi--Yau geometries. Then, these facts are applied in appendix \ref{app:specialunitary} to match the reduction from $U(N)$ to $SU(N)$ gauge theories between geometry and partition function.
								
				\subsection{Geometric description}
				\label{sec:geometric}
					
					A $5d$ $\mN=1$ gauge theory $\mathscr{T}_X$ on its Coulomb branch can be read off from the geometry of a crepant resolution\footnote{A resolution is crepant if it preserves the canonical bundle.} $\widetilde{X} \to X$ of a singular local Calabi--Yau threefold $X$. Likewise, a $3d$ $\mathcal{N}=2$ theory is obtained from crepant resolutions of singular local Calabi--Yau fourfolds. We focus on a threefold $X$.\par
					Most of the theories we consider correspond to $\widetilde{X}$ containing compact divisors formed by $N$ intersecting $\mathbb{P}^{1}$s fibered over a single holomorphic genus zero curve, that we denote $\mathbb{P}^1 _0$. Besides, there are $N_f$ divisors that are $\mathbb{P}^1$ fibrations over a non-compact curve inside $\widetilde{X}$. In the models of subsection \ref{sec:adjhyper} the fibrations are instead over a genus one curve, while for the quivers of section \ref{sec:quiverGT} the compact divisors are fibered over a collection of intersecting $\mathbb{P}^{1} _{0,j}$, $j=1,\dots, L$.\par
					K\"{a}hler moduli $\left\{ \phi_a \right\}$ of holomorphic curves that are Poincar\'{e} dual to compact divisors $\left\{ S_{a} \right\} $ give rise to dynamical fields, while K\"{a}hler moduli $\left\{ m_{\alpha} \right\}$ of curves Poincar\'{e} dual to non-compact divisors $\left\{ D_{\alpha} \right\} $ give rise to background fields. The identification of the extended Coulomb branch with the extended K\"{a}hler cone $\mathscr{C} (X)$ of $X$ stems from these relations.\par
					The gauge theory is characterized by a \YM~coupling $\frac{1}{g_{\text{\tiny YM}} ^2} = \text{vol} \left( \mathbb{P}^1 _{0} \right)$, thus the 't Hooft limit \eqref{eq:tHooft} increases the volume of $\mathbb{P}^1 _{0}$ linearly with the number of compact divisors fibered over it. Besides, the Veneziano limit \eqref{eq:Veneziano} corresponds to take the number of both compact and non-compact exceptional divisors in $\widetilde{X}$ large.\par
					\medskip
					We have reviewed how to read off a gauge theory $\mathscr{T}_X$ from a resolution $\widetilde{X} \to X$. In turn, supersymmetric localization provides an explicit dictionary between the field content of a supersymmetric field theory and a matrix model representation of certain observables in such theory on a compact manifold. Therefore, a two-step procedure yields a map 
					\begin{equation}
						\mathscr{C} (X) \ni \widetilde{X} \longleftrightarrow  \mathscr{C} (X) \ni (\phi , m ) \mapsto Z_{\mathbb{S}^{5}} ^{\mathscr{T}_X } (\phi , m ) ,
					\end{equation}
					whose image is the measure in the matrix model on $\mathbb{S}^{5}$. The sphere partition function is the average over $\mathscr{C}_{\mathrm{gauge}}$ of this quantity: 
					\begin{equation}
						\mathcal{Z}_{\mathbb{S}^{5}}  ^{\mathscr{T}_X } (m) =  \int_{\mathscr{C}_{\mathrm{gauge}}} \dd \phi ~ Z_{\mathbb{S}^{5}} ^{\mathscr{T}_X } (\phi , m ) .
					\end{equation}\par
					In the above setup, these last steps define a dictionary whose entries include 
					\begin{align*}
						\text{existence of $S_a$} \ & \Longrightarrow  \ \text{integrate over $\phi_a$} \\
						\text{existence of $D_{\alpha}$} \ & \Longrightarrow \ \text{hypermultiplet of mass $m_{\alpha}$} \\
						\text{intersections} \ & \Longrightarrow  \ Z_{\text{class}} (\phi) Z^{\text{vec}}_{\text{1-loop}} (\phi) Z^{\text{hyp}}_{\text{1-loop}} (\phi) .
					\end{align*}
					Varying a K\"{a}hler parameter $m_{\alpha}$, we have found a phase transition each time the corresponding volume crosses a threshold determined by $\frac{1}{N} \mathrm{vol} \left( \mathbb{P}^{1} _0 \right)$. Importantly, these transitions take place at strictly infinite rank of the gauge group, and differ in nature from the flop transitions among two birationally equivalent resolved geometries $\widetilde{X}$. An exception to this statement is discussed in subsection \ref{sec:UNF3YMinfty}.\par

				\subsection{St\"{u}ckelberg mechanism from localization}
				\label{app:specialunitary}

					In this appendix we comment on the matrix model interpretation of the St\"{u}ckelberg mechanism presented in \cite{c} to pass from unitary to special unitary quiver gauge theories. The argument has been shown to hold for three-dimensional gauge theories \cite{CollinucciValandro:3d}.\par
					The partition function on either $\mathbb{S}^3$ or $\mathbb{S}^5$ when Fayet--Iliopoulos parameters $\left\{ \xi_j \right\} $ are turned on is written schematically as 
					\begin{equation}
						\mathcal{Z}_{\mathbb{S}^{d}} ^{\mathscr{T}_X } (m) = \frac{1}{\prod_{j} N_j !} \int_{\mathbb{R}^{\text{rank} (G)}} \dd \phi ~ e^{i 4 \pi r^2 \sum_{j=1} ^{L} \xi_j \left( \sum_{a=1} ^{N_j} \phi_{a,j} \right) } ~ Z_{\mathbb{S}^{d}} ^{\mathscr{T}_X } (\phi , m )  ,
					\end{equation}
					for gauge group $G= U (N_1) \times \cdots \times U(N_L)$. The parameters $\xi_{j}$ are dual to the fibre of $X$. Therefore, according to the dictionary in appendix \ref{sec:geometric}, compactifying the $\C^{\ast}$-fibre we have to integrate over the scalars $\xi_j$, producing a $\delta$-constraint at each gauge node. In $3d$, this corresponds to gauging the $U(1)_j$ global symmetry at each node, in agreement with \cite{CollinucciValandro:3d}.\par
					The situation is very similar in $5d$, although we first have to address a subtlety with the integration contour. In section \ref{sec:CBloc} we have inserted $\xi$ as a Lagrange multiplier, while now we want to treat it as the lowest component of a full-fledged dynamical Abelian vector multiplet. Hence, localization dictates to rotate its integration contour $\xi \mapsto i \xi $ \cite{KZ}, eventually producing the correct factor.

			\section{Other vanishing curvature limits}
			\label{app:otherdeco}
				In section \ref{sec:decompactification} we have shown how to reproduce the prepotential of a $5d$ $\mathcal{N}=1$ theory \cite{Intriligator:1997pq} from the localized partition function on $\mathbb{S}^5$ of radius $r$, taking the limit $\frac{1}{r} \to 0$. A slightly different prepotential is obtained as follows \cite{a,Hayashi:2019jvx}: first compactify on $\mathbb{S}^1 \times \R^4 $ with circle of radius $\ell$, take the $4d$ $\mN=2$ prepotential and dress it with the full tower of KK modes, and eventually take the limit $\ell \to \infty$.\par
				The discrepancy between the two prepotentials is translated in the present context as the ambiguity in first putting the theory on curved space, localize, and then take the vanishing curvature limit. In the light of the universality result \cite{Crichigno:2020ouj} 
				\begin{equation}
					\lim_{N \to \infty} \mathcal{F}_{\mathbb{S}^4 \times \mathbb{S}^1} \propto \lim_{N \to \infty} \mathcal{F}_{\mathbb{S}^5} ,
				\end{equation}
				localizing on $\mathbb{S}^1 \times \mathbb{S}^4$ and then taking the decompactification limit leads to a phase structure identical to the one we have obtained.\par

					\section{Eigenvalue densities}
		\label{app:secrho}
			The eigenvalue densities we have found in the main text have the generic form
			\begin{equation}
				\rho (\phi) = c_A \delta \left( \phi - A  \right) + c_B \delta \left( \phi - B  \right) + \sum_{\alpha=1}^{F} c_{\alpha} \delta \left( \phi + m_{\alpha} \right) , \qquad \text{supp} \rho = [A,B] .
			\end{equation}
			In this appendix we collect the explicit expressions of the parameters $c_A$, $c_B$, and the endpoints $A$, $B$ of the support. As explained in subsection \ref{sec:largeNlimit}, $c_{\alpha}= \frac{\zeta_{\alpha}}{2}$ if $ -m_{\alpha} \in [A,B]$ and $c_{\alpha}=0$ otherwise. We do not report these coefficients below.\par
			Recall that $A$ and $B$ are found solving quadratic equations. In all the subsequent expressions the correct choice of sign in front of each square root has already been made. For instance, $\sqrt{\lambda^2 (\cdots)}= + \vert \lambda \vert \sqrt{ (\cdots)}$ is understood.

		\subsection{Eigenvalue densities: ${U(N)}$ theories}	
		\label{app:rhoUN}
			In this appendix we collect the coefficients $c_A$, $c_B$ and the endpoints $A$, $B$ determining the eigenvalue density $\rho (\phi)$ in the various $U(N)$ gauge theories studied in section \ref{sec:UN}.
			
			\subsubsection*{${F=1}$, ${\vert \lambda \vert  \to \infty}$}
				\begin{subequations}
				\label{eq:rhoUNF1YMinf}
				\begin{equation}
					c_A  = \begin{cases}  \frac{(2-\zeta ) t-2}{4 t}  & m<0  \\  \frac{(2+\zeta ) t-2}{4 t}  & m>0 , \end{cases}   \qquad c_B = \begin{cases}  \frac{(2+\zeta ) t+2}{4 t}  & m<0  \\  \frac{(2-\zeta ) t+2}{4 t}  & m>0 ,  \end{cases}
				\end{equation}
				\begin{align}
					A & = \begin{cases}  \frac{m}{\zeta  t+2} \left(\sqrt{2} \sqrt{\frac{\zeta  t ((\zeta +2) t+2)}{(\zeta -2) t+2}}-\zeta t\right) & m<0  \\ \frac{\zeta  m t-\frac{\sqrt{2} m \sqrt{\zeta  (-t) \left(\left(\zeta ^2-4\right) t^2-4\zeta  t+4\right)}}{(\zeta +2) t-2}}{2-\zeta  t}  & m>0 , \end{cases}  \\
					 B & = \begin{cases} \frac{\sqrt{2} m \sqrt{\zeta  t ((\zeta -2) t+2) ((\zeta +2) t+2)}-\zeta  m t ((\zeta +2) t+2)}{(\zeta  t+2) ((\zeta +2) t+2)} & m<0  \\ -\frac{m \left(\zeta  t ((\zeta -2) t-2)+\sqrt{2} \sqrt{\zeta  (-t) ((\zeta -2) t-2)((\zeta +2) t-2)}\right)}{((\zeta -2) t-2) (\zeta  t-2)}  & m>0 . \end{cases}
				\end{align}		
				\end{subequations}
				
				\subsubsection*{${F=1}$, ${\vert \lambda \vert < \infty}$}
				\begin{subequations}
				\label{eq:rhoUNF1gen}
				\begin{equation}
					c_A = \begin{cases}  \frac{(2-\zeta ) t-2}{4 t}  & m< m_{\text{cr},1} \\   \frac{(2-\zeta ) t-2}{4 t} & m_{\text{cr},1} < m < m_{\text{cr},2}  \\  \frac{(2+\zeta ) t-2}{4 t}  & m>m_{\text{cr},2} , \end{cases}   \qquad c_B = \begin{cases}  \frac{(2+\zeta ) t+2}{4 t}  & m<m_{\text{cr},1} \\ \frac{(2-\zeta ) t+2}{4 t}   & m_{\text{cr},1} < m < m_{\text{cr},2} \\   \frac{(2-\zeta ) t+2}{4 t}  & m>m_{\text{cr},2}  , \end{cases}
				\end{equation}
				\begin{align}
					A & = \begin{cases} {\scriptstyle \frac{\sqrt{2} \sqrt{\lambda ^2 (-t) ((\zeta -2) t+2) ((\zeta +2) t+2) \left(2 t (\zeta   \lambda  m+1)-\zeta  \lambda ^2 m^2\right)}-\lambda  t ((\zeta -2) t+2) (\zeta  \lambda  m+2)}{\lambda ^2 ((\zeta -2) t+2) (\zeta  t+2)} }  & {\scriptstyle m<m_{\text{cr},1} } \\  {\scriptstyle \frac{\sqrt{\lambda ^2 t^2 \left((\zeta -2)^2 t^2-4\right)}}{\lambda \left( (\zeta -2) t+2 \right) } - \frac{t}{\lambda } } & {\scriptstyle m_{\text{cr},1} < m < m_{\text{cr},2} } \\ {\scriptstyle \frac{\sqrt{2} \sqrt{\lambda ^2 t ((\zeta -2) t-2) ((\zeta +2) t-2) \left(2 t (\zeta \lambda  m-1)-\zeta  \lambda ^2 m^2\right)}+\lambda  t ((\zeta +2) t-2) (2-\zeta  \lambda  m)}{\lambda ^2 (\zeta  t-2) ((\zeta +2) t-2)} } & {\scriptstyle m>m_{\text{cr},2} } , \end{cases}  \\
					 B & = \begin{cases} {\scriptstyle -\frac{\sqrt{2} \sqrt{\lambda ^2 (-t) ((\zeta -2) t+2) ((\zeta +2) t+2) \left(2 t (\zeta  \lambda  m+1)-\zeta  \lambda ^2 m^2\right)}+\lambda  t ((\zeta +2) t+2) (\zeta \lambda  m+2)}{\lambda ^2 (\zeta  t+2) ((\zeta +2) t+2)} } & {\scriptstyle m<m_{\text{cr},1} } \\  {\scriptstyle \frac{\sqrt{\lambda ^2 t^2 \left((\zeta -2)^2 t^2-4\right)}}{\lambda  \left( (\zeta -2) t-2 \right)}- \frac{ t}{\lambda } } & {\scriptstyle m_{\text{cr},1} < m < m_{\text{cr},2} } \\ {\scriptstyle -\frac{\sqrt{2} \sqrt{\lambda ^2 t \left(\left(\zeta ^2-4\right) t^2-4 \zeta  t+4\right) \left(2 t (\zeta  \lambda  m-1)-\zeta  \lambda ^2 m^2\right)}+\lambda  t ((\zeta -2) t-2) (\zeta  \lambda  m-2)}{\lambda ^2 ((\zeta -2) t-2) (\zeta  t-2)} } & {\scriptstyle m>m_{\text{cr},2} } . \end{cases}
				\end{align}		
				\end{subequations}

				\subsubsection*{${F=1}$, ${\vert t \vert \to \infty}$}
					
				\begin{subequations}
				\label{eq:rhoUNF1tinfty}
				\begin{equation}
					c_A = \begin{cases}  \frac{2-\zeta }{4}  & m< m_{\text{cr},1} \\   \frac{2-\zeta }{4} & m_{\text{cr},1} < m < m_{\text{cr},2}  \\  \frac{2+\zeta }{4}  & m>m_{\text{cr},2} , \end{cases}   \qquad c_B = \begin{cases}  \frac{2+\zeta }{4 }  & m<m_{\text{cr},1} \\ \frac{2-\zeta }{4}   & m_{\text{cr},1} < m < m_{\text{cr},2} \\   \frac{2-\zeta }{4 }  & m>m_{\text{cr},2}  , \end{cases}  
				\end{equation}
				\begin{equation}
					A  = \begin{cases} {\scriptstyle \frac{2 \left((2-\zeta ) \lambda +\sqrt{\left(4-\zeta ^2\right) \lambda ^2 (\zeta \lambda  m+1)}\right)}{(\zeta -2) \zeta  \lambda ^2}-m } & {\scriptstyle  m<m_{\text{cr},1} } \\ {\scriptstyle  \frac{2}{(2-\zeta) \lambda } } & {\scriptstyle  m_{\text{cr},1} < m < m_{\text{cr},2} } \\  {\scriptstyle -\frac{2 \left((\zeta +2) \lambda +\sqrt{\left(4-\zeta ^2\right) \lambda ^2 (1-\zeta  \lambda  m)}\right)}{(-\zeta -2) \zeta  \lambda ^2}-m } & {\scriptstyle  m>m_{\text{cr},2} } , \end{cases}  
				\end{equation}
				\begin{equation}
					 B = \begin{cases} {\scriptstyle  -\frac{2 \left((\zeta +2) \lambda +\sqrt{\left(4-\zeta ^2\right) \lambda ^2 (\zeta \lambda  m+1)}\right)}{\zeta  (\zeta +2) \lambda ^2}-m } & {\scriptstyle  m<m_{\text{cr},1} } \\ {\scriptstyle - \frac{2}{(2-\zeta) \lambda } } & {\scriptstyle m_{\text{cr},1} < m < m_{\text{cr},2} } \\ {\scriptstyle \frac{2 \left((2-\zeta ) \lambda +\sqrt{\left(4-\zeta ^2\right) \lambda ^2 (1-\zeta \lambda  m)}\right)}{(2-\zeta ) \zeta  \lambda ^2}-m } & {\scriptstyle m>m_{\text{cr},2} }. \end{cases}
				\end{equation}
				\end{subequations}

			\subsubsection*{${F=2}$, ${\vert \lambda \vert \to \infty}$. Symmetric case}
			If $t>\frac{1}{1-\zeta}$:
			\begin{subequations}
			\begin{align}
				c_A & =  \frac{t-1}{2t} , \qquad c_B = \frac{t+1}{2t} ,  \\
				A  & = \begin{cases} \zeta  m t \left(\frac{t+1}{\sqrt{t^2-1}}-1\right) & m<0 \\ - \zeta m t \left(\frac{t+1}{\sqrt{t^2-1}}-1\right)  & m>0 , \end{cases} \ \qquad \  B  = \begin{cases} \zeta  m t \left(\frac{t-1}{\sqrt{t^2-1}}-1\right) & m<0 \\ - \zeta  m t \left(\frac{t-1}{\sqrt{t^2-1}}-1\right) & m>0. \end{cases} 
			\end{align}
			\label{eq:rhoUNF2symYMinftpos}
			\end{subequations}
			If $t<-\frac{1}{1-\zeta}$:
			\begin{subequations}
			\begin{align}
				c_A & =  \frac{t-1}{2t} , \qquad c_B = \frac{t+1}{2t} ,  \\
				A & = \begin{cases} -\frac{\zeta  m t \left(\sqrt{t^2-1}+t-1\right)}{t-1}  & m<0 \\  \frac{\zeta  m t \left(\sqrt{t^2-1}+t-1\right)}{t-1} & m>0 , \end{cases} \ \qquad \  B = \begin{cases} -\frac{\zeta  m t \left(\sqrt{t^2-1}+t+1\right)}{t+1} & m<0 \\ \frac{\zeta  m t \left(\sqrt{t^2-1}+t+1\right)}{t+1} & m>0. \end{cases}
			\end{align}
			\label{eq:rhoUNF2symYMinftneg}
			\end{subequations}

			\subsubsection*{${F=2}$, ${\vert \lambda \vert < \infty}$. Symmetric case}
			\begin{subequations}
			\label{eq:rhoUNF2symtpos}
			\begin{equation}
			c_A = \begin{cases} \frac{t-1}{2t} & {\scriptstyle m>m_{\text{cr,1}}} \\ \frac{t (1- \zeta) -1 }{2t}   & {\scriptstyle m_{\text{cr,2}} < m < m_{\text{cr,1}} \text{ and } t>0 } \\ \frac{t-1}{2t} & {\scriptstyle m_{\text{cr,2}} <  m<m_{\text{cr,1}} \text{ and } t <0 } \\  \frac{t (1- \zeta) -1 }{2t}  & {\scriptstyle m<m_{\text{cr,2}}} ,  \end{cases} \qquad c_B = \begin{cases} \frac{t+1}{2t}  & {\scriptstyle m>m_{\text{cr,1}}} \\ \frac{t+1}{2t}  &  {\scriptstyle m_{\text{cr,2}} < m < m_{\text{cr,1}} \text{ and } t>0 } \\ \frac{t(1 - \zeta) +1}{2t} & {\scriptstyle m_{\text{cr,2}} < m < m_{\text{cr,1}} \text{ and } t<0} \\  \frac{t(1 - \zeta) +1}{2t}  & {\scriptstyle m<m_{\text{cr,2}} } ,  \end{cases} 
			\end{equation}
			\begin{equation}
				A  = \begin{cases}  {\scriptstyle \frac{\sqrt{t^2 \left(t^2-1\right) \left(\frac{1}{\lambda }-\zeta  m\right)^2}}{1-t}+\zeta  m t-\frac{t}{\lambda } } & {\scriptstyle m>m_{\text{cr},1} } \\ 
				 	 {\scriptstyle  \frac{\sqrt{2} \lambda  \sqrt{\frac{t (t+1) ((\zeta -1) t+1) \left(\zeta  \lambda ^2 m^2+2 t (\zeta  \lambda  m-1)\right)}{\lambda ^2}}+(\zeta -1) t^2 (\zeta  \lambda  m-2)+t (\zeta  \lambda  m-2)}{\lambda  ((\zeta -1) t+1) (\zeta  t+2)} } & {\scriptstyle m_{\text{cr},2}<m<m_{\text{cr},1} \text{\footnotesize and } t>0 }\\
				   	{\scriptstyle -\frac{\sqrt{2} \sqrt{(t-1) t ((\zeta -1) t-1) \left(2 t (\zeta  \lambda  m-1)-\zeta  \lambda ^2 m^2\right)}+(t-1) t (\zeta  \lambda  m-2)}{\lambda  (t-1) (\zeta  t-2)} } & {\scriptstyle m_{\text{cr},2}<m<m_{\text{cr},1} \text{\footnotesize and } t<0 } \\
				     {\scriptstyle \frac{\sqrt{\frac{t^2 \left((\zeta -1)^2 t^2-1\right)}{\lambda ^2}}}{(\zeta -1) t+1}-\frac{t}{\lambda } }  & {\scriptstyle m<m_{\text{cr},2} } , \end{cases} 
			\end{equation}
			\begin{equation}
				 B = \begin{cases}  {\scriptstyle  -\frac{\sqrt{t^2 \left(t^2-1\right) \left(\frac{1}{\lambda }-\zeta  m\right)^2}}{t+1}+\zeta  m t-\frac{t}{\lambda } } & {\scriptstyle m>m_{\text{cr},1} } \\ 
				  	 {\scriptstyle  \frac{\sqrt{2} \sqrt{t (t+1) ((\zeta -1) t+1) \left(\zeta  \lambda ^2 m^2+2 t (\zeta  \lambda  m-1)\right)}+t (t+1) (\zeta  \lambda  m-2)}{\lambda  (t+1) (\zeta  t+2)} }  & {\scriptstyle m_{\text{cr},2}<m<m_{\text{cr},1} \text{\footnotesize and } t>0 } \\
					 {\scriptstyle \frac{\sqrt{2} \sqrt{(t-1) t ((\zeta -1) t-1) \left(2 t (\zeta  \lambda  m-1)-\zeta  \lambda ^2 m^2\right)}-t ((\zeta -1) t-1) (\zeta  \lambda  m-2)}{\lambda  ((\zeta -1) t-1) (\zeta  t-2)} }  & {\scriptstyle m_{\text{cr},2}<m<m_{\text{cr},1} \text{\footnotesize and } t<0 } \\ 
				  	 {\scriptstyle -\frac{\frac{\sqrt{t^2 \left((\zeta -1)^2 t^2-1\right)}}{(\zeta -1) t-1}+t}{\lambda } }  & {\scriptstyle m<m_{\text{cr},2} }  . \end{cases} 
			\end{equation}
			\end{subequations}

			\subsubsection*{${F=2}$, ${\vert \lambda \vert \to  \infty}$. Generic case}
				We impose $m_2=m$ and $m_1 = - \frac{\zeta_2}{\zeta_1} m$.
				\begin{subequations}
				\label{eq:rhoUNF2gen}
				\begin{equation}
				c_A = \begin{cases}  \frac{t (2-\zeta _1+\zeta_2  )-2}{4 t}   & m<0  \\ \frac{t (2+\zeta _1-\zeta_2 )-2}{4 t}   & m>0 , \end{cases} \qquad c_B = \begin{cases} \frac{t (2 + \zeta _1 -\zeta _2 ) +2}{4 t}  & m<0  \\ \frac{t (2 - \zeta _1 + \zeta _2 ) +2}{4 t}     & m>0 , \end{cases} 
				\end{equation}
				\begin{equation}
				 A  = \begin{cases}  {\scriptstyle \frac{m \left(2 \zeta _1 \zeta _2 t \left(\left(\zeta _1-\zeta _2-2\right) t+2\right)+\sqrt{\zeta _1 \zeta _2 (-t) \left(\left(\zeta _1-\zeta _2-2\right) t+2\right) \left(\left(\zeta _1-\zeta _2+2\right) t+2\right) \left(\zeta _1^2 t+2 \zeta _1 \left(\zeta _2 t+1\right)+\zeta _2 \left(\zeta _2 t-2\right)\right)}\right)}{\zeta _1 \left(\left(\zeta _1-\zeta _2-2\right) t+2\right) \left(\left(\zeta _1-\zeta _2\right) t+2\right)}  }  & {\scriptstyle m<0 } \\ 
			       {\scriptstyle   -\frac{m \left(2 \zeta _1 \zeta _2 t \left(\left(-\zeta _1+\zeta _2-2\right) t+2\right)+\sqrt{\zeta _1 \zeta _2 (-t) \left(\left(-\zeta_1+\zeta _2-2\right) t+2\right) \left(\left(-\zeta _1+\zeta _2+2\right) t+2\right) \left(\zeta _1^2 t+2 \zeta _1 \left(\zeta _2t-1\right)+\zeta _2 \left(\zeta _2 t+2\right)\right)}\right)}{\zeta _1 \left(\zeta _1 (-t)+\zeta _2 t+2\right) \left(\left(-\zeta_1+\zeta _2-2\right) t+2\right)} }   &  {\scriptstyle m>0 } , \end{cases} 
			    \end{equation}
			    \begin{equation}
				  B = \begin{cases} {\scriptstyle   \frac{2 \zeta _1 \zeta _2 m t \left(\left(\zeta _1-\zeta _2+2\right) t+2\right)-m \sqrt{\zeta _1 \zeta _2 (-t) \left(\left(\zeta _1-\zeta_2-2\right) t+2\right) \left(\left(\zeta _1-\zeta _2+2\right) t+2\right) \left(\zeta _1^2 t+2 \zeta _1 \left(\zeta _2 t+1\right)+\zeta _2 \left(\zeta _2 t-2\right)\right)}}{\zeta _1 \left(\left(\zeta _1-\zeta _2\right) t+2\right) \left(\left(\zeta _1-\zeta _2+2\right) t+2\right)}  }   & {\scriptstyle m<0 } \\ 
				         {\scriptstyle   \frac{m \left(-2 \zeta _2 t-\frac{\sqrt{\zeta _1 \zeta _2 (-t) \left(\left(-\zeta _1+\zeta _2-2\right) t+2\right) \left(\left(-\zeta_1+\zeta _2+2\right) t+2\right) \left(\zeta _1^2 t+2 \zeta _1 \left(\zeta _2 t-1\right)+\zeta _2 \left(\zeta _2 t+2\right)\right)}}{\zeta_1 \left(\zeta _1 t-\zeta _2 t-2 (t+1)\right)}\right)}{\zeta _1 (-t)+\zeta _2 t+2} }   &  {\scriptstyle m >0 }  . \end{cases}  
				\end{equation}
				\end{subequations}

			\subsubsection*{${F=2}$, ${\vert \lambda \vert \to  \infty}$. Generic case revisited}
				When $-m_1>B, -m_2<A$, 
				\begin{subequations}
				\label{eq:rhoUNF2YMinftyI}
				\begin{equation}
					c_A = \frac{ 2 -\zeta _1 +\zeta _2 }{4 } - \frac{1}{2t}, \qquad c_B = \frac{2 + \zeta _1 -\zeta _2 }{4 } + \frac{1}{2t}  , 
				\end{equation}
				\begin{align}
					A & = {\scriptstyle \frac{\sqrt{-t \left(\left(\zeta _1-\zeta _2-2\right) t+2\right) \left(\left(\zeta_1-\zeta _2+2\right) t+2\right) \left(-2 \zeta _1 m_1^2+2 \zeta _2 m_2^2+\zeta _1\zeta _2 \left(m_1-m_2\right){}^2 t\right)}-t \left(\zeta _1 m_1-\zeta _2 m_2\right)\left(\left(\zeta _1-\zeta _2-2\right) t+2\right)}{\left(\left(\zeta _1-\zeta_2-2\right) t+2\right) \left(\left(\zeta _1-\zeta _2\right) t+2\right)} } , \\
					B & =  {\scriptstyle -\frac{t \left(\zeta _1 m_1-\zeta _2 m_2\right) \left(\left(\zeta _1-\zeta _2+2\right)t+2\right)+\sqrt{-t \left(\left(\zeta _1-\zeta _2-2\right) t+2\right)\left(\left(\zeta _1-\zeta _2+2\right) t+2\right) \left(-2 \zeta _1 m_1^2+2 \zeta _2m_2^2+\zeta _1 \zeta _2 \left(m_1-m_2\right){}^2 t\right)}}{\left(\left(\zeta _1-\zeta_2\right) t+2\right) \left(\left(\zeta _1-\zeta _2+2\right) t+2\right)} }.
				\end{align}
				\end{subequations}
				When $-m_2<A<-m_1<B$, 
				\begin{subequations}
				\label{eq:rhoUNF2YMinftyIIm1}
				\begin{equation}
					c_A = \frac{2 -\zeta _1 +\zeta _2 }{4 } - \frac{1}{2t}, \qquad c_B = \frac{2-\zeta _1 -\zeta _2 }{4} + \frac{1}{2t} , 
				\end{equation}
				\begin{align}
					A & =  {\scriptstyle \frac{\zeta _2 m_2 t \left(\left(-\zeta _1+\zeta _2+2\right) t-2\right)-\sqrt{2} m_2\sqrt{\zeta _2 t \left(\left(\zeta _1-\zeta _2-2\right) t+2\right) \left(\left(\zeta_1+\zeta _2-2\right) t-2\right)}}{\left(\left(\zeta _1-\zeta _2-2\right) t+2\right) \left(\zeta _2 t-2\right)} } , \\
					B & =  {\scriptstyle \frac{m_2 \left(\zeta _2 (-t)-\frac{\sqrt{2} \sqrt{\zeta _2 t \left(\left(\zeta _1-\zeta_2-2\right) t+2\right) \left(\left(\zeta _1+\zeta _2-2\right) t-2\right)}}{\left(\zeta_1+\zeta _2-2\right) t-2}\right)}{\zeta _2 t-2} } .
				\end{align}
				\end{subequations}

			\subsubsection*{${F=2}$, ${-\infty< \lambda <0}$. Generic case}
				When $A < -m_1,-m_2 <B$,  
				\begin{subequations}
				\label{eq:rhoUNF2genmiddle}
				\begin{equation}
					c_A = \frac{2-\zeta_1 - \zeta_2}{4} - \frac{1}{2t} , \qquad c_B = \frac{2-\zeta_1 - \zeta_2}{4} + \frac{1}{2t} , 
				\end{equation}
				\begin{equation}
					A = \frac{t}{\lambda} \left(\sqrt{1-\frac{4}{\left(\zeta _1+\zeta _2\right) t-2 t+2}}-1\right) ,  \ \qquad \ B =  \frac{t}{\lambda } \left(\sqrt{\frac{4}{\left(\zeta _1+\zeta _2\right) t-2(t+1)}+1}-1\right)  .
				\end{equation}
				\end{subequations}
				When $-m_2 <A$ and $A<-m_1<B$, 
				\begin{subequations}
				\label{eq:rhoUNF2genintA}
				\begin{equation}
					c_A = \frac{2-\zeta_1 + \zeta_2}{4} - \frac{1}{2t} , \qquad c_B = \frac{2-\zeta_1 - \zeta_2}{4} + \frac{1}{2t} , 				
				\end{equation}
				\begin{align}
					A & = {\scriptstyle \frac{\zeta _2 (-\lambda ) m_2 t+\frac{\sqrt{2} \sqrt{-t \left(\zeta _1 t-\zeta _2 t-2 t+2\right) \left(\left(\zeta _1+\zeta _2\right) t-2 (t+1)\right) \left(\zeta _2 \lambda  m_2 \left(2 t-\lambda  m_2\right)-2 t\right)}}{\zeta _1 t-\zeta _2 t-2 t+2}+2t}{\lambda  \left(\zeta _2 t-2\right)}   } , \\
					B & = {\scriptstyle \frac{t \left(\left(\zeta _1+\zeta _2-2\right) t-2\right) \left(2-\zeta _2 \lambda  m_2\right)+\sqrt{2} \sqrt{-t \left(\left(\zeta _1-\zeta _2-2\right) t+2\right) \left(\left(\zeta _1+\zeta _2-2\right) t-2\right) \left(\zeta _2 \lambda  m_2 \left(2 t-\lambda  m_2\right)-2 t\right)}}{\lambda  \left(\zeta _2 t-2\right) \left(\left(\zeta _1+\zeta _2-2\right) t-2\right)} }.
				\end{align}
				\end{subequations}
				When $-m_1>B$ and $A<-m_2<B$, 
				\begin{subequations}
				\label{eq:rhoUNF2genintB}
				\begin{equation}
					c_A = \frac{2-\zeta_1 - \zeta_2}{4} - \frac{1}{2t} , \qquad c_B = \frac{2+\zeta_1 - \zeta_2}{4} + \frac{1}{2t} , 
				\end{equation}
				\begin{align}
					A & = {\scriptstyle \frac{1}{\lambda ^2 \left(\zeta _1 t+2\right)}   \left[ \frac{\sqrt{2} \sqrt{\lambda ^2 t \left(\left(\zeta _1-\zeta _2+2\right) t+2\right) \left(\left(\zeta _1+\zeta _2-2\right) t+2\right) \left(\zeta _1 \lambda  m_1 \left(\lambda  m_1-2 t\right)-2 t\right)}}{\left(\zeta _1+\zeta _2-2\right) t+2}-\lambda  t \left(\zeta _1 \lambda  m_1+2\right) \right]  } , \\
					B & = {\scriptstyle -\frac{1}{\lambda ^2 \left(\zeta _1 t+2\right)}  \left[ \frac{\sqrt{2} \sqrt{\lambda ^2 t \left(\left(\zeta _1-\zeta _2+2\right) t+2\right) \left(\left(\zeta _1+\zeta _2-2\right) t+2\right) \left(\zeta _1 \lambda  m_1 \left(\lambda  m_1-2 t\right)-2 t\right)}}{\left(\zeta _1-\zeta _2+2\right) t+2}+\lambda  t \left(\zeta _1 \lambda  m_1+2\right)\right]  } .
				\end{align}
				\end{subequations}
				When $-m_1 >B$ and $-m_2<A$, 
				\begin{subequations}
				\label{eq:rhoUNF2genextA}
				\begin{equation}
					c_A = \frac{2-\zeta_1 - \zeta_2}{4} - \frac{1}{2t} , \qquad c_B = \frac{2+\zeta_1 + \zeta_2}{4} + \frac{1}{2t} , 
				\end{equation}
				\begin{equation}
				\begin{aligned}
					A & = \left[  {\scriptstyle  \sqrt{\lambda ^2 (-t) \left(\left(\zeta _1-\zeta _2-2\right) t+2\right) \left(\left(\zeta _1-\zeta _2+2\right) t+2\right) \left(-2 \zeta _1 \lambda  m_1 \left(\lambda  m_1-2 t\right)+\zeta _2 \lambda  \left(\zeta _1 \lambda  \left(m_1-m_2\right){}^2 t+2 m_2 \left(\lambda  m_2-2 t\right)\right)+4t\right) } }  \right.   \\  
					&  \quad \left.  {\scriptstyle  -\lambda  t \left(\left(\zeta _1-\zeta _2-2\right) t+2\right) \left(\zeta _1 \lambda  m_1-\zeta _2 \lambda  m_2+2\right) } \right] {\scriptstyle \frac{ 1 }{ \lambda ^2 \left(\left(\zeta _1-\zeta_2-2\right) t+2\right) \left(\zeta _1 t-\zeta _2 t+2\right)}  }  , 
				\end{aligned}
				\end{equation}
				\begin{equation}
				\begin{aligned}
					B & = - \left[ { \scriptstyle   \sqrt{\lambda ^2 (-t) \left(\left(\zeta _1-\zeta _2-2\right) t+2\right) \left(\left(\zeta _1-\zeta _2+2\right) t+2\right) \left(-2 \zeta _1 \lambda  m_1 \left(\lambda  m_1-2 t\right)+\zeta _2 \lambda  \left(\zeta _1 \lambda  \left(m_1-m_2\right){}^2 t+2 m_2 \left(\lambda  m_2-2 t\right)\right)+4 t\right)} } \right. \\
						& \quad \quad \left.   { \scriptstyle  +\lambda  t \left(\left(\zeta _1-\zeta _2+2\right) t+2\right) \left(\zeta _1 \lambda  m_1-\zeta _2 \lambda  m_2+2\right)  } \right]  { \scriptstyle \frac{1}{\lambda ^2 \left(\left(\zeta _1-\zeta _2+2\right) t+2\right) \left(\zeta _1 t-\zeta _2 t+2\right)}  } .
				\end{aligned}
				\end{equation}
				\end{subequations}
				When $-m_1 <A$ and $-m_2>B$, 
				\begin{subequations}
				\label{eq:rhoUNF2genextB}
				\begin{equation}
					c_A = \frac{2-\zeta_1 - \zeta_2}{4} - \frac{1}{2t} , \qquad c_B = \frac{2+\zeta_1 + \zeta_2}{4} + \frac{1}{2t} , 
				\end{equation}
				\begin{equation}
				\begin{aligned}
					A & = \left[ {\scriptstyle    \sqrt{\lambda ^2 t \left(\left(\zeta _1+\zeta _2-2\right) t-2\right) \left(\left(\zeta _1+\zeta _2+2\right) t-2\right) \left(-2 \zeta _1 \lambda  m_1 \left(\lambda  m_1-2 t\right)+\zeta _2 \lambda  \left(\zeta _1 \lambda  \left(m_1-m_2\right){}^2 t-2 m_2 \left(\lambda  m_2-2 t\right)\right)-4 t\right)} } \right. \\
						& \quad \left.   {\scriptstyle    -\lambda  t \left(\left(\zeta _1+\zeta _2+2\right) t-2\right) \left(\zeta _1 \lambda  m_1+\zeta _2 \lambda  m_2-2\right) } \right] { \scriptstyle \frac{1}{\lambda ^2 \left(\left(\zeta _1+\zeta_2\right) t-2\right) \left(\left(\zeta _1+\zeta _2+2\right) t-2\right)}  } ,  
				\end{aligned}
				\end{equation}
				\begin{equation}
				\begin{aligned}
					B & = \left[ {\scriptstyle    \sqrt{t \left(\left(\zeta _1+\zeta _2-2\right) t-2\right) \left(\left(\zeta_1+\zeta _2+2\right) t-2\right) \left(-2 \zeta _1 \lambda  m_1 \left(\lambda  m_1-2t\right)+\zeta _2 \lambda  \left(\zeta _1 \lambda  \left(m_1-m_2\right){}^2 t-2 m_2 \left(\lambda  m_2-2 t\right)\right)-4 t\right)}  } \right. \\    
						& \quad \left. {\scriptstyle  -t \left(\left(\zeta _1+\zeta _2-2\right) t-2\right) \left(\zeta _1 \lambda  m_1+\zeta _2 \lambda  m_2-2\right) } \right] { \scriptstyle \frac{1}{\lambda  \left(\left(\zeta _1+\zeta _2-2\right) t-2\right) \left(\left(\zeta _1+\zeta _2\right) t-2\right)} } .
				\end{aligned}
				\end{equation}
				\end{subequations}

		\subsection{Eigenvalue densities: ${SU(N)}$ theories}
		\label{app:rhoSuN}
				In this appendix we collect the endpoints $A$, $B$ determining the eigenvalue density $\rho (\phi)$ in the various $SU(N)$ gauge theories studied in section \ref{sec:SuN}. The coefficients $c_A$ and $c_B$ are equal to the ones in the corresponding $U(N)$ theory, and we do not report them as they already appear in appendix \ref{app:rhoUN}.\par
				
		\subsubsection*{${F=1}$, ${\vert \lambda \vert  \to \infty}$}
				
				\begin{equation}
					A = \begin{cases} -\frac{\zeta  m t}{(\zeta -2) t+2}  & m<0  \\ 
					 	     -\frac{\zeta  m t}{(\zeta +2) t-2}  & m>0 , \end{cases}  \ \qquad \ 
					 B = \begin{cases}-\frac{\zeta  m t}{(\zeta +2) t+2} & m<0  \\ 
					 	 \frac{\zeta  m t}{2-(\zeta -2) t}  & m>0 . \end{cases}
				\label{eq:rhoSuNF1YMinf}
				\end{equation}

			\subsubsection*{${F=1}$, ${\vert \lambda \vert < \infty}$}
				
				\begin{equation}
					A = \begin{cases}   -\frac{t (\zeta  \lambda  m+2)}{\lambda  ((\zeta -2) t+2)}  & m<m_{\text{cr,1}}  \\ 
					 		     -\frac{t (\zeta  \lambda  m+2)}{\lambda  ((\zeta -2) t+2)}  & m_{\text{cr,1}} < m < m_{\text{cr,2}}   \\
					 	        \frac{t (2-\zeta  \lambda  m)}{\lambda  ((\zeta +2) t-2)}  & m>m_{\text{cr,2}} , \end{cases}   \ \qquad \ 
					 B = \begin{cases}  \frac{t (\zeta  \lambda  m-2)}{\lambda  ((\zeta +2) t+2)} 	& m<m_{\text{cr,1}}  \\ 
					 		    \frac{t (\zeta  \lambda  m-2)}{\lambda  ((\zeta -2) t+2)}  & m_{\text{cr,1}} < m < m_{\text{cr,2}}   \\
					 	          \frac{t (2+\zeta  \lambda  m)}{\lambda  ((\zeta -2) t-2)}  & m>m_{\text{cr,2}} . \end{cases} 
				\label{eq:rhoSuNF1full}
				\end{equation}
			
			\subsubsection*{${F=1}$, ${\vert t \vert  \to \infty}$}

				\label{eq:rhoSuNF1CSinf}
				\begin{equation}
					A = \begin{cases}  \frac{m \zeta \lambda + 2 }{(2-\zeta ) \lambda }   &  m< \lambda^{-1}  \\ 
					 	     \frac{2 +\lambda m \zeta }{\lambda(2-\zeta)} &  \lambda^{-1} < m < - \lambda^{-1}  \\
					 	     \frac{2-m \zeta \lambda}{(2+\zeta)\lambda }  & m>- \lambda^{-1}  , \end{cases}  \qquad  B = \begin{cases} - \frac{m \zeta \lambda + 2 }{(2+\zeta ) \lambda }  &  m< \lambda^{-1}  \\ 
					 	 - \frac{2 - \lambda m \zeta }{\lambda(2-\zeta)}  & \lambda^{-1} < m < - \lambda^{-1}  \\
					 	     \frac{m \zeta \lambda - 2 }{(2-\zeta ) \lambda } &  m>- \lambda^{-1}  . \end{cases}
				\end{equation}

				\subsubsection*{${F=2}$, ${\vert \lambda \vert  \to \infty}$. Symmetric case}
				
				\begin{equation}
					A = \begin{cases} \frac{2 \zeta  m t}{t-1} & m< 0  \\ 
					 	       -\frac{2 \zeta  m t}{t-1} & m>0 . \end{cases}   \ \qquad \ 
					 B = \begin{cases} -\frac{2 \zeta  m t}{t+1}   & m< 0  \\ 
					 	     \frac{2 \zeta  m t}{t+1}  & m>0 . \end{cases}
				\label{eq:rhoSuNF2sYMinf}
				\end{equation}

				\subsubsection*{${F=2}$, ${\vert \lambda \vert  < \infty}$. Symmetric case}

				\begin{equation}
				\label{eq:rhoSuNF2sfull}
					A = \begin{cases}  \frac{t \left(\frac{1}{\lambda }-\zeta  m\right)}{t-1}  & {\scriptstyle m> m_{\text{cr},1} } \\ 
					 	              -\frac{t}{\lambda +(\zeta -1) \lambda  t}  & {\scriptstyle m_{\text{cr},2} < m < m_{\text{cr},1} \text{ and } t\lambda <0 } \\
					 	         	 \frac{t \left(\frac{1}{\lambda }-\zeta  m\right)}{t-1}  & {\scriptstyle m_{\text{cr},2} < m < m_{\text{cr},1} \text{ and } t\lambda > 0   } \\
					 	             -\frac{t}{\lambda +(\zeta -1) \lambda  t}   & {\scriptstyle m< m_{\text{cr},2} }, \end{cases}  \qquad  B = \begin{cases}  \frac{t (\zeta  \lambda  m-1)}{\lambda  (t+1)}  & {\scriptstyle  m > m_{\text{cr},1}  } \\ 
					 		     	 \frac{t (\zeta  \lambda  m-1)}{\lambda  (t+1)}  & {\scriptstyle m_{\text{cr},2} < m < m_{\text{cr},1} \text{ and }  t\lambda <0  } \\
					 		      	 -\frac{t}{\lambda (1+ t (1- \zeta) )}   & {\scriptstyle m_{\text{cr},2} < m < m_{\text{cr},1} \text{ and } t\lambda > 0 } \\
					 	         	 -\frac{t}{\lambda -\zeta  \lambda  t+\lambda  t}   &  {\scriptstyle m< m_{\text{cr},2} } . \end{cases}
				\end{equation}

		\end{appendix}

	\bibliography{phases5d,5dlargeN,5dLoc,3dDecompactification}

\providecommand{\href}[2]{#2}\begingroup\raggedright\begin{thebibliography}{10}

\bibitem{Seiberg96}
N.~Seiberg, \emph{{Five-dimensional SUSY field theories, nontrivial fixed
  points and string dynamics}},
  \href{https://doi.org/10.1016/S0370-2693(96)01215-4}{\emph{Phys. Lett. B}
  {\bfseries 388} (1996) 753}
  [\href{https://arxiv.org/abs/hep-th/9608111}{{\ttfamily hep-th/9608111}}].

\bibitem{Cordova:2016xhm}
C.~Cordova, T.~T. Dumitrescu and K.~Intriligator, \emph{{Deformations of
  Superconformal Theories}},
  \href{https://doi.org/10.1007/JHEP11(2016)135}{\emph{JHEP} {\bfseries 11}
  (2016) 135} [\href{https://arxiv.org/abs/1602.01217}{{\ttfamily
  1602.01217}}].

\bibitem{Chang:2018xmx}
C.-M. Chang, \emph{{5d and 6d SCFTs Have No Weak Coupling Limit}},
  \href{https://doi.org/10.1007/JHEP09(2019)016}{\emph{JHEP} {\bfseries 09}
  (2019) 016} [\href{https://arxiv.org/abs/1810.04169}{{\ttfamily
  1810.04169}}].

\bibitem{Witten}
E.~Witten, \emph{{Phase transitions in M theory and F theory}},
  \href{https://doi.org/10.1016/0550-3213(96)00212-X}{\emph{Nucl. Phys. B}
  {\bfseries 471} (1996) 195}
  [\href{https://arxiv.org/abs/hep-th/9603150}{{\ttfamily hep-th/9603150}}].

\bibitem{Morrison:1996xf}
D.~R. Morrison and N.~Seiberg, \emph{{Extremal transitions and five-dimensional
  supersymmetric field theories}},
  \href{https://doi.org/10.1016/S0550-3213(96)00592-5}{\emph{Nucl. Phys. B}
  {\bfseries 483} (1997) 229}
  [\href{https://arxiv.org/abs/hep-th/9609070}{{\ttfamily hep-th/9609070}}].

\bibitem{Intriligator:1997pq}
K.~A. Intriligator, D.~R. Morrison and N.~Seiberg, \emph{{Five-dimensional
  supersymmetric gauge theories and degenerations of Calabi-Yau spaces}},
  \href{https://doi.org/10.1016/S0550-3213(97)00279-4}{\emph{Nucl. Phys. B}
  {\bfseries 497} (1997) 56}
  [\href{https://arxiv.org/abs/hep-th/9702198}{{\ttfamily hep-th/9702198}}].

\bibitem{Hayashi:2014kca}
H.~Hayashi, C.~Lawrie, D.~R. Morrison and S.~Schafer-Nameki, \emph{{Box Graphs
  and Singular Fibers}},
  \href{https://doi.org/10.1007/JHEP05(2014)048}{\emph{JHEP} {\bfseries 05}
  (2014) 048} [\href{https://arxiv.org/abs/1402.2653}{{\ttfamily 1402.2653}}].

\bibitem{FF0}
F.~Apruzzi, C.~Lawrie, L.~Lin, S.~Sch\"afer-Nameki and Y.-N. Wang, \emph{{5d
  Superconformal Field Theories and Graphs}},
  \href{https://doi.org/10.1016/j.physletb.2019.135077}{\emph{Phys. Lett. B}
  {\bfseries 800} (2020) 135077}
  [\href{https://arxiv.org/abs/1906.11820}{{\ttfamily 1906.11820}}].

\bibitem{a}
C.~Closset, M.~Del~Zotto and V.~Saxena, \emph{{Five-dimensional SCFTs and gauge
  theory phases: an M-theory/type IIA perspective}},
  \href{https://doi.org/10.21468/SciPostPhys.6.5.052}{\emph{SciPost Phys.}
  {\bfseries 6} (2019) 052} [\href{https://arxiv.org/abs/1812.10451}{{\ttfamily
  1812.10451}}].

\bibitem{FF1}
F.~Apruzzi, C.~Lawrie, L.~Lin, S.~Schäfer-Nameki and Y.-N. Wang, \emph{{Fibers
  add Flavor, Part I: Classification of 5d SCFTs, Flavor Symmetries and BPS
  States}}, \href{https://doi.org/10.1007/JHEP11(2019)068}{\emph{JHEP}
  {\bfseries 11} (2019) 068}
  [\href{https://arxiv.org/abs/1907.05404}{{\ttfamily 1907.05404}}].

\bibitem{FF2}
F.~Apruzzi, C.~Lawrie, L.~Lin, S.~Schäfer-Nameki and Y.-N. Wang, \emph{{Fibers
  add Flavor, Part II: 5d SCFTs, Gauge Theories, and Dualities}},
  \href{https://doi.org/10.1007/JHEP03(2020)052}{\emph{JHEP} {\bfseries 03}
  (2020) 052} [\href{https://arxiv.org/abs/1909.09128}{{\ttfamily
  1909.09128}}].

\bibitem{Bhardwaj:2020gyu}
L.~Bhardwaj and G.~Zafrir, \emph{{Classification of 5d N=1 gauge theories}},
  \href{https://doi.org/10.1007/JHEP12(2020)099}{\emph{JHEP} {\bfseries 12}
  (2020) 099} [\href{https://arxiv.org/abs/2003.04333}{{\ttfamily
  2003.04333}}].

\bibitem{Eckhard:2020jyr}
J.~Eckhard, S.~Sch\"afer-Nameki and Y.-N. Wang, \emph{{Trifectas for T$_{N}$ in
  5d}}, \href{https://doi.org/10.1007/JHEP07(2020)199}{\emph{JHEP} {\bfseries
  07} (2020) 199} [\href{https://arxiv.org/abs/2004.15007}{{\ttfamily
  2004.15007}}].

\bibitem{c}
A.~Collinucci and R.~Valandro, \emph{{The role of U(1)'s in 5d theories, Higgs
  branches, and geometry}},
  \href{https://doi.org/10.1007/JHEP10(2020)178}{\emph{JHEP} {\bfseries 10}
  (2020) 178} [\href{https://arxiv.org/abs/2006.15464}{{\ttfamily
  2006.15464}}].

\bibitem{Bhardwaj:2020avz}
L.~Bhardwaj, \emph{{Flavor symmetry of 5$d$ SCFTs. Part II. Applications}},
  \href{https://doi.org/10.1007/JHEP04(2021)221}{\emph{JHEP} {\bfseries 04}
  (2021) 221} [\href{https://arxiv.org/abs/2010.13235}{{\ttfamily
  2010.13235}}].

\bibitem{Pestun:2016jze}
V.~Pestun and M.~Zabzine, \emph{{Introduction to localization in quantum field
  theory}}, \href{https://doi.org/10.1088/1751-8121/aa5704}{\emph{J. Phys. A}
  {\bfseries 50} (2017) 443001}
  [\href{https://arxiv.org/abs/1608.02953}{{\ttfamily 1608.02953}}].

\bibitem{KZ}
J.~Källén and M.~Zabzine, \emph{{Twisted supersymmetric 5D Yang-Mills theory
  and contact geometry}},
  \href{https://doi.org/10.1007/JHEP05(2012)125}{\emph{JHEP} {\bfseries 05}
  (2012) 125} [\href{https://arxiv.org/abs/1202.1956}{{\ttfamily 1202.1956}}].

\bibitem{KQZ}
J.~Källén, J.~Qiu and M.~Zabzine, \emph{{The perturbative partition function
  of supersymmetric 5D Yang-Mills theory with matter on the five-sphere}},
  \href{https://doi.org/10.1007/JHEP08(2012)157}{\emph{JHEP} {\bfseries 08}
  (2012) 157} [\href{https://arxiv.org/abs/1206.6008}{{\ttfamily 1206.6008}}].

\bibitem{Hosomichi:2012ek}
K.~Hosomichi, R.-K. Seong and S.~Terashima, \emph{{Supersymmetric Gauge
  Theories on the Five-Sphere}},
  \href{https://doi.org/10.1016/j.nuclphysb.2012.08.007}{\emph{Nucl. Phys. B}
  {\bfseries 865} (2012) 376}
  [\href{https://arxiv.org/abs/1203.0371}{{\ttfamily 1203.0371}}].

\bibitem{Kim2}
H.-C. Kim and S.~Kim, \emph{{M5-branes from gauge theories on the 5-sphere}},
  \href{https://doi.org/10.1007/JHEP05(2013)144}{\emph{JHEP} {\bfseries 05}
  (2013) 144} [\href{https://arxiv.org/abs/1206.6339}{{\ttfamily 1206.6339}}].

\bibitem{Lockhart:2012vp}
G.~Lockhart and C.~Vafa, \emph{{Superconformal Partition Functions and
  Non-perturbative Topological Strings}},
  \href{https://doi.org/10.1007/JHEP10(2018)051}{\emph{JHEP} {\bfseries 10}
  (2018) 051} [\href{https://arxiv.org/abs/1210.5909}{{\ttfamily 1210.5909}}].

\bibitem{Mezei:2018url}
M.~Mezei, S.~S. Pufu and Y.~Wang, \emph{{Chern-Simons theory from M5-branes and
  calibrated M2-branes}},
  \href{https://doi.org/10.1007/JHEP08(2019)165}{\emph{JHEP} {\bfseries 08}
  (2019) 165} [\href{https://arxiv.org/abs/1812.07572}{{\ttfamily
  1812.07572}}].

\bibitem{Mreview}
J.~A. Minahan, \emph{{Matrix models for 5d super Yang--Mills}},
  \href{https://doi.org/10.1088/1751-8121/aa5cbe}{\emph{J. Phys. A} {\bfseries
  50} (2017) 443015} [\href{https://arxiv.org/abs/1608.02967}{{\ttfamily
  1608.02967}}].

\bibitem{JafferisPufu}
D.~L. Jafferis and S.~S. Pufu, \emph{{Exact results for five-dimensional
  superconformal field theories with gravity duals}},
  \href{https://doi.org/10.1007/JHEP05(2014)032}{\emph{JHEP} {\bfseries 05}
  (2014) 032} [\href{https://arxiv.org/abs/1207.4359}{{\ttfamily 1207.4359}}].

\bibitem{Kallen:2012zn}
J.~Källén, J.~A. Minahan, A.~Nedelin and M.~Zabzine, \emph{{$N^3$-behavior
  from 5D Yang-Mills theory}},
  \href{https://doi.org/10.1007/JHEP10(2012)184}{\emph{JHEP} {\bfseries 10}
  (2012) 184} [\href{https://arxiv.org/abs/1207.3763}{{\ttfamily 1207.3763}}].

\bibitem{Minahan:2013jwa}
J.~A. Minahan, A.~Nedelin and M.~Zabzine, \emph{{5D super Yang-Mills theory and
  the correspondence to AdS$_7$/CFT$_6$}},
  \href{https://doi.org/10.1088/1751-8113/46/35/355401}{\emph{J. Phys. A}
  {\bfseries 46} (2013) 355401}
  [\href{https://arxiv.org/abs/1304.1016}{{\ttfamily 1304.1016}}].

\bibitem{Assel:2012nf}
B.~Assel, J.~Estes and M.~Yamazaki, \emph{{Wilson Loops in 5d N=1 SCFTs and
  AdS/CFT}}, \href{https://doi.org/10.1007/s00023-013-0249-5}{\emph{Annales
  Henri Poincar\'{e}} {\bfseries 15} (2014) 589}
  [\href{https://arxiv.org/abs/1212.1202}{{\ttfamily 1212.1202}}].

\bibitem{Giasemidis:2013oea}
G.~Giasemidis, R.~J. Szabo and M.~Tierz, \emph{{Supersymmetric gauge theories,
  Coulomb gases and Chern-Simons matrix models}},
  \href{https://doi.org/10.1103/PhysRevD.89.025016}{\emph{Phys. Rev. D}
  {\bfseries 89} (2014) 025016}
  [\href{https://arxiv.org/abs/1310.3122}{{\ttfamily 1310.3122}}].

\bibitem{Chang:2017mxc}
C.-M. Chang, M.~Fluder, Y.-H. Lin and Y.~Wang, \emph{{Romans Supergravity from
  Five-Dimensional Holograms}},
  \href{https://doi.org/10.1007/JHEP05(2018)039}{\emph{JHEP} {\bfseries 05}
  (2018) 039} [\href{https://arxiv.org/abs/1712.10313}{{\ttfamily
  1712.10313}}].

\bibitem{Fluder:2018chf}
M.~Fluder and C.~F. Uhlemann, \emph{{Precision Test of AdS$_6$/CFT$_5$ in Type
  IIB String Theory}},
  \href{https://doi.org/10.1103/PhysRevLett.121.171603}{\emph{Phys. Rev. Lett.}
  {\bfseries 121} (2018) 171603}
  [\href{https://arxiv.org/abs/1806.08374}{{\ttfamily 1806.08374}}].

\bibitem{Crichigno:2018adf}
P.~M. Crichigno, D.~Jain and B.~Willett, \emph{{5d Partition Functions with A
  Twist}}, \href{https://doi.org/10.1007/JHEP11(2018)058}{\emph{JHEP}
  {\bfseries 11} (2018) 058}
  [\href{https://arxiv.org/abs/1808.06744}{{\ttfamily 1808.06744}}].

\bibitem{Uhlemann}
C.~F. Uhlemann, \emph{{Exact results for 5d SCFTs of long quiver type}},
  \href{https://doi.org/10.1007/JHEP11(2019)072}{\emph{JHEP} {\bfseries 11}
  (2019) 072} [\href{https://arxiv.org/abs/1909.01369}{{\ttfamily
  1909.01369}}].

\bibitem{UhlemannWL}
C.~F. Uhlemann, \emph{{Wilson loops in 5d long quiver gauge theories}},
  \href{https://doi.org/10.1007/JHEP09(2020)145}{\emph{JHEP} {\bfseries 09}
  (2020) 145} [\href{https://arxiv.org/abs/2006.01142}{{\ttfamily
  2006.01142}}].

\bibitem{Minahan:2014hwa}
J.~A. Minahan and A.~Nedelin, \emph{{Phases of planar 5-dimensional
  supersymmetric Chern-Simons theory}},
  \href{https://doi.org/10.1007/JHEP12(2014)049}{\emph{JHEP} {\bfseries 12}
  (2014) 049} [\href{https://arxiv.org/abs/1408.2767}{{\ttfamily 1408.2767}}].

\bibitem{Minahan:2020ifb}
J.~A. Minahan and A.~Nedelin, \emph{{Five-dimensional gauge theories on spheres
  with negative couplings}},
  \href{https://doi.org/10.1007/JHEP02(2021)102}{\emph{JHEP} {\bfseries 02}
  (2021) 102} [\href{https://arxiv.org/abs/2007.13760}{{\ttfamily
  2007.13760}}].

\bibitem{Russo:2012ay}
J.~Russo and K.~Zarembo, \emph{{Large N Limit of N=2 SU(N) Gauge Theories from
  Localization}}, \href{https://doi.org/10.1007/JHEP10(2012)082}{\emph{JHEP}
  {\bfseries 10} (2012) 082} [\href{https://arxiv.org/abs/1207.3806}{{\ttfamily
  1207.3806}}].

\bibitem{Russo:2013qaa}
J.~G. Russo and K.~Zarembo, \emph{{Evidence for Large-N Phase Transitions in
  N=2* Theory}}, \href{https://doi.org/10.1007/JHEP04(2013)065}{\emph{JHEP}
  {\bfseries 04} (2013) 065} [\href{https://arxiv.org/abs/1302.6968}{{\ttfamily
  1302.6968}}].

\bibitem{Russo:2013kea}
J.~Russo and K.~Zarembo, \emph{{Massive N=2 Gauge Theories at Large N}},
  \href{https://doi.org/10.1007/JHEP11(2013)130}{\emph{JHEP} {\bfseries 11}
  (2013) 130} [\href{https://arxiv.org/abs/1309.1004}{{\ttfamily 1309.1004}}].

\bibitem{Chen:2014vka}
X.~Chen-Lin, J.~Gordon and K.~Zarembo, \emph{{$ \mathcal{N}={2}^{\ast } $
  super-Yang-Mills theory at strong coupling}},
  \href{https://doi.org/10.1007/JHEP11(2014)057}{\emph{JHEP} {\bfseries 11}
  (2014) 057} [\href{https://arxiv.org/abs/1408.6040}{{\ttfamily 1408.6040}}].

\bibitem{Zarembo:2014ooa}
K.~Zarembo, \emph{{Strong-Coupling Phases of Planar N=2* Super-Yang-Mills
  Theory}}, \href{https://doi.org/10.1007/s11232-014-0232-4}{\emph{Theor. Math.
  Phys.} {\bfseries 181} (2014) 1522}
  [\href{https://arxiv.org/abs/1410.6114}{{\ttfamily 1410.6114}}].

\bibitem{Chen-Lin:2015dfa}
X.~Chen-Lin and K.~Zarembo, \emph{{Higher Rank Wilson Loops in N = 2*
  Super-Yang-Mills Theory}},
  \href{https://doi.org/10.1007/JHEP03(2015)147}{\emph{JHEP} {\bfseries 03}
  (2015) 147} [\href{https://arxiv.org/abs/1502.01942}{{\ttfamily
  1502.01942}}].

\bibitem{Hollowood:2015oma}
T.~J. Hollowood and S.~P. Kumar, \emph{{Partition function of $
  \mathcal{N}={2}^{\ast } $ SYM on a large four-sphere}},
  \href{https://doi.org/10.1007/JHEP12(2015)016}{\emph{JHEP} {\bfseries 12}
  (2015) 016} [\href{https://arxiv.org/abs/1509.00716}{{\ttfamily
  1509.00716}}].

\bibitem{Russo:2019ipg}
J.~G. Russo, \emph{{Properties of the partition function of ${\cal N}=2$
  supersymmetric QCD with massive matter}},
  \href{https://doi.org/10.1007/JHEP07(2019)125}{\emph{JHEP} {\bfseries 07}
  (2019) 125} [\href{https://arxiv.org/abs/1905.05267}{{\ttfamily
  1905.05267}}].

\bibitem{Barranco}
A.~Barranco and J.~G. Russo, \emph{{Large N phase transitions in supersymmetric
  Chern-Simons theory with massive matter}},
  \href{https://doi.org/10.1007/JHEP03(2014)012}{\emph{JHEP} {\bfseries 03}
  (2014) 012} [\href{https://arxiv.org/abs/1401.3672}{{\ttfamily 1401.3672}}].

\bibitem{Russo:2014bda}
J.~G. Russo, G.~A. Silva and M.~Tierz, \emph{{Supersymmetric U(N)
  Chern\textendash{}Simons-Matter Theory and Phase Transitions}},
  \href{https://doi.org/10.1007/s00220-015-2399-4}{\emph{Commun. Math. Phys.}
  {\bfseries 338} (2015) 1411}
  [\href{https://arxiv.org/abs/1407.4794}{{\ttfamily 1407.4794}}].

\bibitem{Anderson:2014hxa}
L.~Anderson and K.~Zarembo, \emph{{Quantum Phase Transitions in Mass-Deformed
  ABJM Matrix Model}},
  \href{https://doi.org/10.1007/JHEP09(2014)021}{\emph{JHEP} {\bfseries 09}
  (2014) 021} [\href{https://arxiv.org/abs/1406.3366}{{\ttfamily 1406.3366}}].

\bibitem{Anderson:2015ioa}
L.~Anderson and J.~G. Russo, \emph{{ABJM Theory with mass and FI deformations
  and Quantum Phase Transitions}},
  \href{https://doi.org/10.1007/JHEP05(2015)064}{\emph{JHEP} {\bfseries 05}
  (2015) 064} [\href{https://arxiv.org/abs/1502.06828}{{\ttfamily
  1502.06828}}].

\bibitem{Anderson:2017xrv}
L.~Anderson and N.~Drukker, \emph{{More Large $N$ limits of 3d gauge
  theories}}, \href{https://doi.org/10.1088/1751-8121/aa7e11}{\emph{J. Phys. A}
  {\bfseries 50} (2017) 345401}
  [\href{https://arxiv.org/abs/1701.04409}{{\ttfamily 1701.04409}}].

\bibitem{STWL}
L.~Santilli and M.~Tierz, \emph{{Phase transitions and Wilson loops in
  antisymmetric representations in Chern--Simons-matter theory}},
  \href{https://doi.org/10.1088/1751-8121/ab335c}{\emph{J. Phys. A} {\bfseries
  52} (2019) 385401} [\href{https://arxiv.org/abs/1808.02855}{{\ttfamily
  1808.02855}}].

\bibitem{Nedelin}
A.~Nedelin, \emph{{Phase transitions in 5D super Yang-Mills theory}},
  \href{https://doi.org/10.1007/JHEP07(2015)004}{\emph{JHEP} {\bfseries 07}
  (2015) 004} [\href{https://arxiv.org/abs/1502.07275}{{\ttfamily
  1502.07275}}].

\bibitem{Pan:2017zie}
Y.~Pan and W.~Peelaers, \emph{{Chiral Algebras, Localization and Surface
  Defects}}, \href{https://doi.org/10.1007/JHEP02(2018)138}{\emph{JHEP}
  {\bfseries 02} (2018) 138}
  [\href{https://arxiv.org/abs/1710.04306}{{\ttfamily 1710.04306}}].

\bibitem{Wang:2020seq}
Y.~Wang, \emph{{Taming defects in $ \mathcal{N} $ = 4 super-Yang-Mills}},
  \href{https://doi.org/10.1007/JHEP08(2020)021}{\emph{JHEP} {\bfseries 08}
  (2020) 021} [\href{https://arxiv.org/abs/2003.11016}{{\ttfamily
  2003.11016}}].

\bibitem{Qiu:2013pta}
J.~Qiu and M.~Zabzine, \emph{{5D Super Yang-Mills on $Y^{p,q}$ Sasaki-Einstein
  manifolds}}, \href{https://doi.org/10.1007/s00220-014-2194-7}{\emph{Commun.
  Math. Phys.} {\bfseries 333} (2015) 861}
  [\href{https://arxiv.org/abs/1307.3149}{{\ttfamily 1307.3149}}].

\bibitem{SST}
L.~Santilli, R.~J. Szabo and M.~Tierz, \emph{{Five-dimensional cohomological
  localization and squashed $q$-deformations of two-dimensional Yang-Mills
  theory}}, \href{https://doi.org/10.1007/JHEP06(2020)036}{\emph{JHEP}
  {\bfseries 20} (2020) 036}
  [\href{https://arxiv.org/abs/2003.09411}{{\ttfamily 2003.09411}}].

\bibitem{Minahan:2015any}
J.~A. Minahan, \emph{{Localizing gauge theories on $S^{d}$}},
  \href{https://doi.org/10.1007/JHEP04(2016)152}{\emph{JHEP} {\bfseries 04}
  (2016) 152} [\href{https://arxiv.org/abs/1512.06924}{{\ttfamily
  1512.06924}}].

\bibitem{Gorantis:2017vzz}
A.~Gorantis, J.~A. Minahan and U.~Naseer, \emph{{Analytic continuation of
  dimensions in supersymmetric localization}},
  \href{https://doi.org/10.1007/JHEP02(2018)070}{\emph{JHEP} {\bfseries 02}
  (2018) 070} [\href{https://arxiv.org/abs/1711.05669}{{\ttfamily
  1711.05669}}].

\bibitem{Jefferson:2017ahm}
P.~Jefferson, H.-C. Kim, C.~Vafa and G.~Zafrir, \emph{{Towards Classification
  of 5d SCFTs: Single Gauge Node}},
  \href{https://arxiv.org/abs/1705.05836}{{\ttfamily 1705.05836}}.

\bibitem{Klebanov:2011gs}
I.~R. Klebanov, S.~S. Pufu and B.~R. Safdi, \emph{{F-Theorem without
  Supersymmetry}}, \href{https://doi.org/10.1007/JHEP10(2011)038}{\emph{JHEP}
  {\bfseries 10} (2011) 038} [\href{https://arxiv.org/abs/1105.4598}{{\ttfamily
  1105.4598}}].

\bibitem{Chang:2017cdx}
C.-M. Chang, M.~Fluder, Y.-H. Lin and Y.~Wang, \emph{{Spheres, Charges,
  Instantons, and Bootstrap: A Five-Dimensional Odyssey}},
  \href{https://doi.org/10.1007/JHEP03(2018)123}{\emph{JHEP} {\bfseries 03}
  (2018) 123} [\href{https://arxiv.org/abs/1710.08418}{{\ttfamily
  1710.08418}}].

\bibitem{Fluder:2020pym}
M.~Fluder and C.~F. Uhlemann, \emph{{Evidence for a 5d F-theorem}},
  \href{https://doi.org/10.1007/JHEP02(2021)192}{\emph{JHEP} {\bfseries 02}
  (2021) 192} [\href{https://arxiv.org/abs/2011.00006}{{\ttfamily
  2011.00006}}].

\bibitem{Gross:1980he}
D.~J. Gross and E.~Witten, \emph{{Possible Third Order Phase Transition in the
  Large N Lattice Gauge Theory}},
  \href{https://doi.org/10.1103/PhysRevD.21.446}{\emph{Phys. Rev. D} {\bfseries
  21} (1980) 446}.

\bibitem{Wadia:1980cp}
S.~R. Wadia, \emph{{$N$ = Infinity Phase Transition in a Class of Exactly
  Soluble Model Lattice Gauge Theories}},
  \href{https://doi.org/10.1016/0370-2693(80)90353-6}{\emph{Phys. Lett. B}
  {\bfseries 93} (1980) 403}.

\bibitem{Wadia:2012fr}
S.~R. Wadia, \emph{{A Study of U(N) Lattice Gauge Theory in 2-dimensions}},
  \href{https://arxiv.org/abs/1212.2906}{{\ttfamily 1212.2906}}.

\bibitem{Neuberger:1989kd}
H.~Neuberger, \emph{{Scaling Regime at the Large $N$ Phase Transition of
  Two-dimensional Pure Gauge Theories}},
  \href{https://doi.org/10.1016/0550-3213(90)90465-P}{\emph{Nucl. Phys. B}
  {\bfseries 340} (1990) 703}.

\bibitem{Hartnoll:2006}
S.~A. Hartnoll and S.~Kumar, \emph{{Higher rank Wilson loops from a matrix
  model}}, \href{https://doi.org/10.1088/1126-6708/2006/08/026}{\emph{JHEP}
  {\bfseries 08} (2006) 026}
  [\href{https://arxiv.org/abs/hep-th/0605027}{{\ttfamily hep-th/0605027}}].

\bibitem{Russo:2017ngf}
J.~G. Russo and K.~Zarembo, \emph{{Wilson loops in antisymmetric
  representations from localization in supersymmetric gauge theories}},
  \href{https://doi.org/10.1142/S0129055X18400147}{\emph{Rev. Math. Phys.}
  {\bfseries 30} (2018) 1840014}
  [\href{https://arxiv.org/abs/1712.07186}{{\ttfamily 1712.07186}}].

\bibitem{Santilli:2020ueh}
L.~Santilli and M.~Tierz, \emph{{Exact equivalences and phase discrepancies
  between random matrix ensembles}},
  \href{https://doi.org/10.1088/1742-5468/aba594}{\emph{J. Stat. Mech.}
  {\bfseries 2008} (2020) 083107}
  [\href{https://arxiv.org/abs/2003.10475}{{\ttfamily 2003.10475}}].

\bibitem{Morrison:2020ool}
D.~R. Morrison, S.~Schafer-Nameki and B.~Willett, \emph{{Higher-Form Symmetries
  in 5d}}, \href{https://doi.org/10.1007/JHEP09(2020)024}{\emph{JHEP}
  {\bfseries 09} (2020) 024}
  [\href{https://arxiv.org/abs/2005.12296}{{\ttfamily 2005.12296}}].

\bibitem{Albertini:2020mdx}
F.~Albertini, M.~Del~Zotto, I.~Garc\'{i}a~Etxebarria and S.~S. Hosseini,
  \emph{{Higher Form Symmetries and M-theory}},
  \href{https://doi.org/10.1007/JHEP12(2020)203}{\emph{JHEP} {\bfseries 12}
  (2020) 203} [\href{https://arxiv.org/abs/2005.12831}{{\ttfamily
  2005.12831}}].

\bibitem{BenettiGenolini:2020doj}
P.~Benetti~Genolini and L.~Tizzano, \emph{{Instantons, symmetries and anomalies
  in five dimensions}},
  \href{https://doi.org/10.1007/JHEP04(2021)188}{\emph{JHEP} {\bfseries 04}
  (2021) 188} [\href{https://arxiv.org/abs/2009.07873}{{\ttfamily
  2009.07873}}].

\bibitem{Bergman:2013koa}
O.~Bergman, D.~Rodr\'{i}guez-G\'{o}mez and G.~Zafrir, \emph{{5d superconformal
  indices at large N and holography}},
  \href{https://doi.org/10.1007/JHEP08(2013)081}{\emph{JHEP} {\bfseries 08}
  (2013) 081} [\href{https://arxiv.org/abs/1305.6870}{{\ttfamily 1305.6870}}].

\bibitem{Hwang:2014uwa}
C.~Hwang, J.~Kim, S.~Kim and J.~Park, \emph{{General instanton counting and 5d
  SCFT}}, \href{https://doi.org/10.1007/JHEP07(2015)063}{\emph{JHEP} {\bfseries
  07} (2015) 063} [\href{https://arxiv.org/abs/1406.6793}{{\ttfamily
  1406.6793}}].

\bibitem{Bourget:2020gzi}
A.~Bourget, J.~F. Grimminger, A.~Hanany, M.~Sperling and Z.~Zhong,
  \emph{{Magnetic Quivers from Brane Webs with O5 Planes}},
  \href{https://doi.org/10.1007/JHEP07(2020)204}{\emph{JHEP} {\bfseries 07}
  (2020) 204} [\href{https://arxiv.org/abs/2004.04082}{{\ttfamily
  2004.04082}}].

\bibitem{Bergman:2020myx}
O.~Bergman and D.~Rodr\'\i{}guez-G\'omez, \emph{{The Cat\textquoteright{}s
  Cradle: deforming the higher rank E$_{1}$ and $ {\tilde{E}}_1 $ theories}},
  \href{https://doi.org/10.1007/JHEP02(2021)122}{\emph{JHEP} {\bfseries 02}
  (2021) 122} [\href{https://arxiv.org/abs/2011.05125}{{\ttfamily
  2011.05125}}].

\bibitem{Li:2021rqr}
X.~Li and F.~Yagi, \emph{{Thermodynamic limit of Nekrasov partition function
  for 5-brane web with O5-plane}},
  \href{https://doi.org/10.1007/JHEP06(2021)004}{\emph{JHEP} {\bfseries 06}
  (2021) 004} [\href{https://arxiv.org/abs/2102.09482}{{\ttfamily
  2102.09482}}].

\bibitem{Coccia:2020cku}
L.~Coccia, \emph{{Topologically twisted index of $T[SU(N)]$ at large $N$}},
  \href{https://doi.org/10.1007/JHEP05(2021)264}{\emph{JHEP} {\bfseries 05}
  (2021) 264} [\href{https://arxiv.org/abs/2006.06578}{{\ttfamily
  2006.06578}}].

\bibitem{Coccia:2020wtk}
L.~Coccia and C.~F. Uhlemann, \emph{{On the planar limit of 3d
  $T_\rho^\sigma[SU(N)]$}},  \href{https://arxiv.org/abs/2011.10050}{{\ttfamily
  2011.10050}}.

\bibitem{Hayashi:2014hfa}
H.~Hayashi, Y.~Tachikawa and K.~Yonekura, \emph{{Mass-deformed T$_{N}$ as a
  linear quiver}}, \href{https://doi.org/10.1007/JHEP02(2015)089}{\emph{JHEP}
  {\bfseries 02} (2015) 089} [\href{https://arxiv.org/abs/1410.6868}{{\ttfamily
  1410.6868}}].

\bibitem{Gaiotto:2014ina}
D.~Gaiotto and H.-C. Kim, \emph{{Surface defects and instanton partition
  functions}}, \href{https://doi.org/10.1007/JHEP10(2016)012}{\emph{JHEP}
  {\bfseries 10} (2016) 012} [\href{https://arxiv.org/abs/1412.2781}{{\ttfamily
  1412.2781}}].

\bibitem{Ashok:2017bld}
S.~K. Ashok, M.~Bill\`{o}, E.~Dell'Aquila, M.~Frau, V.~Gupta, R.~R. John and
  A.~Lerda, \emph{{Surface operators in 5d gauge theories and duality
  relations}}, \href{https://doi.org/10.1007/JHEP05(2018)046}{\emph{JHEP}
  {\bfseries 05} (2018) 046}
  [\href{https://arxiv.org/abs/1712.06946}{{\ttfamily 1712.06946}}].

\bibitem{Gutperle:2020rty}
M.~Gutperle and C.~F. Uhlemann, \emph{{Surface defects in holographic 5d
  SCFTs}}, \href{https://doi.org/10.1007/JHEP04(2021)134}{\emph{JHEP}
  {\bfseries 04} (2021) 134}
  [\href{https://arxiv.org/abs/2012.14547}{{\ttfamily 2012.14547}}].

\bibitem{Esole:2015xfa}
M.~Esole and S.-H. Shao, \emph{{M-theory on Elliptic Calabi-Yau Threefolds and
  6d Anomalies}},  \href{https://arxiv.org/abs/1504.01387}{{\ttfamily
  1504.01387}}.

\bibitem{CollinucciValandro:3d}
A.~Collinucci and R.~Valandro, \emph{{A string theory realization of special
  unitary quivers in 3 dimensions}},
  \href{https://doi.org/10.1007/JHEP11(2020)157}{\emph{JHEP} {\bfseries 11}
  (2020) 157} [\href{https://arxiv.org/abs/2008.10689}{{\ttfamily
  2008.10689}}].

\bibitem{Hayashi:2019jvx}
H.~Hayashi, S.-S. Kim, K.~Lee and F.~Yagi, \emph{{Complete prepotential for 5d
  $ \mathcal{N} $ = 1 superconformal field theories}},
  \href{https://doi.org/10.1007/JHEP02(2020)074}{\emph{JHEP} {\bfseries 02}
  (2020) 074} [\href{https://arxiv.org/abs/1912.10301}{{\ttfamily
  1912.10301}}].

\bibitem{Crichigno:2020ouj}
P.~M. Crichigno and D.~Jain, \emph{{The 5d Superconformal Index at Large $N$
  and Black Holes}}, \href{https://doi.org/10.1007/JHEP09(2020)124}{\emph{JHEP}
  {\bfseries 09} (2020) 124}
  [\href{https://arxiv.org/abs/2005.00550}{{\ttfamily 2005.00550}}].

\end{thebibliography}\endgroup

\end{document}